\documentclass[letterpaper]{article} % DO NOT CHANGE THIS
\usepackage{aaai2026}  % DO NOT CHANGE THIS
\usepackage{times}  % DO NOT CHANGE THIS
\usepackage{helvet}  % DO NOT CHANGE THIS
\usepackage{courier}  % DO NOT CHANGE THIS
\usepackage[hyphens]{url}  % DO NOT CHANGE THIS
\usepackage{graphicx} % DO NOT CHANGE THIS
\usepackage{natbib}  % DO NOT CHANGE THIS AND DO NOT ADD ANY OPTIONS TO IT
\usepackage{caption} % DO NOT CHANGE THIS AND DO NOT ADD ANY OPTIONS TO IT
\usepackage{algorithm}
\usepackage{algorithmic}
\usepackage{amsmath, amssymb}

\usepackage{xcolor}       
\usepackage{multirow}     
\usepackage{array}        
\usepackage{dblfloatfix}   
\usepackage{subfig}
\usepackage{booktabs}
\usepackage[table]{xcolor}

\usepackage{newfloat}
\usepackage{listings}
\DeclareCaptionStyle{ruled}{labelfont=normalfont,labelsep=colon,strut=off} % DO NOT CHANGE THIS
\floatstyle{ruled}
\newfloat{listing}{tb}{lst}{}
\floatname{listing}{Listing}
\title{CometNet: Contextual Motif-guided Long-term Time Series Forecasting}
\author{
    Weixu Wang,
    Xiaobo Zhou\textsuperscript{\rm *},
    Xin Qiao,
    Lei Wang,
    Tie Qiu
}
\affiliations{
    School of Computer Science and Technology, Tianjin University, China\\

    \{weixuwang, xiaobo.zhou, xinqiao, wanglei2019\}@tju.edu.cn, qiutie@ieee.org\\
    \textsuperscript{\rm *}Corresponding Author 
}

\begin{document}

\maketitle

\begin{abstract}
Long-term Time Series Forecasting is crucial across numerous critical domains, yet its accuracy remains fundamentally constrained by the receptive field bottleneck in existing models. Mainstream Transformer- and Multi-layer Perceptron (MLP)-based methods mainly rely on finite look-back windows, limiting their ability to model long-term dependencies and hurting forecasting performance. Naively extending the look-back window proves ineffective, as it not only introduces prohibitive computational complexity, but also drowns vital long-term dependencies in historical noise. To address these challenges, we propose CometNet, a novel Contextual Motif-guided Long-term Time Series Forecasting framework. CometNet first introduces a Contextual Motif Extraction module that identifies recurrent, dominant contextual motifs from complex historical sequences, providing extensive temporal dependencies far exceeding limited look-back windows; Subsequently, a Motif-guided Forecasting module is proposed, which integrates the extracted dominant motifs into forecasting. By dynamically mapping the look-back window to its relevant motifs, CometNet effectively harnesses their contextual information to strengthen long-term forecasting capability. Extensive experimental results on eight real-world datasets have demonstrated that CometNet significantly outperforms current state-of-the-art (SOTA) methods, particularly on extended forecast horizons.
\end{abstract}

% Uncomment the following to link to your code, datasets, an extended version or similar.
% You must keep this block between (not within) the abstract and the main body of the paper.
% \begin{links}
%     \link{Code}{https://aaai.org/example/code}
%     \link{Datasets}{https://aaai.org/example/datasets}
%     \link{Extended version}{https://aaai.org/example/extended-version}
% \end{links}

\section{Introduction}
\label{sec1_introduction}
Time series forecasting (TSF) is a fundamental task in modern data science, driving critical decisions in domains ranging from meteorology~\cite{bi2023accurate}, energy management~\cite{balkanski2023energy} to transportation~\cite{fang2025efficient} and finance~\cite{arsenault2025survey}. With technological advancements, TSF methodologies have evolved from traditional statistical models~\cite{ariyo2014stock, bahdanau2015neural} to sophisticated deep learning architectures, including Transformers~\cite{zhang2023crossformer, nie2023patchtst} and MLPs~\cite{das2023tide, tang2025patchmlp}.
However, achieving accurate and robust long-term time series forecasting (LTSF) remains a persistent and formidable challenge.
\begin{figure}[h!]
\centering
    \subfloat[MSE across various forecast horizons (L = 96)]{\includegraphics[width=0.45\textwidth, height=0.12\textheight]{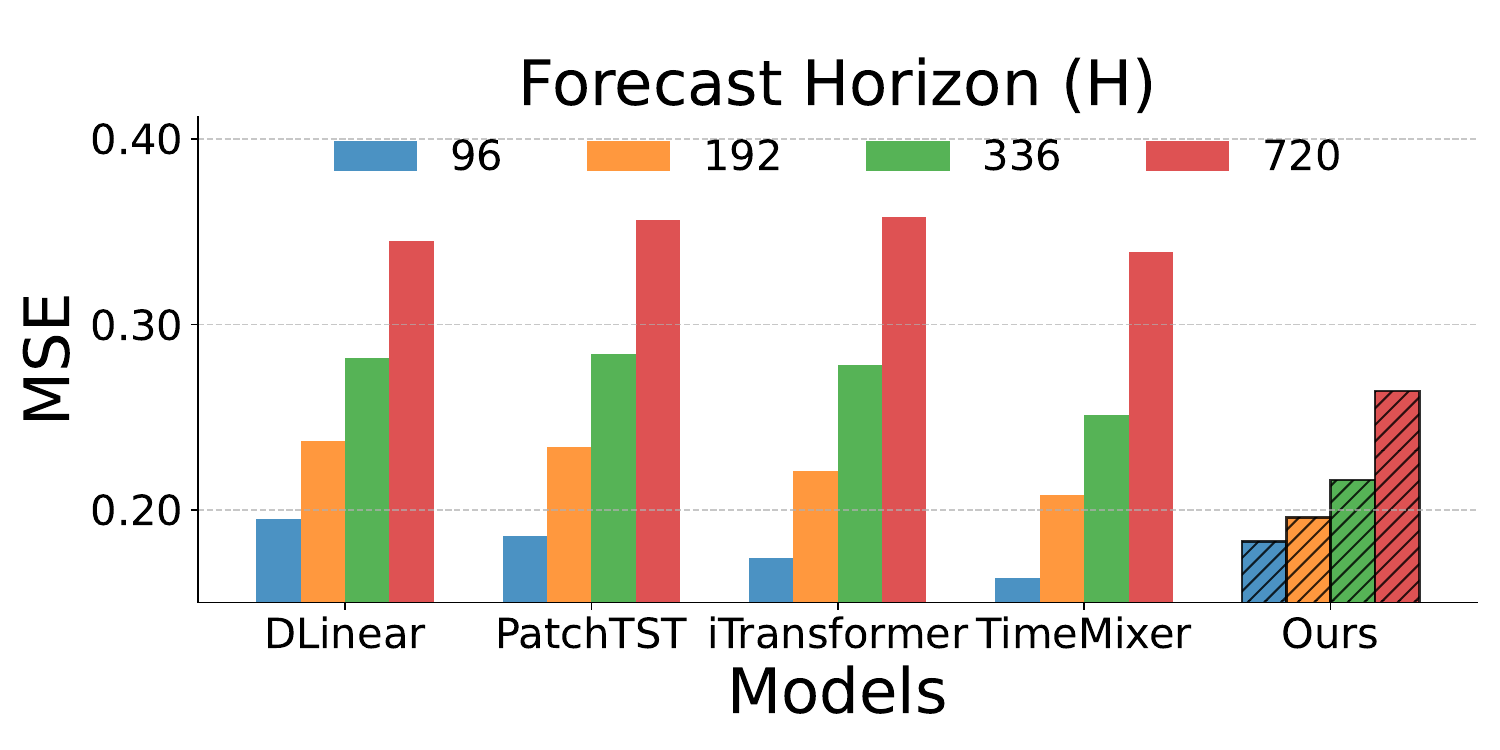}
    \label{fig1:forecast}
    }
\hfil
    \subfloat[MSE across various look-back windows (H = 720)]{\includegraphics[width=0.45\textwidth, height=0.12\textheight]{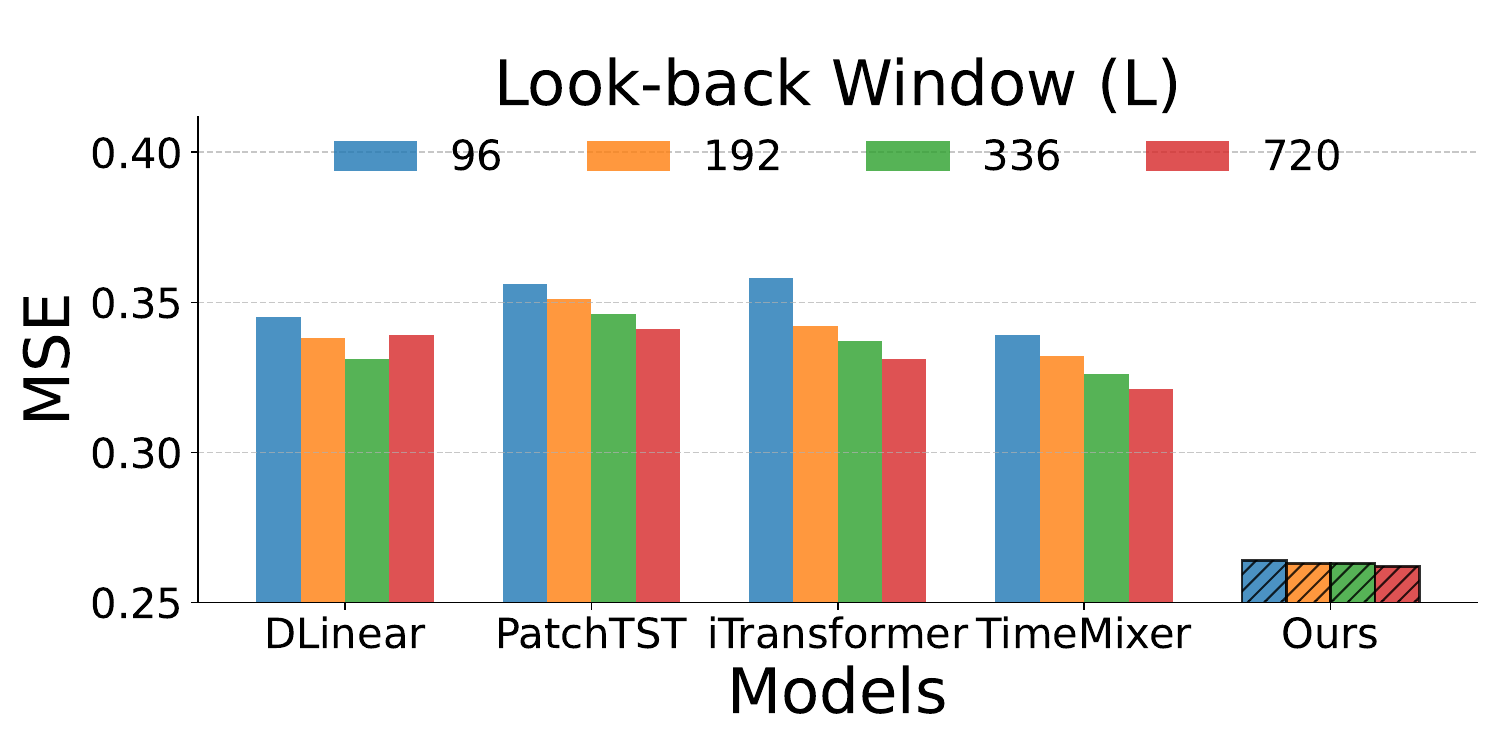}
    \label{fig1:look_back}
    }
\caption{The Receptive Field Bottleneck in LTSF.}
\label{fig1:problem}
\end{figure}

\begin{figure*}[ht]
    \centering
    \includegraphics[width=0.85\textwidth, height=0.17\textheight]{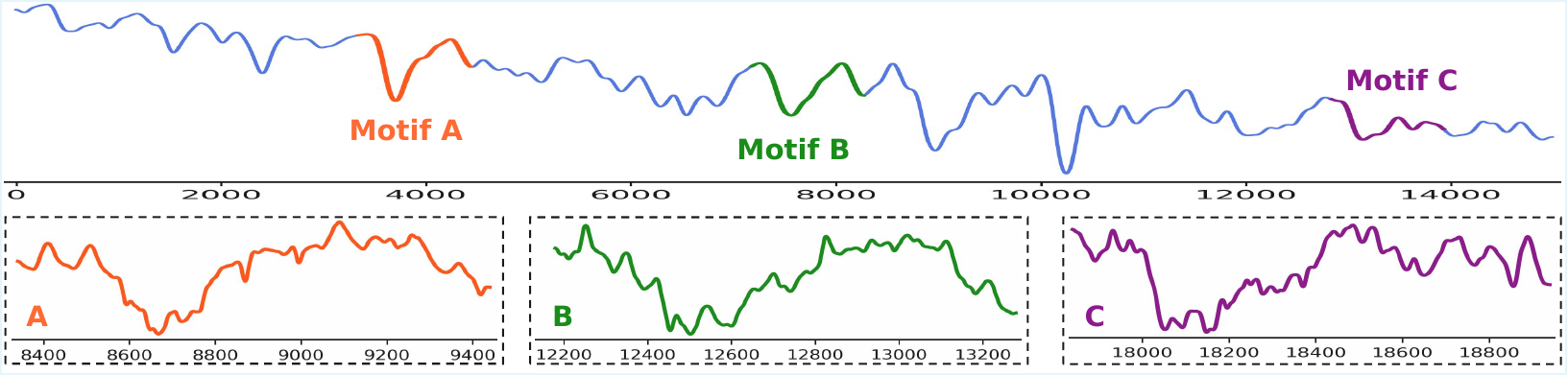}
    \caption{An example of a contextual motfi and its recurring instances in ETTm1 dataset.}
    \label{fig2:motivation}
\end{figure*}
While recent advances have introduced powerful models such as PatchTST~\cite{nie2023patchtst}, iTransformer~\cite{liu2024itransformer} and TimeMixer++~\cite{wang2025timemixer++}, they suffer from a fundamental architectural constraint: \textbf{the Receptive Field Bottleneck}. Specifically, contemporary models typically operate within finite-length look-back windows, attempting to predict future sequences from this limited temporal context~\cite{wu2023timesnet, chen2025probabilistic}. This design inherently fragments continuous time series into discrete, context-independent segments, thereby limiting the models' ability to learn long-term dependencies beyond the restricted look-back window~\cite{zeng2023transformers}. Although sliding windows theoretically traverse the entire time series during training, gradient backpropagation remains confined within individual windows~\cite{kang2024introducing}, preventing differentiable learning of long-term dependencies. Consequently, this bottleneck leads to severe performance degradation, as Fig.~\ref{fig1:forecast} shows, with a limited look-back window $(L=96)$, the models exhibit substantially growing prediction errors as the forecast horizon extends.

An intuitive approach is to extend the look-back window, enabling models to capture longer-range dependencies. While Fig.~\ref{fig1:look_back} confirms that enlarged windows yield improved prediction performance, it also suffers from exponential growth in computational complexity and training time~\cite{tan24llm}. Moreover, overly extended look-back windows can drown out meaningful temporal dependencies within irrelevant historical noise, leading to diminishing returns or even performance degradation~\cite{zeng2023transformers}. Although the recent approach like Batched Spectral Attention (BSA)~\cite{kang2024introducing} enhances long-term dependency modeling by preserving cross-sample temporal correlations, it still struggles to capture meaningful temporal dependencies across thousands of time steps. 

Fortunately, the evolution of real-world time series is not disordered. Instead, it is often governed by some recurring, representative long-context patterns~\cite{huang23pattern} that we term contextual motifs. As illustrated in Fig.~\ref{fig2:motivation}, these motifs can span thousands of time steps and reappear at distant points in the series, while maintaining similar temporal patterns. Their extended temporal horizons encode rich contextual information, like factory production cycles or seasonal climate shifts, which is vital for understanding long-term dynamics. Building on this insight, we argue that these contextual motifs are key to transcending the limitations of local look-back windows. By discovering and leveraging these motifs, we can provide the long-term temporal dependencies needed for long-term forecasting.

To operationalize this insight, we propose \textbf{CometNet}, a novel \textbf{CO}ntextual \textbf{M}o\textbf{t}if-guided \textbf{NET}work for Long-term Time Series Forecasting. CometNet introduces a new forecasting paradigm that integrates contextual motifs into the forecasting process, enabling accurate long-term predictions. The framework operates through two core modules. First, a \textbf{Contextual Motif Extraction} module is designed to analyze the entire historical time series and establish a comprehensive library of dominant contextual motifs. Second, we further develop a \textbf{Motif-guided Forecasting} module that dynamically identify look-back window's most relevant motif from the pre-established library and uses it to inform the prediction of the distant future. By doing so, our model effectively circumvents the receptive field bottleneck, grounding its predictions not just on the limited look-back window, but also on rich long-term contextual information covered by the dominant motifs. The main contributions of this paper are summarized as follows:

\begin{itemize}

\item We propose ComtNet, a novel contextual motif-guided network for long-term time series forecasting that leverages contextual motifs to overcome the receptive field bottleneck. 

\item We design an efficient contextual motif extraction module that systematically analyzes historical time series to establish a comprehensive library comprising dominant contextual motifs.

\item We develop a motif-guided forecasting module that leverages our pre-established motif library to inform future predictions.

\item Comprehensive experiments have been conducted across a broad spectrum of LTSF benchmarks, demonstrating the superiority of ComtNET over current SOTA methods.

\end{itemize}

\begin{figure*}[ht]
    \centering
    \includegraphics[width=0.95\textwidth, height=0.35\textheight]{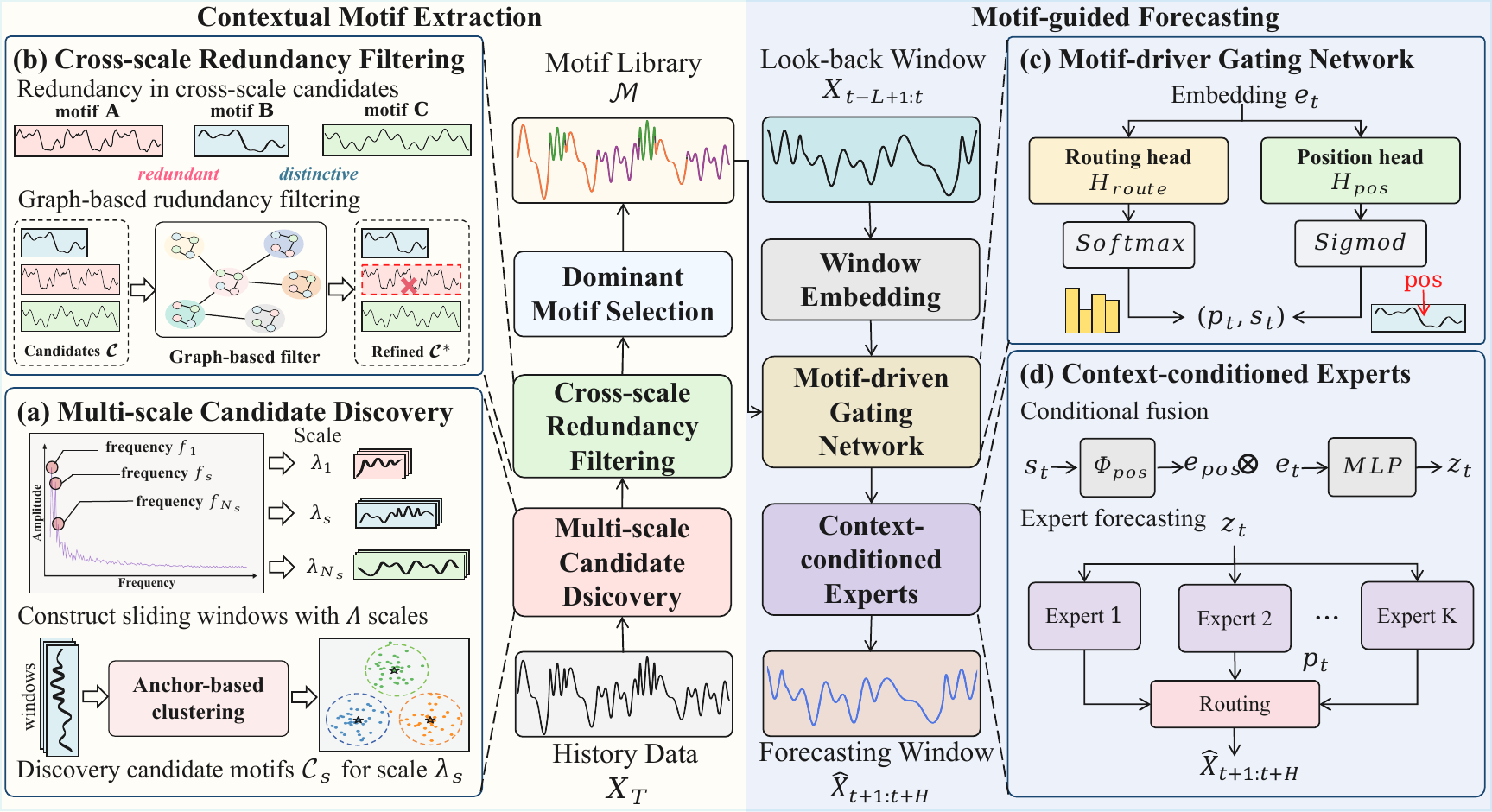}
    \caption{The architecture of CometNet.}
    \label{fig:architecture}
\end{figure*}

\section{Related Work}
\label{sec2_related_work}
In recent years, significant progress in LTSF has been predominantly driven by advancements in deep learning, with Transformer-based architectures and MLP-based models emerging as the two predominant paradigms in the field.

\subsection{Transformer-based Methods}
The Transformer architecture, with its powerful attention mechanism, has been extensively adapted for modeling temporal dependencies. With early variants like Reformer~\cite{kitaev2020reformer} and Informer~\cite{zhou21informer} focused on tackling the critical issue of quadratic complexity. To better capture complex dependencies, recent prominent works have moved beyond standard attention, exploring alternative structures to model dependencies. For example, Autoformer~\cite{chen21autoformer} captures more robust temporal patterns by embedding series decomposition within its attention mechanism, while Fedformer~\cite{zhou2022fedformer} leverage frequency-domain analysis for capturing periodic characteristics. A pivotal advancement came with PatchTST~\cite{nie2023patchtst}, which introduced a channel-independent patching mechanism. This approach segments each time series variable into patches and focuses on capturing temporal relationships across these patches, significantly improving performance. This paradigm was further evolved by iTransformer~\cite{liu2024itransformer}, which inverted the typical role of attention and feed-forward networks, applying attention to model correlations across variables while using MLPs to capture temporal dynamics within each channel. 

\subsection{MLP-based Methods}
In response to the high complexity and potential training instability of Transformers, a competing paradigm centered on simpler MLP and linear structures has gained significant traction. DLinear~\cite{zeng2023transformers} pioneered this trend by demonstrating that a simple linear model applied to decomposed trend and seasonal components can achieve competitive forecasting performance. Subsequent works further advanced this paradigm. TSMixer~\cite{ekambaram2023tsmixer} introduced a pure MLP-based architecture that iteratively mixes temporal and feature dimensions, while FreTS~\cite{yi2023frequency} and FilterTS~\cite{wang25filterts} incorporated frequency-domain filtering to enhance MLP-based predictions. More recently, models like TimeMixer~\cite{wang2024timemixer} and TimeMixer++~\cite{wang2025timemixer++} explored more sophisticated mixing strategies to further improve performance.

Despite their notable progress, these methods primarily concentrate on optimizing information extraction within localized temporal windows, which lack a dedicated mechanism to explicitly leverage the rich contextual dependencies embedded in the vast history beyond this finite scope. Though recent approach like BSA~\cite{kang2024introducing} attempts to alleviate this by preserving cross-sample temporal correlations through spectral attention, it still challenged in identifying meaningful temporal dependencies from complex historical sequences.

\section{Methodology}

\subsubsection{Problem Statement.} 
The objective of LTSF is to learn a mapping function $f: \mathbb{R}^{L \times N} \rightarrow \mathbb{R}^{H \times N}$ that predicts future sequence $X_{t:t+H} \in \mathbb{R}^{H \times N}$ based on the observed look-back window $X_{t-L+1:t} \in \mathbb{R}^{L \times N}$, where $N$ denotes the number of channels, $L$ and $H$ represent the observation window length and the prediction horizon, respectively. The function $f$ is typically learned from the historical time series $X_{1:T}$ to capture temporal dependencies and generalize to unseen sequences.

\subsection{Overall Architecture}

To overcome the receptive field bottleneck inherent in limited look-back windows, we propose \textbf{CometNet}, a novel framework that leverages dominant contextual motifs to capture temporal dependencies far beyond the look-back window, significantly enhancing long-term forecasting performance. As illustrated in Fig.~\ref{fig:architecture}, CometNet introduces a two-module paradigm: first, a Contextual Motif Extraction module establishes a dominant contextual motif library from historical data, and second, a Motif-guided Forecasting module utilizes this library to make accurate long-term predictions.

The Contextual Motif Extraction module is designed to analyze the entire historical time series $X_{1:T} \in \mathbb{R}^{T \times N}$, to extract dominant contextual motifs. To handle channel-wise heterogeneity in multivariate time series (MTS), we adopt a channel-independent strategy~\cite{nie2023patchtst, han2024tradeoff}, decomposing $X_{1:T}$ into $N$ univariate series:
% \begin{equation}
% X_{1:T} \to \sum_{n=1}^{N} X^n_{1:T},
% \end{equation}

\begin{equation}
X_{1:T} \to \{X^1_{1:T}, \cdots, X^n_{1:T}, \cdots, X^N_{1:T}\},
\end{equation}
where $X^n_{1:T} \in \mathbb{R}^{T \times 1}$ represents the $n$-th channel. For brevity, we omit the channel superscript $n$ hereafter. A cascaded motif extraction process is then applied to each univariate series to establish a library of its dominant motifs: 
\begin{equation}
\mathcal{M} = CMExtractor(X_{1:T}),
\end{equation}
where $\mathcal{M} = \{m_1, m_2, \cdots, m_K\}$ denotes the extracted library of $K$ dominant contextual motifs, and $CMExtractor(\cdot)$ represents the contextual motif extraction function.

The Motif-guided Forecasting module is engineered to leverage this pre-established motif library $\mathcal{M}$ for long-term prediction. We implement this using a Mixture-of-Experts (MoE) architecture, where each dominant motif $m_k \in \mathcal{M}$ is conceptually associated with a dedicated expert network $E_k$, which specializes in capturing the temporal dynamics inherent to that motif.
Specifically, given a look-back window $X_{t-L+1:t}$, the forecasting module employs a motif-guided gating mechanism to dynamically associate the input with the most relevant dominant motif $m_k$ in the library $\mathcal{M}$. This association not only selects the appropriate expert $E_k$, but also provides crucial contextual information about the window's position within the motif. The activated expert then generates the pattern-specific forecasts. This forecasting process is formally defined as:
\begin{equation}
\hat{X}_{t+1:t+H} = MGForecaster(X_{t-L+1:t} \mid \mathcal{M}),      
\end{equation}
where $MGForecaster(\cdot)$ represents the motif-guided forecasting function, which makes predictions conditioned on the pre-established motif library $\mathcal{M}$. 

This design effectively circumvents the receptive field bottleneck, enabling the model to ground its predictions in both the local context of the look-back window and the long-term dependencies embedded in the dominant motifs. Next, we will introduce each module in detail.

\subsection{Contextual Motif Extraction}

The primary goal of this module is to automatically establish a library of dominant contextual motifs from the entire historical time series. The key challenge lies in the unknown scales of these motifs. A naive multi-scale exploration would not only create an exponentially large candidate space but also introduce significant cross-scale redundancy. To address this, we propose a cascaded motif extraction framework that combines multi-scale candidate discovery, cross-scale redundancy filtering, and benefit-driven selection to efficiently construct a high-quality library of dominant motifs.

\subsubsection{Multi-scale Candidate Discovery.} 
This initial step aims to efficiently discover a diverse set of candidate motifs across multiple time scales. Given a historical univariate time series $X_{1:T}$, we first apply moving average smoothing~\cite{chen21autoformer} to obtain a detrended series $X^{det}_{1:T}$ with more stable periodic characteristics. Subsequently, the Fast Fourier Transform (FFT)~\cite{duhamel1990fast} is employed to identify the dominant frequencies in the data:
\begin{equation}\label{eq1}
    A_f = Amp \Big (FFT(X^{det}_{1:T} \Big ),
\end{equation}
where $Amp(\cdot)$ calculates the amplitude spectrum. The corresponding period $\lambda_f$ for each frequency component $f$ is derived as $\lambda_f = \left\lceil\frac{T}{f}\right\rceil$. Based on the amplitude intensity, we select the top-$N_s$ periods $\Lambda=\{\lambda_1, \lambda_2, \cdots, \lambda_{N_s}\}$ with the highest spectral magnitudes to serve as the set of relevant scales. For each identified scale $\lambda_s \in \Lambda$, we first downsample the sequence to improve computational efficiency. On this downsampled series, we employ an anchor-based clustering approach to efficiently group similar subsequences. Specifically, to avoid exhaustive pairwise comparisons, we randomly sample $N_r$ subsequences of length $\lambda_s$ as candidate anchors $\mathcal{A}_s = \{a_1,...,a_{N_r}\}$ and construct their correlation matrix $R^{s} \in \mathbb{R}^{N_r \times N_r}$, where each element $R^{s}_{ij}$ is the Pearson correlation coefficient between anchors $a_i$ and $a_j$. The density score for each anchor $a_i$ at scale $\lambda_s$ is then calculated as:
\begin{equation}
\rho_i^{s} = \sum_{j=1}^{N_r} R_{ij}^{s} \cdot \mathbb{I}(R_{ij}^{s} > \tau_s),
\end{equation}
where $\tau_s$ is a similarity threshold. After selecting the top $N_c$ anchors with the highest density scores as cluster centroids, all subsequences of length $\lambda_s$ are assigned to their nearest centroid. From each resulting cluster, we select the most representative subsequence (the medoid) as a candidate motif. This process yields a comprehensive set of candidate motifs from all scales, $\mathcal{C}=\bigcup_{s=1}^{N_s}C_s$, for the subsequent refinement stages.

\subsubsection{Cross-scale Redundancy Filtering.}
To refine the candidate set $\mathcal{C}$ while preserving its diversity, we develop a cross-scale filtering strategy that eliminates redundant motifs using graph-based analysis. We first model the relationships between all candidate motifs in $\mathcal{C}$ as a weighted, undirected graph $\mathcal{G} = (\mathcal{V}, \mathcal{E})$, where each vertex $v_i \in \mathcal{V}$ corresponds to a candidate motif $c_i \in \mathcal{C}$. Edge weights are determined by a normalized Dynamic Time Warping (DTW)~\cite{sakoe2003dynamic} similarity score, with a sparsity constraint:
\begin{equation}
w_{ij} = S_{DTW}(c_i, c_j) \cdot \mathbb{I}(S_{DTW}(c_i, c_j) > \tau_g),
\end{equation}
where $S_{DTW}(\cdot)$ is the normalized DTW similarity, and the threshold $\tau_g$ controls the granularity of redundancy filtering.

The graph $\mathcal{G}$ is then decomposed into a set of connected components $\{\mathcal{G}_1, \mathcal{G}_2, \cdots, \mathcal{G}_{N_g}\}$. Each components $\mathcal{G}_k = (\mathcal{V}_k, \mathcal{E}_k)$ represents a cluster of semantically similar motifs that may span different scales. We obtain a refined candidate set $\mathcal{C}^*$ by selecting a single prototype from each component. The prototype $c_k^*$ for component $\mathcal{G}_k$ is identified as the motif with the highest intra-cluster affinity:   
\begin{equation}
c_k^* = \underset{c \in \mathcal{V}_k}{\text{argmax}} \frac{1}{|\mathcal{V}_k|} \sum_{c' \in \mathcal{V}_k} S_{\text{DTW}}(c, c').
\end{equation}
This process effectively filters redundancy while preserving a diverse set of representative candidates.

\subsubsection{Dominant Motif Selection.}
An ideal library of dominant motifs must be representative, comprehensive, and diverse, as these properties fundamentally determine its capacity to provide precise contextual guidance. To this end, we propose a benefit-driven selection strategy to select the final dominant motifs from the refined set $\mathcal{C}^*$. First, each candidate $c_i^* \in \mathcal{C}^*$ is evaluated for its intrinsic quality using a composite metric:
\begin{equation}
Q(c^*_i) = \alpha_S \cdot Q_S(c^*_i) + \alpha_P \cdot Q_P(c^*_i) + \alpha_A \cdot Q_A(c^*_i),
\end{equation}
where $Q_S(\cdot)$, $Q_P(\cdot)$, $Q_A(\cdot)$ quantify the saliency, prevalence and atomicity of patten $c^*_i$, respectively. While $\alpha_S$, $\alpha_P$, and $\alpha_A$ are their corresponding weighting coefficients.

Next, we formulate a benefit score that jointly optimizes for pattern quality, overall temporal coverage, and library diversity. The benefit of adding a candidate $c^*_i$ to our current selection $\mathcal{S}$ is calculated as: 
\begin{equation}
B(c^*_i \mid \mathcal{S}) = Q(c^*_i) \cdot Cov(c^*_i \mid \mathcal{S}) \cdot Div(c^*_i \mid \mathcal{S}),
\end{equation}
where $Cov(\cdot)$ measures the marginal temporal coverage gained by adding $c^*_i$, and $Div(\cdot)$ measures its marginal contribution to the library's diversity. By iteratively selecting the top-$K$ candidates with the highest benefit scores, we construct the final library of dominant contextual motifs, $\mathcal{M} = \{m_1, m_2, \cdots, m_K\}$, that optimally balances these crucial properties. The detailed formulations of these metrics are provided in Appendix.

\subsection{Motif-guided Forecasting}

After establishing the library of dominant contextual motifs $\mathcal{M}$, we introduce the Motif-guided Forecasting module. This module is designed to leverage the rich, long-term contextual information embedded in the motif library to guide its predictions. Its core is a motif-aware MoE architecture, where the number of experts is equal to the number of dominant motifs, $K = |\mathcal{M}|$. The forecasting process consists of three sequential stages: window embedding, motif-driven gating, and context-conditioned prediction by the experts.

\subsubsection{Window Embedding.}
Given any look-back window $X_{t-L+1:t}$, we first map it into a compact latent representation $e_t \in R^{d_e}$ via a window embedding module (implemented as a MLP), preserving key temporal information while reducing dimensionality,
\begin{equation}
\mathbf{e}_t = LayerNorm(MLP(\mathcal{X}_{t-L+1:t})).
\end{equation}

\subsubsection{Motif-driven Gating Network.}
We then develop a motif-driven gating network to dynamically integrate our pre-established motif library $\mathcal{M}$ into the forecasting process. This network acts as an intelligent router, receiving the window embedding $e_t$ as input and dynamically associating the current local observation with the relevant contextual motif. Unlike standard MoE gates, which only selects which expert to use, our gate also provides window position information in its relevant motif, offering fine-grained prediction guidance.

Specifically, our gating network processes the embedding $e_t$ through two parallel heads,  both implemented as lightweight MLPs. The first is a routing head $H_{route}$, which maps $e_t$ to a logit vector. This vector is then passed through a Softmax function to generate a probability distribution $p_t \in \mathbb{R}^K$, representing the selection probabilities for the $K$ experts,
\begin{equation}
p_t = Softmax(H_{route}(e_t)),
\end{equation}
the $k$-th element $p_{t, k}$ represents the confidence in selecting the $k$-th expert, which corresponds to the $k$-th dominant motif $m_k \in \mathcal{M}$. 

Simultaneously, a parallel position head $H_{pos}$ maps the same embedding $e_t$ into a scalar value $s_t \in [0, 1]$ via a Sigmoid activation function. This scalar encodes the relative position of the current observation within the life cycle of its associated motif. 
\begin{equation}
s_{t} = \sigma(H_{pos}(e_t)).
\end{equation}

Together, $p_t$ and $s_t$ form a dual instructional signal that specifies both the associated motif and the temporal progression within it. This provides rich, dual-faceted guidance for the downstream experts.

\subsubsection{Context-conditioned Experts.}
Our framework contains $K$ independent context-conditioned expert networks $\{E_1, E_2, \cdots, E_K\}$, each designed to specialize in the dynamics of a corresponding dominant motif $m_k \in \mathcal{M}$. The core contribution of these experts lies in their ability to perform context-conditioned forecasting by leveraging the dual instructional signal from the gating network.

For each expert $E_k$, the shared inputs are the window embedding $e_t$ and its positional score $s_t$. The process begins with a position encoder $\Phi_{pos}$, implemented as a MLP. It elevates the low-dimensional positional score $s_t$ into a high-dimensional position embedding $e_{pos} \in \mathbb{R}^{d_e}$. Subsequently, a conditional fusion layer concatenates the window embedding $e_t$ with the position embedding $e_{pos}$ and processes them through another MLP to learn their non-linear interactions. This produces a unified, contextually-aware conditional representation $z_t \in \mathbb{R}^{d_e}$.
\begin{equation}
z_t = F_{fuse}(concat(e_t, \Phi_{pos}(s_t))).
\end{equation}
Then, the highly conditional representation $z_t$ is fed into $K$ parallel, expert-specific prediction heads $\{P_1, P_2, \cdots, P_K\}$, to generate a set of expert-specific forecasts $\{\hat{X}_{1, {t+1:t+H}}, \hat{X}_{2, {t+1:t+H}}, \cdots, \hat{X}_{K, {t+1:t+H}}\}$, where $\hat{X}_{k, {t+1:t+H}} \in \mathbb{R}^{H \times 1}$. The final prediction of the entire model $\hat{X}_{t+1:t+H}$ is the weighted sum of all expert outputs, with the weights provided by the gating network's probability distribution $p_t$.
\begin{equation}
\hat{X}_{t+1:t+H} = \sum_{k=1}^{K} p_{t,k} \cdot \hat{X}_{k, {t+1:t+H}} = \sum_{k=1}^{K} p_{t,k} \cdot P_k(z_t).
\end{equation}
This design enables our model to dynamically and granularly adjust its predictive strategy based on both the high-level contextual motif and the subtle variations of the stage within it.

\section{Experiments}

\subsubsection{Experimental Details}
% --- 自定义颜色和命令 ---
% 定义最佳和次佳结果的颜色
\definecolor{bestcolor}{RGB}{204, 0, 0}
\definecolor{secondbestcolor}{RGB}{0, 0, 204}

% 定义用于格式化数字的命令
% \best 用于最佳结果 (红色粗体)
% \secondbest 用于次佳结果 (蓝色下划线)
\newcommand{\best}[1]{\textbf{\textcolor{bestcolor}{#1}}}
\newcommand{\secondbest}[1]{\textcolor{secondbestcolor}{\underline{#1}}}

\begin{table*}[h!]
% --- 表格格式设置 ---
\centering
\small % Use 9 point type as requested.
\setlength{\tabcolsep}{3pt} % Compress columns to fit.

\begin{tabular}{c@{\hspace{1pt}}c|cc|cc|cc|cc|cc|cc|cc|cc}
\hline
% --- 表头 ---
&\multirow{2}{*}{Models} & \multicolumn{2}{c|}{\textbf{CometNet}} & \multicolumn{2}{c|}{\textbf{TimeMixer++}} & \multicolumn{2}{c|}{\textbf{FilterTS}} & \multicolumn{2}{c|}{\textbf{PatchMLP}} & \multicolumn{2}{c|}{\textbf{BSA}} & \multicolumn{2}{c|}{\textbf{iTransformer}} & \multicolumn{2}{c|}{\textbf{PatchTST}} & \multicolumn{2}{c}{\textbf{DLinear}} \\
& & \multicolumn{2}{c|}{(Ours)} & \multicolumn{2}{c|}{(2025)} & \multicolumn{2}{c|}{(2025)} & \multicolumn{2}{c|}{(2025)} & \multicolumn{2}{c|}{(2024)} & \multicolumn{2}{c|}{(2024)} & \multicolumn{2}{c|}{(2023)} & \multicolumn{2}{c}{(2023)} \\ \cline{3-18}
&Metric & MSE & MAE & MSE & MAE & MSE & MAE & MSE & MAE & MSE & MAE & MSE & MAE & MSE & MAE & MSE & MAE \\ \hline

% --- ETTh1 ---
% CometNet, TimeMixer++, FilterTS, PatchMLP, BSA, iTransformer, PatchTST, DLinear
\multirow{5}{*}{\rotatebox{90}{ETTh1}} 
& 96 & \best{0.345} & \secondbest{0.398} & 0.361 & 0.403 & 0.374 & 0.391 & 0.391 & 0.403 & 0.427 & 0.442 & 0.386 & 0.405 & 0.460 & 0.447 & 0.397 & 0.412 \\

& 192 & \best{0.368} & \best{0.412} & \secondbest{0.416} & 0.441 & 0.424 & \secondbest{0.421} & 0.444 & 0.431 & 0.480 & 0.481 & 0.441 & 0.512 & 0.477 & 0.429 & 0.446 & 0.441 \\ 

& 336 & \best{0.387} & \best{0.423} & \secondbest{0.430} & \secondbest{0.434} & 0.464 & 0.441 & 0.490 & 0.456 & 0.537 & 0.521 & 0.487 & 0.458 & 0.546 & 0.496 & 0.489 & 0.467 \\

& 720 & \best{0.391} & \best{0.434} & \secondbest{0.467} & \secondbest{0.451} & 0.470 & 0.466 & 0.528 & 0.496 & 0.697 & 0.620 & 0.503 & 0.491 & 0.544 & 0.517 & 0.513 & 0.510  \\ \cline{2-18}

& Avg & \best{0.373} & \best{0.417} & \secondbest{0.419} & 0.432 & 0.433 & \secondbest{0.430} & 0.463 & 0.447 & 0.536 & 0.516 & 0.454 & 0.447 & 0.516 & 0.484 & 0.461 & 0.457 \\ \hline

% --- ETTh2 ---
\multirow{5}{*}{\rotatebox{90}{ETTh2}} 
& 96 & \best{0.205} & \best{0.281} & 0.276 & 0.328 & 0.290 & 0.338 & 0.305 & 0.353 & \secondbest{0.234} & \secondbest{0.323} & 0.297 & 0.349 & 0.308 & 0.355 & 0.340 & 0.394 \\

& 192 & \best{0.212} & \best{0.305} & 0.342 & 0.379 & 0.374 & 0.390 & 0.402 & 0.410 & \secondbest{0.290} & \secondbest{0.362} & 0.380 & 0.400 & 0.393 & 0.405 & 0.482 & 0.479 \\

& 336 & \best{0.219} & \best{0.318} & 0.346 & 0.398 & 0.406 & 0.420 & 0.436 & 0.437 & \secondbest{0.327} & \secondbest{0.388} & 0.428 & 0.432 & 0.427 & 0.436 & 0.591 & 0.541 \\

& 720 & \best{0.228} & \best{0.332} & \secondbest{0.392} & \secondbest{0.415} & 0.418 & 0.437 & 0.437 & 0.449 & 0.414 & 0.439 & 0.427 & 0.445 & 0.436 & 0.450 & 0.839 & 0.661 \\ \cline{2-18}

& Avg & \best{0.216} & \best{0.309} & 0.339 & 0.380 & 0.372 & 0.396 & 0.395 & 0.412 & \secondbest{0.316} & \secondbest{0.378} & 0.383 & 0.407 & 0.391 & 0.411 & 0.563 & 0.519 \\ \hline

% --- ETTm1 ---
\multirow{5}{*}{\rotatebox{90}{ETTm1}} 
& 96 & \best{0.210} & \best{0.315} & \secondbest{0.310} & \secondbest{0.334} & 0.321 & 0.360 & 0.324 & 0.362 & 0.382 & 0.398 & 0.334 & 0.368 & 0.352 & 0.374 & 0.346 & 0.374 \\ 

& 192 & \best{0.211} & \best{0.322} & \secondbest{0.348} & \secondbest{0.362} & 0.363 & 0.382 & 0.369 & 0.385 & 0.423 & 0.428 & 0.390 & 0.393 & 0.374 & 0.387  & 0.391 & 0.391 \\

& 336 & \best{0.282} & \best{0.368} & \secondbest{0.376} & \secondbest{0.391} & 0.395 & 0.403 & 0.402 & 0.407 & 0.492 & 0.466 & 0.426 & 0.420 & 0.421 & 0.414 & 0.415 & 0.415 \\ 

& 720 & \best{0.303} & \best{0.387} & \secondbest{0.440} & \secondbest{0.423} & 0.462 & 0.438 & 0.474 & 0.446 & 0.567 & 0.516 & 0.491 & 0.459 & 0.462 & 0.449 & 0.473 & 0.451 \\ \cline{2-18}

& Avg & \best{0.252} & \best{0.348} & \secondbest{0.369} & \secondbest{0.378} & 0.385 & 0.396 & 0.392 & 0.400 & 0.466 & 0.452 & 0.407 & 0.410 & 0.406 & 0.407 & 0.404 & 0.408 \\ \hline 

% --- ETTm2 ---
\multirow{5}{*}{\rotatebox{90}{ETTm2}} 
& 96 & \best{0.117} & \best{0.244} & 0.170 & \secondbest{0.245} & 0.172 & 0.255 & 0.180 & 0.263 & \secondbest{0.153} & 0.258 & 0.180 & 0.264 & 0.183 & 0.270 & 0.193 & 0.293 \\ 

& 192 & \best{0.148} & \best{0.280} & 0.229 & 0.291 & 0.237 & 0.299 & 0.244 & 0.306 & \secondbest{0.189} & \secondbest{0.290} & 0.250 & 0.309 & 0.255 & 0.314 & 0.284 & 0.361 \\ 

& 336 & \best{0.156} & \best{0.285} & 0.303 & 0.343 & 0.299 & 0.398 & 0.312 & 0.349 & \secondbest{0.230} & \secondbest{0.321} & 0.311 & 0.348 & 0.309 & 0.347 & 0.382 & 0.329 \\ 

& 720 & \best{0.171} & \best{0.299} & 0.373 & 0.399 & 0.397 & 0.394 & 0.411 & 0.407 & \secondbest{0.303} & \secondbest{0.369} & 0.412 & 0.407 & 0.412 & 0.404 & 0.558 & 0.525 \\ \cline{2-18} 

& Avg & \best{0.148} & \best{0.277} & 0.269 & 0.320 & 0.276 & 0.321 & 0.287 & 0.331 & \secondbest{0.218} & \secondbest{0.309} & 0.288 & 0.332 & 0.290 & 0.334 & 0.354 & 0.402 \\ \hline 

% --- Weather ---
\multirow{5}{*}{\rotatebox{90}{Weather}} 
& 96 & 0.183 & 0.209 & \best{0.155} & \secondbest{0.205} & 0.162 & 0.207 & 0.164 & 0.210 & \secondbest{0.159} & \best{0.203} & 0.174 & 0.214 & 0.186 & 0.227 & 0.195 & 0.252 \\ 

& 192 & \best{0.196} & \best{0.223} & \secondbest{0.201} & \secondbest{0.245} & 0.209 & 0.252 & 0.211 & 0.251 & 0.205 & 0.246 & 0.221 & 0.254 & 0.234 & 0.265 & 0.237 & 0.295 \\

& 336 & \best{0.216} & \best{0.241} & \secondbest{0.237} & \secondbest{0.265} & 0.263 & 0.292 & 0.269 & 0.294 & 0.252 & 0.285 & 0.278 & 0.296 & 0.284 & 0.301 & 0.282 & 0.331 \\

& 720 & \best{0.264} & \best{0.280} & \secondbest{0.312} & \secondbest{0.334} & 0.344 & 0.344 & 0.349 & 0.345 & 0.325 & 0.337 & 0.358 & 0.347 & 0.356 & 0.349 & 0.345 & 0.382 \\ \cline{2-18} 

& Avg & \best{0.215} & \best{0.238} & \secondbest{0.226} & \secondbest{0.262} & 0.244 & 0.274 & 0.248 & 0.275 & 0.235 & 0.268 & 0.258 & 0.278 & 0.265 & 0.285 & 0.265 & 0.315 \\ \hline

% --- Exchange ---
\multirow{5}{*}{\rotatebox{90}{Exchange}} 
& 96 & 0.086 & 0.216 & \secondbest{0.085} & 0.214 & \best{0.081} & \best{0.199} & 0.094 & 0.215 & 0.089 & 0.210 & 0.086 & 0.206 & 0.088 & \secondbest{0.205} & 0.088 & 0.218 \\ 

& 192 & \best{0.137} & \best{0.269} & 0.175 & 0.313 & \secondbest{0.171} & 0.294 & 0.182 & 0.306 & 0.183 & \secondbest{0.289} & 0.177 & 0.299 & 0.176 & 0.299 & 0.176 & 0.315 \\ 

& 336 & \best{0.176} & \best{0.293} & 0.316 & 0.420 & 0.321 & 0.409 & 0.339 & 0.423 & 0.317 & 0.403 & 0.331 & 0.417 & \secondbest{0.301} & \secondbest{0.397} & 0.313 & 0.427 \\ 

& 720 & \best{0.423} & \best{0.437} & 0.851 & 0.689 & 0.837 & 0.688 & 0.873 & 0.701 & \secondbest{0.647} & \secondbest{0.536} & 0.847 & 0.691 & 0.901 & 0.714 & 0.839 & 0.695 \\ \cline{2-18}

& Avg & \best{0.206} & \best{0.304} & 0.357 & 0.391 & 0.352 & 0.397 & 0.372 & 0.411 & \secondbest{0.309} & \secondbest{0.360} & 0.360 & 0.403 & 0.367 & 0.404 & 0.354 & 0.414 \\ \hline

% --- Electricity ---
\multirow{5}{*}{\rotatebox{90}{Electricity}} 
& 96 & \secondbest{0.139} & \secondbest{0.228} & \best{0.135} & \best{0.222} & 0.151 & 0.245 & 0.160 & 0.257 & 0.142 & 0.239 & 0.148 & 0.240 & 0.190 & 0.296 & 0.210 & 0.302 \\ 

& 192 & \secondbest{0.154} & \best{0.233} & \best{0.147} & \secondbest{0.235} & 0.163 & 0.256 & 0.176 & 0.271 & 0.163 & 0.257 & 0.162 & 0.253 & 0.199 & 0.304 & 0.210 & 0.305 \\ 

& 336 & \secondbest{0.167} & \secondbest{0.247} & \best{0.164} & \best{0.245} & 0.180 & 0.274 & 0.197 & 0.292 & 0.176 & 0.274 & 0.178 & 0.269 & 0.217 & 0.319 & 0.223 & 0.319 \\ 

& 720 & \best{0.172} & \best{0.256} & \secondbest{0.212} & 0.310 & 0.224 & 0.311 & 0.249 & 0.333 & 0.221 & \secondbest{0.308} & 0.225 & 0.317 & 0.258 & 0.352 & 0.258 & 0.350 \\ \cline{2-18}

& Avg & \best{0.158} & \best{0.241} & \secondbest{0.165} & \secondbest{0.253} & 0.180 & 0.271 & 0.196 & 0.288 & 0.176 & 0.269 & 0.178 & 0.270 & 0.216 & 0.318 & 0.225 & 0.319 \\ \hline

% --- Traffic ---
\multirow{5}{*}{\rotatebox{90}{Traffic}} 
& 96 & 0.408 & 0.271 & \best{0.392} & \best{0.253} & 0.448 & 0.309 & 0.475 & 0.329 & \secondbest{0.392} & 0.272 & 0.395 & \secondbest{0.268} & 0.526 & 0.347 & 0.650 & 0.396 \\ 
& 192 & 0.419 & \secondbest{0.275} & \best{0.402} & \best{0.258} & 0.455 & 0.307 & 0.483 & 0.330 & \secondbest{0.417} & 0.281 & 0.417 & 0.276 & 0.522 & 0.332 & 0.598 & 0.370 \\ 
& 336 & \best{0.426} & \secondbest{0.279} & \secondbest{0.428} & \best{0.263} & 0.472 & 0.313 & 0.498 & 0.336 & 0.433 & 0.289 & 0.433 & 0.283 & 0.517 & 0.334 & 0.605 & 0.373 \\ 
& 720 & \best{0.436} & \secondbest{0.283} & \secondbest{0.441} & \secondbest{0.282} & 0.508 & 0.332 & 0.542 & 0.359 & 0.470 & 0.310 & 0.467 & 0.302 & 0.552 & 0.352 & 0.645 & 0.394 \\ \cline{2-18} 
& Avg & \secondbest{0.422} & \secondbest{0.277} & \best{0.416} & \best{0.264} & 0.471 & 0.315 & 0.500 & 0.339 & 0.428 & 0.288 & 0.428 & 0.282 & 0.529 & 0.341 & 0.625 & 0.383 \\ \hline 

\end{tabular}
\caption{Full results for the long-term forecasting task. We compare extensive competitive models under different prediction lengths. Avg is averaged from all four prediction lengths, that is \{96, 192, 336, 720\}. Best and second best results are highlighted in \best{red} and \secondbest{blue}.}
\label{tab:full_results}
\end{table*}

\begin{table*}[t]
\centering
\small
{% <--- Start of local settings group
\renewcommand{\arraystretch}{0.85} % Reduce vertical space between rows
\setlength{\tabcolsep}{3pt}      % Reduce horizontal space between columns
\begin{tabular}{ccccc |cc|cc|cc|cc}
\toprule
\multicolumn{5}{c|}{\textbf{CometNet Components}} & \multicolumn{2}{c|}{\textbf{ETTh1}} & \multicolumn{2}{c|}{\textbf{ETTh2}} & \multicolumn{2}{c|}{\textbf{ETTm1}} & \multicolumn{2}{c}{\textbf{ETTm2}} \\
\cmidrule(lr){1-5} \cmidrule(lr){6-7} \cmidrule(lr){8-9} \cmidrule(lr){10-11} \cmidrule(lr){12-13}
Multi-scale & Filtering & Selection & Gating & Position & MSE & MAE & MSE & MAE & MSE & MAE & MSE & MAE \\
\midrule
$\times$ & $\times$ & $\times$ & $\times$ & $\times$ & 0.544 & 0.531 & 0.653 & 0.572 & 0.483 & 0.464 & 0.526 & 0.539 \\
$\checkmark$ & $\times$ & $\times$ & $\checkmark$ & $\checkmark$ & 0.447 & 0.466 & 0.492 & 0.461 & 0.395 & 0.439 & 0.392 & 0.361 \\
$\checkmark$ & $\checkmark$ & $\times$ & $\checkmark$ & $\checkmark$ & 0.430 & 0.470 & 0.331 & 0.376 & 0.345 & 0.401 & 0.384 & 0.332 \\

$\checkmark$ & $\checkmark$ & $\checkmark$ & $\times$ & $\checkmark$ & 0.476 & 0.481 & 0.292 & 0.414 & 0.364 & 0.422 & 0.311 & 0.326 \\
$\checkmark$ & $\checkmark$ & $\checkmark$ & $\checkmark$ & $\times$ & 0.423 & 0.442 & 0.235 & 0.340 & 0.326 & 0.408 & 0.262 & 0.308 \\
\midrule
$\checkmark$ & $\checkmark$ & $\checkmark$ & $\checkmark$ & $\checkmark$ & \best{0.391} & \best{0.434} & \best{0.228} & \best{0.332} & \best{0.303} & \best{0.387} & \best{0.171} & \best{0.299} \\
\bottomrule
\end{tabular}
}% <--- End of local settings group
\caption{Ablation study of CometNet's key components on ETT datasets $(H=720)$. $\checkmark$ denotes that the component is enabled, while $\times$ denotes it is ablated. The best results are highlighted in \best{red}.}
\label{tab:ablation_result}
\end{table*}
\subsubsection{Datasets.}
We evaluate our proposed model on eight real-world benchmark datasets from various domains including energy, climate, economics and transportation for LTSF task. These datasets encompasses four ETT datasets (ETTh1, ETTh2, ETTm1, ETTm2), Weather, Exchange Rate, Traffic, and Electricity, as previously employed in Autoformer~\cite{chen21autoformer}.

\subsubsection{Baselines and Setup.} We select 7 state-of-the-art LTFS models as baselines, including TimeMixer++~\cite{wang2025timemixer++}, FilterTS~\cite{wang25filterts}, PatchMLP~\cite{tang2025patchmlp}, BSA~\cite{kang2024introducing}, iTransformer~\cite{liu2024itransformer}, PatchTST~\cite{nie2023patchtst} and DLinear~\cite{zeng2023transformers}. To ensure fair comparison, all models adopt identical experimental setup, utilizing a finite look-back window of $L=96$ to generate predictions across multiple horizons $H \in \{96, 192, 336, 720\}$. Mean Squared Error (MSE) and Mean Absolute Error (MAE) are selected as evaluation metrics.

For high-dimensional datasets (e.g., Traffic and Electricity), we employ k-dominant frequency hashing (k-DFH)~\cite{kang2025vardrop} to cluster variables into multiple groups, enabling efficient parallel training of variable groups. More details are provided in Appendix.

\subsection{Main Results}
\begin{figure}[t]
    \centering
    \includegraphics[width=0.9\columnwidth]{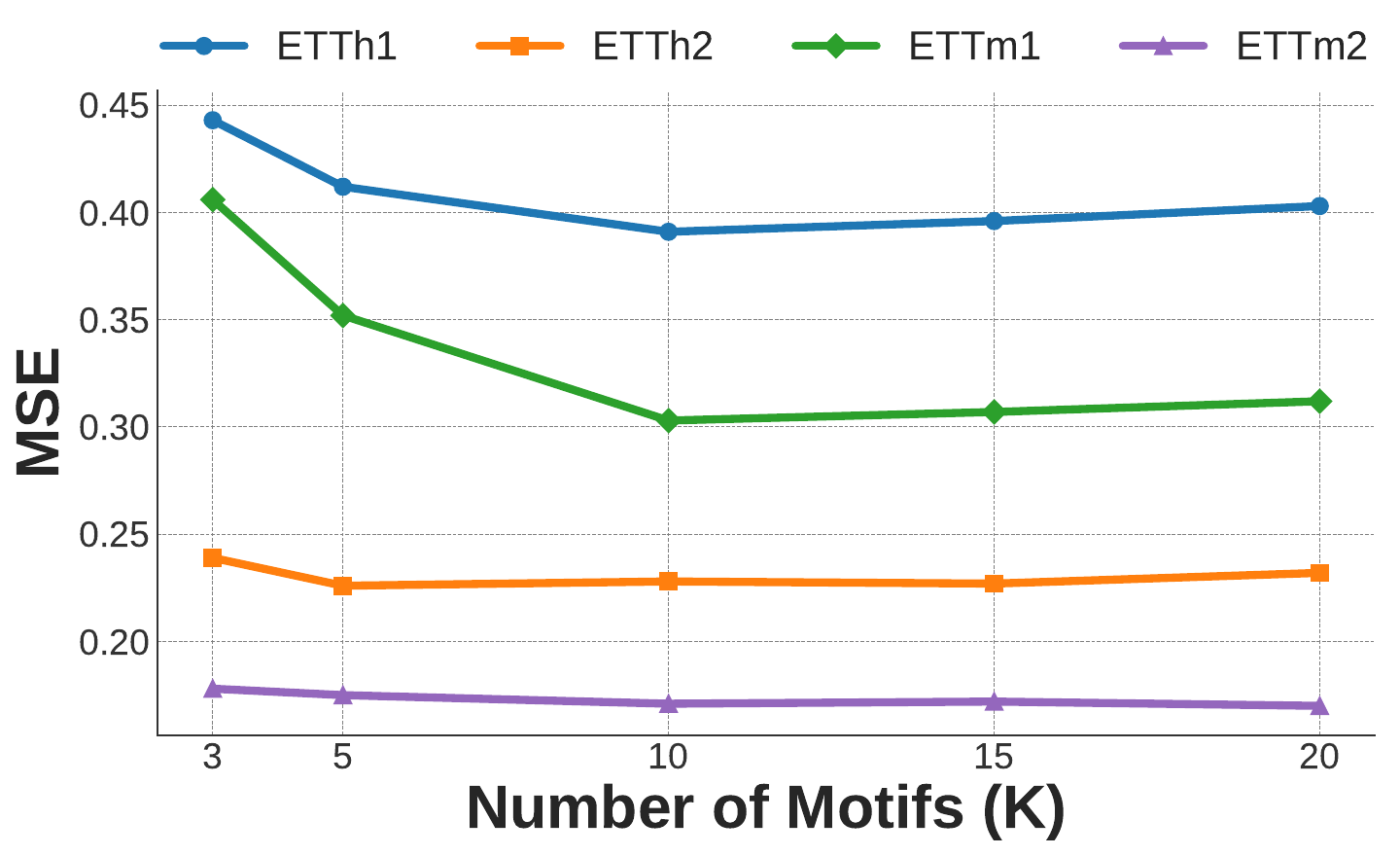}
    \caption{Impact of the number of motifs (K) on the forecasting performance.}
    \label{fig:result_sensity}
\end{figure}
The comprehensive quantitative results of our proposed CometNet against 7 baseline models are presented in Table~\ref{tab:full_results}. The results unequivocally demonstrate the superiority of our approach across a wide array of benchmarks and forecasting horizons. Specifically, across all datasets, CometNet attained either first (7/8) or second (1/8) rank in averaging MSE and MAE performance, establishing robust forecasting superiority.

A key observation from the results is that our approach's competitive advantage becomes more pronounced as the prediction horizon increases. For instance, on the ETTh2 dataset, while CometNet already outperforms baselines at horizon $H=96$, (MSE 0.205 vs 0.276), its margin of victory grows substantially at $H=720$, (MSE 0.228 vs 0.392). This trend holds consistently across nearly all datasets, including ETTh1, ETTh2, ETTm1, ETTm2, Weather, and Exchange, directly verify the validity of our design. Unlike baseline models constrained by receptive field limitations, CometNet leverages learned long-context motifs to guide long-term time series forecasting, thus preventing the error accumulation that plagues long-horizon forecasting.

The slightly weaker performance of CometNet on the Electricity and Traffic datasets can be attributed to the high number of variables in these datasets. This is an expected result since our group-based motif discovery trick trains each group rather than each variable to improve training efficiency. Notably, as prediction horizons extend, CometNet's rank consistently improves, ultimately achieving top performance at the longest horizons. This further confirms the value of our motif-guided approach for long-term forecasting.

\subsection{Ablation Study}
To validate the individual contributions of CometNet's key components, we conduct a comprehensive ablation study on the ETT datasets with a forecast horizon $H=720$. The results, presented in Table~\ref{tab:ablation_result}, unequivocally demonstrate the effectiveness and synergistic cooperation of each designed module. We start with a baseline where all motif-related components are disabled, which yields the poorest performance and confirms the critical need for long-range contextual information. Then, the stepwise reintroduction of Multi-scale discovery, Cross-scale filtering, and motif selection components demonstrates consistent and significant improvements, confirming that each step in our Contextual Motif Extraction process is vital for distilling a high-quality motif library. For the Motif-guided Forecasting module, we verified that removing either the expert routing gate mechanism or the refined position signal leads to a marked drop increase in prediction loss. This highlights the necessity of both dynamically selecting the correct motif-based expert and properly localizing the window's relative position within its motif's lifecycle. Ultimately, the full CometNet model, with all components integrated, achieves the best performance across all metrics, effectively validating our design and proving that each component plays an indispensable role in overcoming the receptive field bottleneck.

\subsection{Sensitive Study}
We conduct a sensitivity analysis to evaluate the impact of the number of dominant motifs $K$ in our library. As illustrated in Fig.~\ref{fig:result_sensity}, the results reveal that when $K$ is too small (e.g., K=3), our model's performance is suboptimal, as an insufficient number of motifs limits the library's expressive power. As $K$ increases, the performance consistently improves, with most datasets achieving their best or near-best results when $K=10$. This suggests that a library of this size strikes an optimal balance, offering a rich set of diverse and representative patterns. Interestingly, further increasing $K$ beyond this point (e.g., to 15 or 20) leads to either stabilization or a slight degradation in performance. This indicates that an overly large library may introduce redundancy or noise, slightly hindering the model's ability to pinpoint the most relevant context. Based on this robust analysis, we set $K=10$ as the default value in all our main experiments.

\section{Conclusion}
This paper introduced CometNet, a novel Contextual Motif-guided framework for Long-term Time Series Forecasting, which enhances prediction accuracy by discovering and leveraging recurrent, long-range contextual motifs. By incorporating a Contextual Motif Extraction module, which builds a library of dominant contextual motifs, with a Motif-guided Forecasting module, CometNet effectively grounds local observations in vital long-term context, thereby infusing predictions with crucial long-term dependencies. Extensive experiments across eight real-world datasets demonstrate that CometNet significantly outperforms current state-of-the-art methods, particularly on extended forecast horizons. These results highlight the benefits of leveraging contextual motifs to guide the modeling of long-term temporal dependencies.  

\section{Acknowledgments.}
This work is supported in part by the National Science Fund for Distinguished Young Scholars (No. 62325208), in part by the National Natural Science Foundation of China (No. 62572342, 62272339), and in part by the Natural Science Foundation of Tianjin (No. 23ZGZNGX00020).

\bibliography{aaai2026}

\section{Implementation Details}
\label{appendix:implementation_details}

\subsection{Dataset Details}
\label{appendix:datasets}

\subsubsection{Dataset Information.} To rigorously evaluate our proposed method, we conduct comprehensive experiments on eight widely-used real-world benchmark datasets, including ETTh1, ETTh2, ETTm1, ETTm2 (collectively known as ETT datasets), Weather, Exchange, Electricity and Traffic. All datasets are publicly available to ensure reproducibility. 

The ETT datasets~\cite{chen21autoformer} comprise seven critical indicators monitoring electricity transformers, with data collected from July 2016 to July 2018. These datasets are further divided into four subsets based on temporal resolution: ETTh1 and ETTh2 record measurements hourly, while ETTm1 and ETTm2 capture finer-grained 15-minute interval data. The Weather dataset~\cite{chen21autoformer} provides high-resolution meteorological records, containing 21 different weather indicators sampled every 10 minutes throughout 2020, collected from the Max Planck Institute for Biogeochemistry weather station. For financial time-series analysis, the Exchange dataset~\cite{chen21autoformer} offers comprehensive daily exchange rate records spanning 26 years (1990-2016) across eight countries. The Electricity dataset~\cite{chen21autoformer} documents hourly power consumption patterns from 321 individual clients. Lastly, the Traffic dataset~\cite{chen21autoformer} contains rich transportation data, including hourly occupancy rates collected from 862 sensors deployed across San Francisco Bay Area freeways between January 2015 and December 2016.

\subsubsection{Dataset Split.}
To ensure rigorous and reproducible evaluation, we meticulously follow the data processing protocols and temporal partitioning scheme established by TimeMixer++~\cite{wang2025timemixer++}. Our dataset splitting strategy employs two distinct configurations: (1) For datasets with sufficient temporal coverage (Weather, Traffic, Electricity, and Exchange), we apply a $\{0.7, 0.1, 0.2\}$ ratio for training, validation, and testing respectively; (2) For datasets with shorter time spans (ETTh1, ETTh2, ETTm1, and ETTm2), we adopt a $\{0.6, 0.2, 0.2\}$ allocation to maintain adequate training samples while preserving robust evaluation capacity. Crucially, all splits are performed in strict chronological order to prevent information leakage and preserve the temporal characteristics of time-series data. This systematic partitioning framework guarantees both methodological consistency and meaningful performance comparison across diverse benchmarks. The details of datasets are provided in Table~\ref{tab:dataset_details}.

\subsubsection{Forecasting setting.}
In alignment with contemporary benchmarking practices (TimeMixer++~\cite{wang2025timemixer++} and iTransformer~\cite{liu2024itransformer}), we establish a standardized evaluation framework with fixed input-output configurations. All experiments utilize a consistent look-back window of 96 time steps, while assessing model performance across four progressively challenging forecasting horizons: $\{96, 192, 336, 720\}$. This comprehensive evaluation scheme enables thorough assessment of model capabilities across varying temporal scales, facilitating direct comparison with state-of-the-art approaches while maintaining experimental rigor.

\begin{table*}[t]
\centering
\caption{Detailed dataset descriptions. \textbf{Dim} denotes the variate number of each dataset. \textbf{Dataset Size} denotes the total number of time points in (Train, Validation, Test) split respectively. \textbf{Prediction Length} denotes the future time points to be predicted and four prediction settings are included in each dataset. \textbf{Frequency} denotes the sampling interval of time points.}
\label{tab:dataset_details}
\begin{tabular}{l| c| c| c| c| l}
\toprule
\textbf{Dataset} & \textbf{Dim} & \textbf{Prediction Length} & \textbf{Dataset Size} & \textbf{Frequency} & \textbf{Information} \\
\midrule
ETTh1 & 7 & \{96, 192, 336, 720\} & (8545, 2881, 2881) & Hourly & Electricity \\
ETTh2 & 7 & \{96, 192, 336, 720\} & (8545, 2881, 2881) & Hourly & Electricity \\
ETTm1 & 7 & \{96, 192, 336, 720\} & (34465, 11521, 11521) & 15min & Electricity \\
ETTm2 & 7 & \{96, 192, 336, 720\} & (34465, 11521, 11521) & 15min & Electricity \\
Weather & 21 & \{96, 192, 336, 720\} & (36792, 5271, 10540) & 10min & Weather \\
Exchange & 8 & \{96, 192, 336, 720\} & (5120, 665, 1422) & Daily & Economy \\
Electricity & 321 & \{96, 192, 336, 720\} & (18317, 2633, 5261) & Hourly & Electricity \\
Traffic & 862 & \{96, 192, 336, 720\} & (12185, 1757, 3509) & Hourly & Transportation \\
\bottomrule
\end{tabular}
\end{table*}

\subsection{Baseline Details}
\label{appendix:baselines}
In this section, we discuss the configurations of the baseline models and specify where the BSA module was integrated into each baseline model. For each dataset, the base model structure configuration was directly replicated from the Time-Series-Library~\cite{wang2024timemixer} scripts when available. Where configurations were not provided, we adjusted them to align closely with the available examples.

\begin{itemize}

\item \textbf{TimeMixer++}~\cite{wang2025timemixer++}: The hyperparameters for the TimeMixer++ model are as follows:

(d\_model = 16, d\_ff = 32, e\_layers = 2, down\_sampling\_layers = 3, down\_sampling\_window = 2, dropout = 0.1, activation = GELU, learning\_rate = 0.01) for the ETTh1 and ETTh2 datasets,

(d\_model = 32, d\_ff = 32, e\_layers = 2, down\_sampling\_layers = 3, down\_sampling\_window = 2, dropout = 0.1, activation = GELU, learning\_rate = 0.01) for the ETTm1 and ETTm2 datasets,

(d\_model = 16, d\_ff = 32, e\_layers = 3, down\_sampling\_layers = 3, down\_sampling\_window = 2, dropout = 0.1, activation = GELU, learning\_rate = 0.01) for the Weather and Electricity datasets,

(d\_model = 32, d\_ff = 64, e\_layers = 3, down\_sampling\_layers = 3, down\_sampling\_window = 2, dropout = 0.1, activation = GELU, learning\_rate = 0.01) for the Traffic dataset.

\item \textbf{iTransformer}~\cite{liu2024itransformer}: The hyperparameters for the iTransformer model are as follows: 

(d\_model = 128, d\_ff = 128, num\_heads = 8, encoder\_layers = 2, dropout = 0.1, activation = GELU, learning\_rate = 0.001) for the ETT and Exchange datasets,

(d\_model = 512, d\_ff = 512, num\_heads = 8, encoder\_layers = 3, dropout = 0.1, activation = GELU, learning\_rate = 0.0005) for the Weather and Electricity datasets,

(d\_model = 512, d\_ff = 512, num\_heads = 8, encoder\_layers = 4, dropout = 0.1, activation = GELU, learning\_rate = 0.001) for the Traffic dataset.

\item \textbf{BSA}~\cite{kang2024introducing}: Following the official implementation of BSA, we adopt the iTransformer + BSA architecture. The hyperparameters for the BSA model are as follows:

(d\_model = 128, d\_ff = 128, num\_heads = 8, encoder\_layers = 2, dropout = 0.1, activation = GELU, learning\_rate = 0.001) for the ETT and Exchange datasets,

(d\_model = 512, d\_ff = 512, num\_heads = 8, encoder\_layers = 3, dropout = 0.1, activation = GELU, learning\_rate = 0.0005) for the Weather and Electricity datasets,

(d\_model = 512, d\_ff = 512, num\_heads = 8, encoder\_layers = 4, dropout = 0.1, activation = GELU, learning\_rate = 0.001) for the Traffic dataset.

\item \textbf{PatchTST}~\cite{nie2023patchtst}: The hyperparameters for the PatchTST model are as follows:

(d\_model = 16, d\_ff = 32, e\_layers = 3, d\_layers = 1, n\_heads = 4, top\_k = 3, factor=3, dropout = 0.1, activation = GELU, learning\_rate = 0.001) for the ETTh1, ETTh2, ETTm1, and ETTm2 datasets,

(d\_model = 16, d\_ff = 32, e\_layers = 2, d\_layers = 1, n\_heads = 4, top\_k = 3, factor=3, dropout = 0.1, activation = GELU, learning\_rate = 0.001) for the Weather and Exchange datasets, 

(d\_model = 16, d\_ff = 32, e\_layers = 2, d\_layers = 1, n\_heads = 8, top\_k = 3, factor=3, dropout = 0.1, activation = GELU, learning\_rate = 0.001) for the Electricity dataset,

(d\_model = 512, d\_ff = 512, e\_layers = 2, d\_layers = 1, n\_heads = 8, top\_k = 5, factor=3, dropout = 0.1, activation = GELU, learning\_rate = 0.001) for the Traffic dataset.

\item \textbf{PatchMLP}~\cite{tang2025patchmlp}: The hyperparameters for the PatchMLP model are as follows:

(d\_model = 1024, d\_ff = 2048, e\_layers = 1, d\_layers = 1, moving\_avg = 13, dropout = 0.1, activation = GELU, learning\_rate = 0.0001) for all datasets.

\item \textbf{FilterTS}~\cite{wang25filterts}: The hyperparameters for the FilterTS model are as follows:

(d\_model = 128, d\_ff = 512, e\_layers = 4, d\_layers = 1, quantile = 0.9, top\_K\_static\_freqs = 10, filter\_type = all, dropout = 0.1, activation=gelu, learning\_rate = 0.001) for the ETTh1, ETTm1 datasets,

(d\_model = 512, d\_ff = 2048, e\_layers = 3, d\_layers = 1, quantile = 0.9, top\_K\_static\_freqs = 10, filter\_type = all, dropout = 0.1, activation=gelu, learning\_rate = 0.0001) for the ETTh2, ETTm2 datasets,

(d\_model = 128, d\_ff = 512, e\_layers = 2, d\_layers = 1, quantile = 0.9, top\_K\_static\_freqs = 10, filter\_type = all, dropout = 0.1, activation=gelu, learning\_rate = 0.005) for the Weather dataset,

(d\_model = 512, d\_ff = 2048, e\_layers = 2, d\_layers = 1, quantile = 0.9, top\_K\_static\_freqs = 10, filter\_type = all, dropout = 0.1, activation=gelu, learning\_rate = 0.0005) for the Exchange dataset,

(d\_model = 256, d\_ff = 2048, e\_layers = 2, d\_layers = 1, quantile = 0.9, top\_K\_static\_freqs = 10, filter\_type = all, dropout = 0.1, activation=gelu, learning\_rate = 0.005) for the Electricity, Traffic datasets.

\item \textbf{DLinear}~\cite{zeng2023transformers}: The only hyperparameter for this baseline model is the (moving\_average = 25) for the series decomposition module from Autoformer~\cite{chen21autoformer}. We set the Individual to True so that there are separate linear models for each number of input variables.

\end{itemize}

\subsection{Metric Details}
\label{appendix:metrics}
For comprehensive evaluation of LTSF performance, we utilize two widely-accepted metrics: Mean Squared Error (MSE) and Mean Absolute Error (MAE). MSE penalizes larger errors more significantly, while MAE provides a more direct measure of the average error magnitude. The formal definitions of these metrics are as follows:
\begin{equation}
\text{MSE} = \frac{1}{H} \sum_{i=1}^{H} (X_i - \hat{X}_i)^2,
\end{equation}
\begin{equation}
\text{MAE} = \frac{1}{H} \sum_{i=1}^{H} |X_i - \hat{X}_i|,
\end{equation}
where $X \in \mathbb{R}^{H \times C}$ and $\hat{X} \in \mathbb{R}^{H \times C}$ represent the ground truth and the predicted sequences, respectively, for a forecast horizon of length $H$ and $C$ channels. $X_i$ and $\hat{X}_i$ denote the values at the $i$-th future time point.

\subsection{Training Details}
\label{appendix:training_details}

All experiments were conducted on NVIDIA RTX 3090 GPU with the PyTorch 1.13 framework. Our model, CometNet, is trained following a multi-stage training strategy to ensure stable convergence and optimal performance of its distinct components.

The training process initiates with the Contextual Motif Extraction module, which performs a one-time execution per dataset to construct the dominant motif library. This critical preprocessing step incorporates several key parameters optimized through empirical validation: we identify the top $N_s=10$ most significant periods via FFT analysis; use a Pearson correlation similarity threshold of $\tau_s=0.7$ for the preliminary anchor-based clustering; each scale maintains a maximum of 10 clusters to ensure diversity. The candidate motifs are further refined using a graph correlation threshold of $\tau_g=0.8$, with the final dominant motif selection determined by a weighted composite quality score ($\alpha_S = 0.6$ for saliency, $\alpha_P=0.2$ for prevalence, and $\alpha_A=0.2$ for atomicity.)  

The main forecasting model is trained in a three-phase manner to ensure each component is effectively optimized.

\subsubsection{Phase 1: Window Embedding Pre-training.} 
To establish a robust feature representation, we first pre-trained the Window Embedding module independently. The module is coupled with a lightweight linear projection head and optimized on the standard forecasting task. This pre-training phase ensures the embeddings capture robust temporal patterns before downstream specialization. The loss function is the MSE loss: 
\begin{equation}
\mathcal{L}_{pre} = L_{MSE}(Linear(Embed(\mathcal{X}_{t-L+1:t})), \mathcal{X}_{t+1:t+H}).
\end{equation}

\subsubsection{Phase 2: Gating Network Specialization.}
With the embedding weights frozen, we then specialize the Motif-driven Gating Network. This phase begins by annotating the training set with ground-truth motif class labels and normalized positional scores, both derived from the pre-established dominant motif library. The gating network is subsequently trained via a joint loss function combining cross-entropy (for class discrimination) loss and MSE (for position refinement) loss, enforcing precise motif-aware routing behavior while maintaining temporal localization accuracy.
\begin{equation}
\mathcal{L}_{gate} = L_{Cross}(p_t, y_{class}) + \alpha_{pos} \cdot L_{MSE}(s_t, y_{pos}),
\end{equation}
where $L_{Cross}$ is the Cross-Entropy loss for the classification of motif selection probabilities $p_t$ against the ground-truth class $y_{class}$, and $L_{MSE}$ measures the MSE error of the predicted positional score $s_t$ against the ground-truth score $y_{pos}$. $\alpha_{pos}$ is a hyperparameter balancing the two terms.

\subsubsection{Phase 3: End-to-End Forecasting.}
Finally, the entire architecture, including the pre-trained embedder, the specialized gating network, and the expert networks, is jointly fine-tuned to harmonize all components for the forecasting task. The final training loss combines MSE (forecasting error) and an MoE load-balancing term (expert utilization) to encourage balanced expert utilization: 
\begin{equation}
\mathcal{L}_{final} = L_{MSE}(\hat{\mathcal{X}}_{t+1:t+H}, \mathcal{X}_{t+1:t+H}) + \gamma \cdot L_{balance},
\end{equation}
where $L_{balance}$ is the MoE load-balancing loss and $\gamma$ is its corresponding weighting coefficient.

\subsubsection{Implementation Details.}
We set the embedding dimension to $d_e=256$. The Window Embedding module consists of a 2-layer MLP with a LeakyReLU activation function, which maps the input look-back window of length $L$ to the latent representation $e_t \in \mathbb{R}^{d_e}$. The Motif-driven Gating Network is designed to be lightweight; both its routing head $H_{route}$ and position head $H_{pos}$ are implemented as single linear layers that map the embedding $e_t$ to the expert logits and the positional score, respectively. Within the Context-conditioned Experts, the position encoder $\Phi_{pos}$ is a 2-layer MLP that elevates the scalar positional score to a high-dimensional embedding. This is then concatenated with the window embedding and processed by a 2-layer conditional fusion MLP. Finally, each of the $K$ expert-specific prediction heads is a 3-layer MLP that maps the final conditional representation to the forecast horizon of length $H$.

During the training process, we use the AdamW optimizer with an initial learning rate of $l_r=0.001$, which is decayed using a cosine annealing scheduler. The batch size is set to 256. The loss-balancing hyperparameters are empirically set to $\alpha_{pos}=1.0$ and $\gamma=0.1$. To prevent overfitting, an early stopping mechanism with a patience of 7 epochs is applied based on the validation loss. To ensure robust and reproducible results, we repeat each experiment 3 times with different random seeds and report the mean performance metrics, reducing the impact of randomness and enhancing statistical reliability.

\section{Motif Extraction Details}
\label{appendix:motif_details}
In this section, we provide a detailed explanation of the design details of our Contextual Motif Extraction module.

\subsection{Multi-scale Candidate Discovery}
\subsubsection{Medoid Calculation.}
After assigning all subsequences at a given scale to their nearest cluster centroids, we identify the most representative subsequence from each cluster, known as the medoid. To enhance the robustness of morphological similarity assessment, we adopt DTW instead of Pearson correlation, as DTW effectively accommodates temporal misalignments while precisely capturing shape-based pattern similarities, thereby ensuring the selected medoid represents the true structural center of its cluster, not merely a point of high density. The computational expense of DTW is strategically managed, as it is applied only within the pre-filtered and significantly smaller clusters, making this precise metric computationally feasible.

The medoid $c_{s,k}$ of a cluster $\mathcal{C}_{s,k}^{sub}$ is formally defined as the subsequence with the minimum average DTW distance to all other members within that cluster:
\begin{equation}
c_{s,k} = \underset{c_i \in \mathcal{C}_{s,k}^{sub}}{\text{argmin}} \frac{1}{|\mathcal{C}_{s,k}^{sub}|} \sum_{c_j \in \mathcal{C}_{s,k}^{sub}} DTW(c_i, c_j),
\end{equation}
where $DTW(\cdot, \cdot)$ represents the standard Dynamic Time Warping distance, $s=\lambda_s$, $k \in [1, N_c]$ denote the selected scale and cluster, respectively. By performing this calculation for each of the $N_c$ clusters at a given scale $\lambda_s$, we compile the set of candidate motifs $\mathcal{C}_s = \{c_{s,1}, \dots, c_{s,N_s}\}$, which encapsulates the most representative temporal patterns discovered at that specific scale.

\subsection{Cross-scale Redundancy Filtering}
\subsubsection{DTW Similarity Score.}
In the Cross-scale Redundancy Filtering stage, we convert the DTW distance, which measures dissimilarity, into a normalized similarity score $S_{DTW} \in [0, 1]$. This is achieved using a Gaussian kernel function, which maps small distances to high similarity scores and large distances to low ones:
\begin{equation}
S_{DTW}(c_i, c_j) = \exp\left(-\frac{DTW(c_i, c_j)^2}{\sigma^2}\right),
\end{equation}
where $\sigma$ is a scaling parameter that controls the sensitivity of the similarity decay. This allows us to model the relationships between motifs in a probabilistic manner.

\subsection{Dominant Motif Selection Metrics.}
\subsubsection{Intrinsic Quality.}
The benefit-driven selection process relies on a set of precise metrics to evaluate the quality, coverage, and diversity of candidate motifs. Specifically, the intrinsic quality $Q$ of each candidate motif $c_i^*$ is a weighted sum of three normalized properties: Saliency, Prevalence, and Atomicity. The individual components are defined as follows:

\begin{itemize}
\item \textbf{Saliency $Q_S$:} Saliency quantifies the prominence of a motif by measuring its most extreme deviation relative to its own internal statistics. It is calculated as the maximum deviation of the motif's peak or trough from its own mean, normalized by its own standard deviation. This captures the motif's relative extremity, ensuring that genuinely prominent patterns receive higher scores. 
\begin{equation}
Q_S(c_i^*) = \frac{\max(|v_{peak} - \mu_{c_i^*}|, |v_{trough} - \mu_{c_i^*}|)}{\sigma_{c_i^*}},
\end{equation}
where $v_{peak}$ and $v_{trough}$ are the maximum and minimum values within the candidate motif $c_i^*$, while $\mu_{c_i^*}$ and $\sigma_{c_i^*}$ are the mean and standard deviation of the motif $c_i^*$ itself.

\item \textbf{Prevalence $Q_P$:} This metric efficiently estimates a motif's recurrence throughout the historical time series. To avoid a computationally prohibitive brute-force search, we efficiently estimate prevalence by leveraging the initial clustering results. The prevalence of a motif $c_i^*$ is defined as the aggregated support from all original subsequence clusters that constitute its representative connected component in the similarity graph.
\begin{equation}
Q_P(c_i^*) = \sum_{c_j \in \mathcal{V}_i} Support(c_j),
\end{equation}
where $\mathcal{V}_i$ is the vertex set of the connected component represented by $c_i^*$, and $Support(c_j)$ is the size of the initial cluster from which the candidate $c_j$ was derived.

\item \textbf{Atomicity $Q_A$:} In our framework, atomicity is implemented as a simplicity score that penalizes longer, more complex patterns. This encourages the selection of concise, atomic motifs. It is formulated as the inverse of the logarithm of the pattern's length.
\begin{equation}
Q_A(c_i^*) = \frac{1}{\log(1 + len(c_i^*))}.
\end{equation}
\end{itemize}

\begin{figure*}[t!]
    \centering
    
    % --- 第一行 ---
    \subfloat[Similarity Threshold $\tau_s$.]{%
        \includegraphics[width=0.45\textwidth]{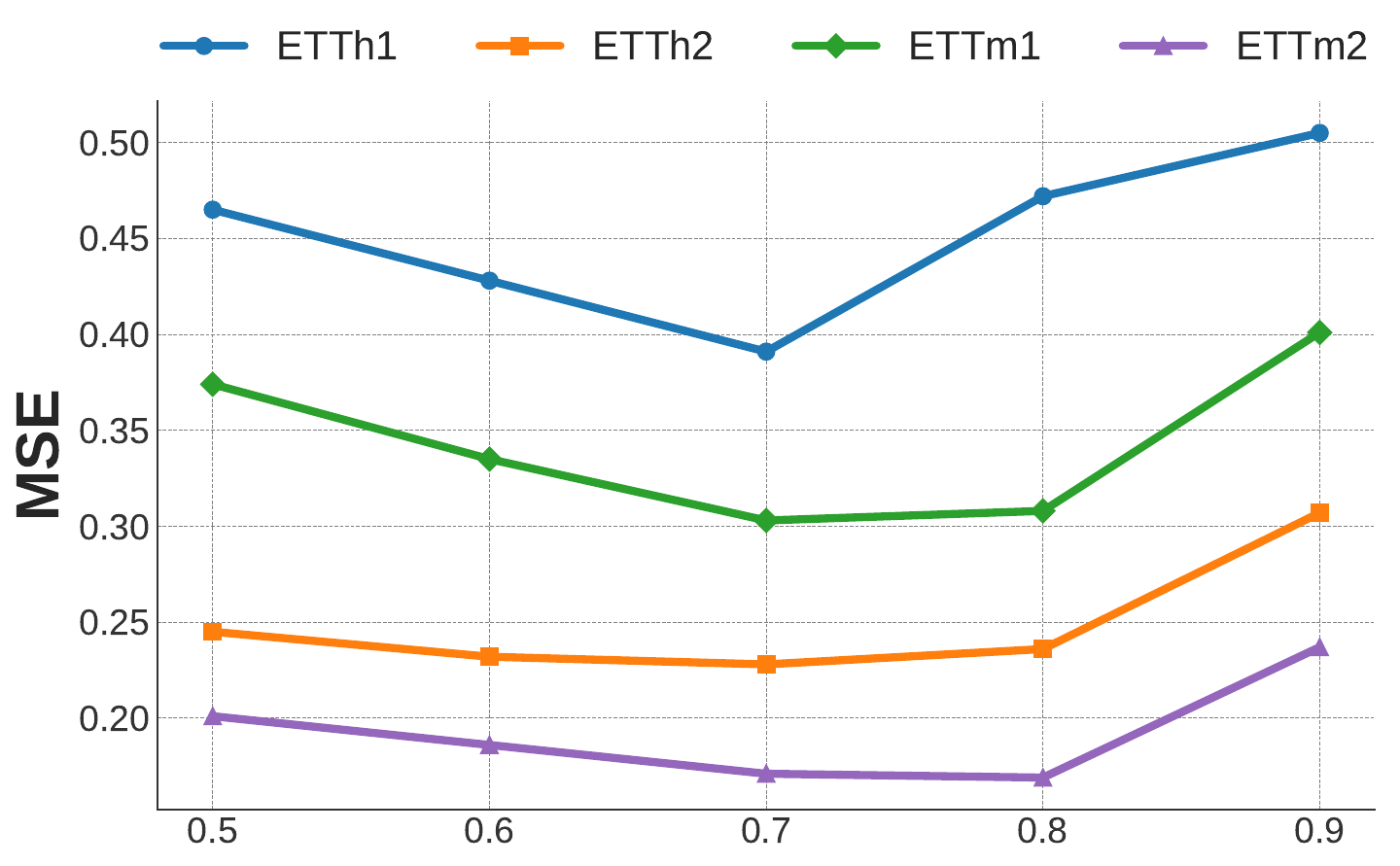}%
        \label{fig:sensitive_s}%
    }
    \hfill % 在图片间添加水平间距
    \subfloat[Redundancy Threshold $\tau_g$.]{%
        \includegraphics[width=0.45\textwidth]{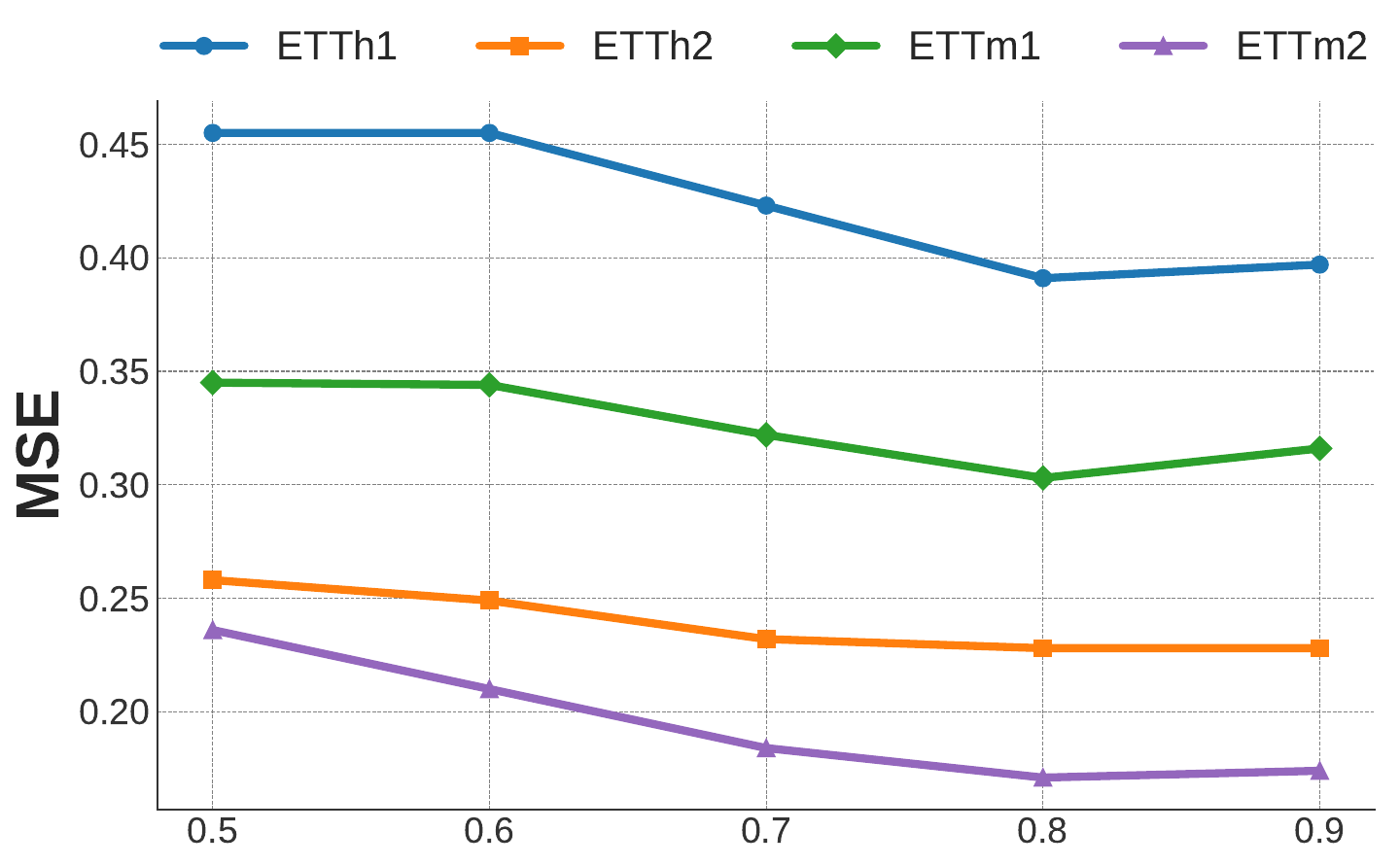}%
        \label{fig:sensitive_g}%
    }
    
    \vspace{0.3cm} % 两行间的垂直间距
    
    % --- 第二行 ---
    \subfloat[Hidden Dimension $d_{e}$.]{%
        \includegraphics[width=0.45\textwidth]{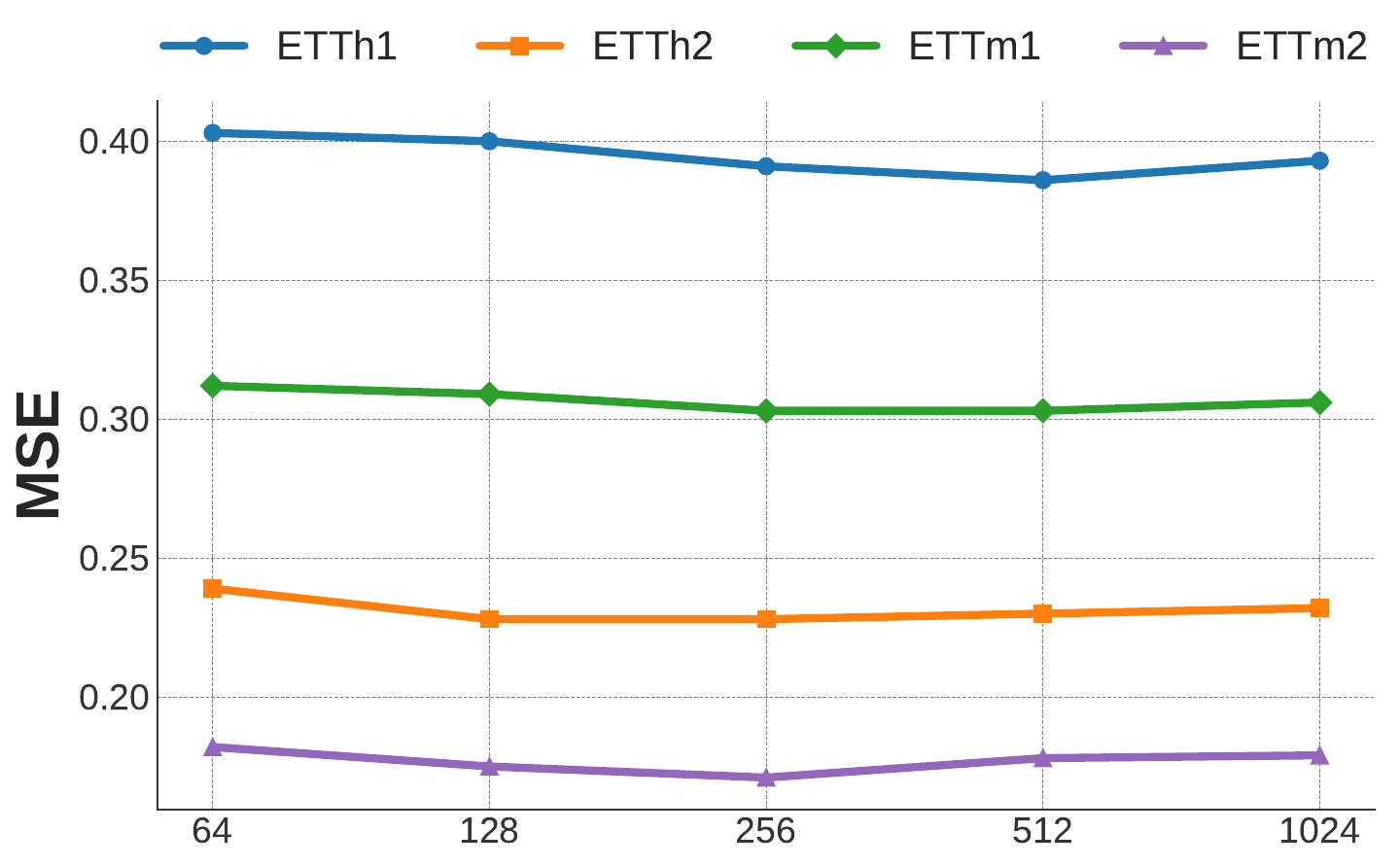}%
        \label{fig:sensitive_d_embedding}%
    }
    \hfill % 在图片间添加水平间距
    \subfloat[Learning Rate $l_r$.]{%
        \includegraphics[width=0.45\textwidth]{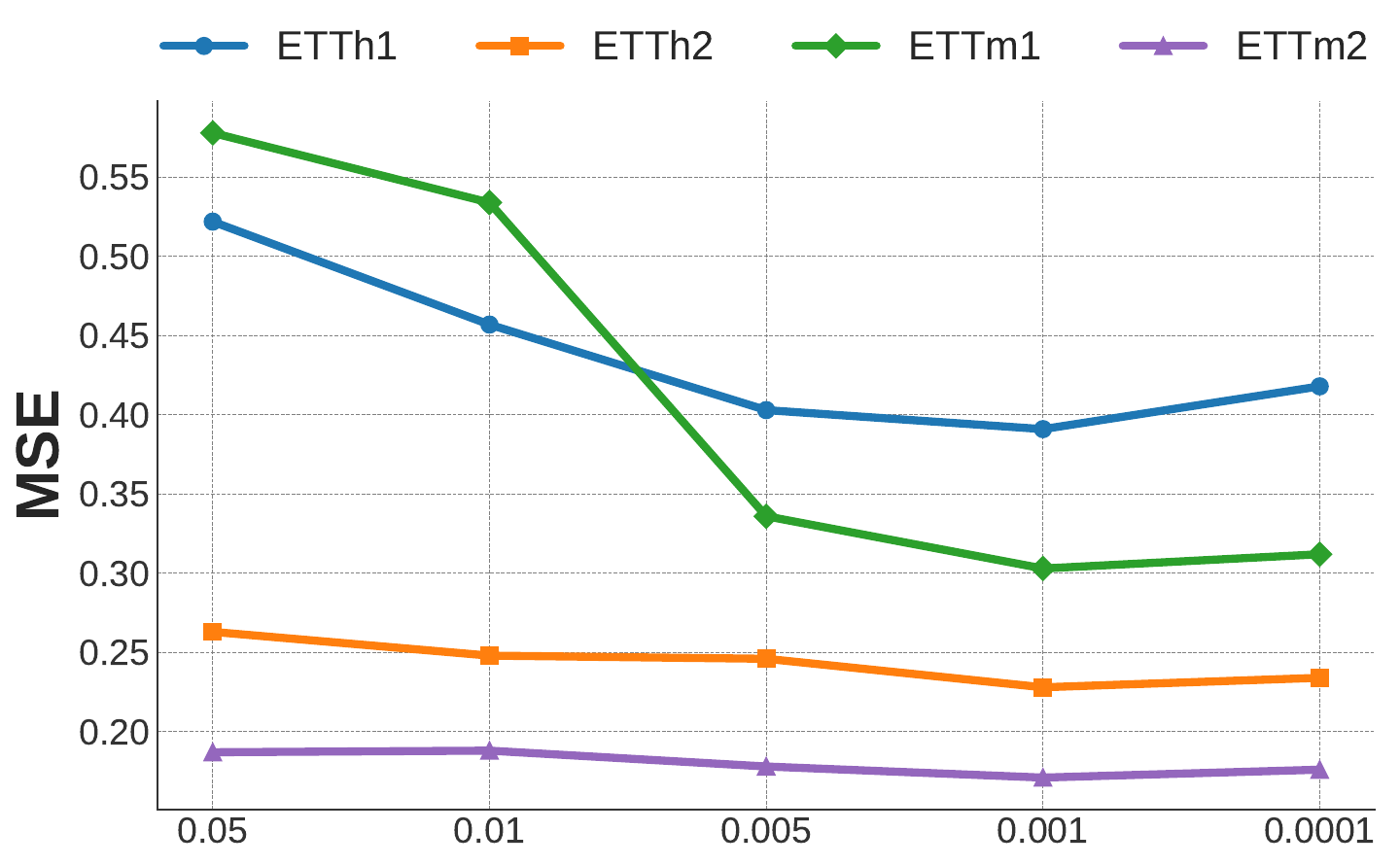}%
        \label{fig:sensitive_lr}%
    }
    
    \caption{Sensitivity analysis of key hyperparameters on the ETT datasets ($L=96$, $H=720$). We report the MSE metric for different values of (a) similarity threshold $\tau_s$, (b) redundancy threshold $\tau_g$, (c) hidden dimension $d_e$, and (d) learning rate $l_r$. The lowest point on each curve indicates the best-performing value for that hyperparameter.}
    \label{fig:sensitive_analysis}
\end{figure*}

\subsubsection{Benefit Score.}
Then, the benefit score dynamically evaluates a candidate by jointly considering its intrinsic quality, its contribution to temporal coverage, and its diversity relative to already selected motifs.

\begin{itemize}
\item \textbf{Marginal Coverage $Cov$:} This metric measures the proportion of a candidate's temporal footprint that is novel. Let $T(c)$ be the set of all time points covered by the occurrences of a motif $c$. The marginal coverage of adding candidate $c_i^*$ to the selected set $\mathcal{S}$ is:
\begin{equation}
Cov(c_i^* \mid \mathcal{S}) = \frac{| T(c_i^*) \setminus \bigcup_{m_k \in \mathcal{S}} T(m_k) |}{| T(c_i^*) |}.
\end{equation}

\item \textbf{Marginal Diversity $Div$}: This metric serves as a penalty factor that discourages the selection of motifs that are too similar to those already in the library $\mathcal{S}$. It is calculated based on the maximum similarity between the candidate $c_i^*$ and any motif $m_k$ already selected.
\begin{equation}
Div(c_i^* \mid \mathcal{S}) = \left(1 - \max_{m_k \in \mathcal{S}} S_{DTW}(c_i^*, m_k)\right)^{\gamma_d}
\end{equation}
where $\gamma_d$ is a hyperparameter to control the penalty intensity, we set $\gamma_d = 2$ in our experiments.
\end{itemize}

\section{Real-world Dataset Experiments}
\label{appendix:experiment_details}

\subsection{Hyperparameter Sensitivity Analysis}
\label{appendix:sensitivity}
To investigate the impact of key hyperparameters on our model's performance and demonstrate its robustness, we conduct a comprehensive sensitivity analysis, with results presented in Fig.~\ref{fig:sensitive_analysis}. The experiments are performed on all four ETT datasets for the long-horizon forecasting task ($L=96$, $H=720$), varying one hyperparameter at a time while keeping others at their optimal values. 

The quality of extracted motifs, which is crucial for our framework, is governed by the similarity threshold $\tau_s$ and the redundancy threshold $\tau_g$. As shown in Fig.~\ref{fig:sensitive_s}, model performance is sensitive to $\tau_s$; a value that is too low (e.g., 0.5) may incorrectly group dissimilar patterns and introduce noise, while a value that is too high (e.g., 0.9) can lead to an overly fragmented pattern set lacking generalization. Performance peaks when $\tau_s$ is set to 0.7. Similarly, Fig.~\ref{fig:sensitive_g} shows that a redundancy threshold $\tau_g$ of 0.8 consistently yields strong results, indicating that our model benefits from a well-curated library of motifs that balances representativeness and diversity. 

Beyond these extraction parameters, we also examine the sensitivity to the hidden dimension $d_e$ and the learning rate $l_r$. As depicted in Fig.~\ref{fig:sensitive_d_embedding}, CometNet exhibits remarkable stability with respect to the hidden dimension. The MSE remains consistently low for a wide range of $d_e$ values from 128 to 1024, suggesting that our model is not prone to overfitting with larger dimensions and can achieve good performance even with a relatively compact architecture. However, as shown in Fig.~\ref{fig:sensitive_lr}, the choice of the learning rate is more critical. An overly large learning rate (e.g., 0.05) can lead to unstable training and degraded performance, while a very small one can slow down convergence. The optimal performance is achieved with a learning rate of 0.001, highlighting the importance of selecting an appropriate learning rate for model optimization. 

In summary, while performance is understandably influenced by the parameters defining the motif discovery process, our model demonstrates excellent robustness to variations in core architectural and optimization hyperparameters, highlighting its stability and ease of use.

\subsection{Hyperparameter Tuning for Baselines}

Since fair hyperparameter tuning is crucial for model comparison, in this section, we conducted a grid search for SOTA baseline models to ensure each was evaluated under its optimal configuration. As shown in Table \ref{tab:d_model_tuning} and Table \ref{tab:lr_tuning}, the optimal hyperparameters (e.g., $d_{model}$ and learning rate $lr$) vary across different models. The results indicate that our model consistently maintains superior performance across various hyperparameter settings, demonstrating its robustness and superiority.
\begin{table}[h]
\centering
\caption{MSE performance under different $d_{model}$ settings.}
\label{tab:d_model_tuning}
\resizebox{\columnwidth}{!}{%
\begin{tabular}{ccccc}
\toprule
$d_{model}$ & Ours & PatchMLP & iTransformer & TimeMixer++ \\ \midrule
128  & 0.396 & 0.870 & 0.720 & 0.628 \\
256  & 0.391 & 0.709 & 0.663 & 0.632 \\
512  & 0.394 & 0.657 & 0.601 & 0.595 \\
1024 & 0.393 & 0.591 & 0.631 & 0.474 \\ \bottomrule
\end{tabular}
}
\end{table}

\begin{table}[h]
\centering
\caption{MSE performance under different lr settings.}
\label{tab:lr_tuning}
\resizebox{\columnwidth}{!}{%
\begin{tabular}{ccccc}
\toprule
lr & Ours & PatchMLP & iTransformer & TimeMixer++ \\ \midrule
$1 \times 10^{-3}$ & 0.402 & 0.585 & 0.640 & 0.603 \\
$5 \times 10^{-3}$ & 0.398 & 0.583 & 0.636 & 0.726 \\
$1 \times 10^{-4}$ & 0.391 & 0.591 & 0.601 & 0.474 \\
$5 \times 10^{-4}$ & 0.396 & 0.581 & 0.623 & 0.535 \\ \bottomrule
\end{tabular}
}
\end{table}

\subsection{Statistical Significance Analysis}
To validate the superiority and stability of our model, we conducted five independent runs with different seeds and performed paired t-tests for significance analysis. As shown in Table \ref{tab:significance_test}, our model significantly outperforms TimeMixer++ on the MSE metric across all four benchmark datasets (ETTh1, ETTh2, ETTm1, and ETTm2), with p-values all being substantially less than 0.001. This result strongly demonstrates the significant advantages and reliability of our method.
\begin{table}[h]
\centering
\caption{Comparison of MSE performance between our model and TimeMixer++ on the ETTh1, ETTh2, ETTm1, and ETTm2 datasets. We report the mean and standard deviation over 5 random seeds and calculate the p-value using a paired t-test.}
\label{tab:significance_test}
\resizebox{\columnwidth}{!}{%
\begin{tabular}{lccc}
\toprule
\textbf{MSE} & \textbf{Ours} & \textbf{TimeMixer++} & \textbf{P value} \\ \midrule
ETTh1 & $0.395 \pm 0.003$ & $0.527 \pm 0.030$ & $3.2 \times 10^{-4}$ \\
ETTh2 & $0.225 \pm 0.004$ & $0.436 \pm 0.004$ & $2.9 \times 10^{-7}$ \\
ETTm1 & $0.306 \pm 0.003$ & $0.467 \pm 0.004$ & $7.0 \times 10^{-8}$ \\
ETTm2 & $0.169 \pm 0.002$ & $0.394 \pm 0.003$ & $1.2 \times 10^{-6}$ \\ \bottomrule
\end{tabular}
}
\end{table}

\subsection{Extracted Motif Visualization}
\label{appendix:motif_visualization}
\begin{figure*}[h!]
    \centering
    
    \subfloat[ETTh1 Dataset Motifs\label{fig:motif_etth1}]{%
        \includegraphics[width=0.95\textwidth]{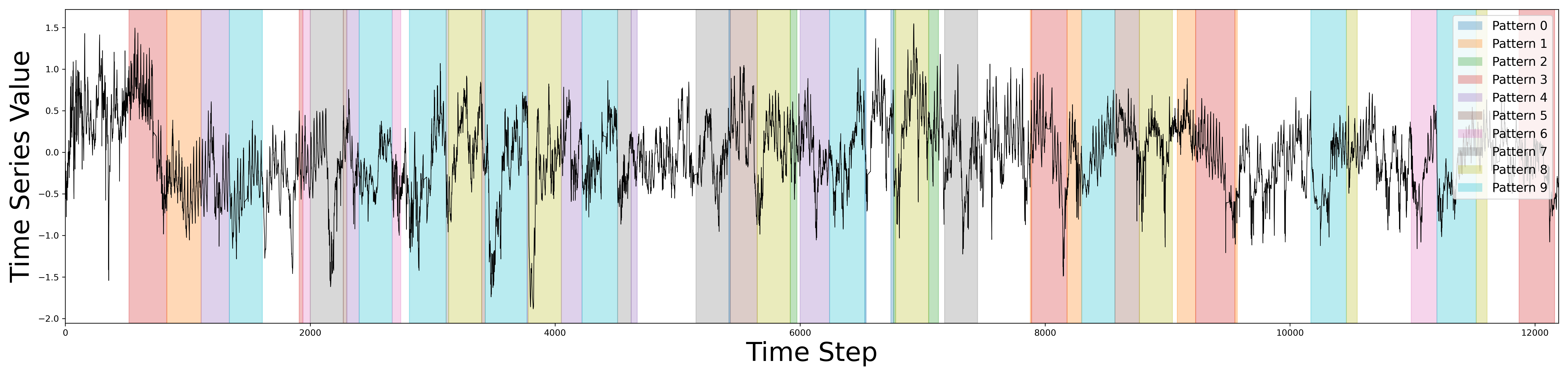}%
    }
    
    \vspace{0.1cm} % Small vertical space between rows
    
    \subfloat[ETTh2 Dataset Motifs\label{fig:motif_etth2}]{%
        \includegraphics[width=0.95\textwidth]{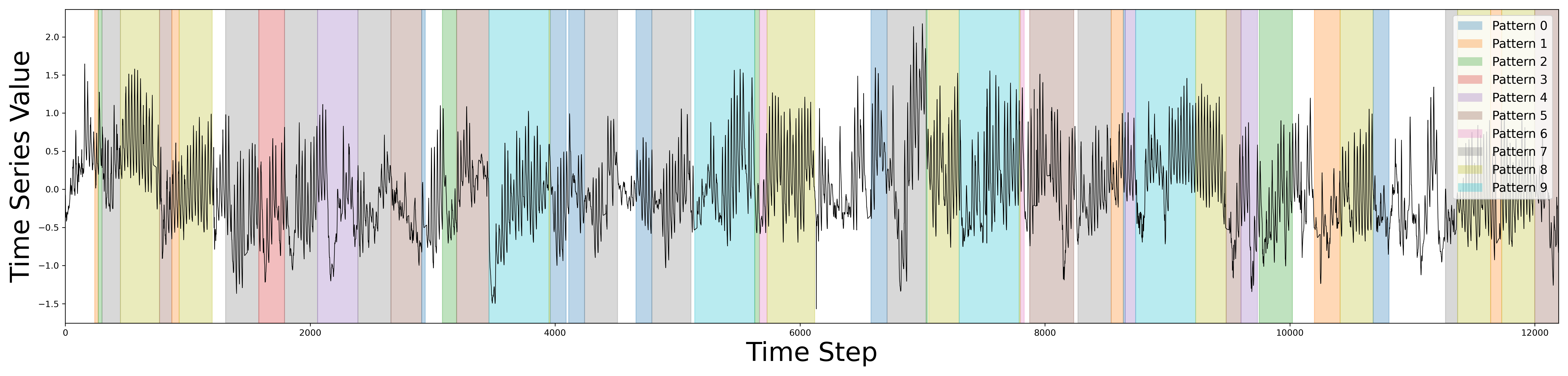}%
    }

    \vspace{0.1cm}
    
    \subfloat[ETTm1 Dataset Motifs\label{fig:motif_ettm1}]{%
        \includegraphics[width=0.95\textwidth]{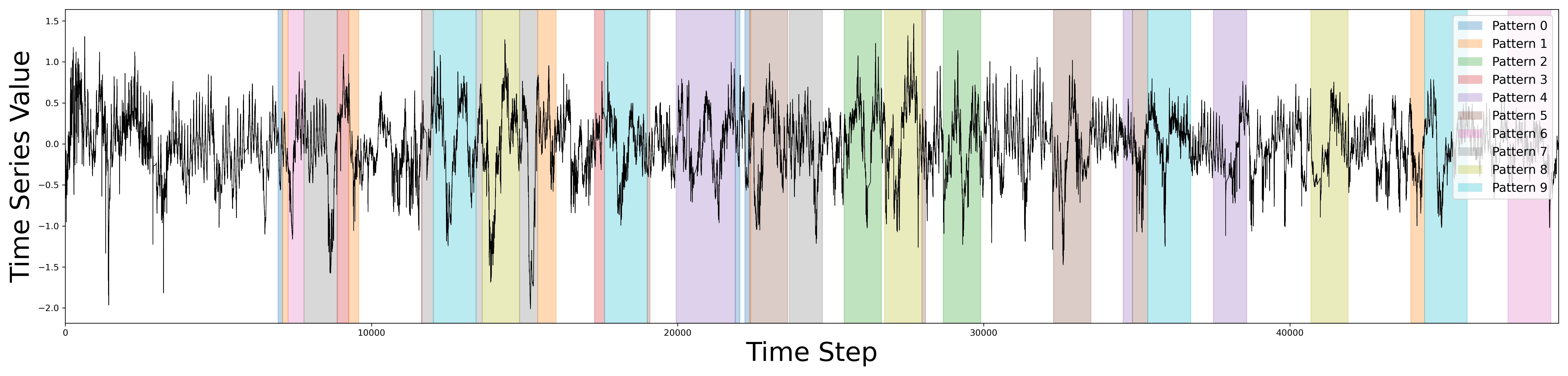}%
    }
    
    \vspace{0.1cm}

    \subfloat[ETTm2 Dataset Motifs\label{fig:motif_ettm2}]{%
        \includegraphics[width=0.95\textwidth]{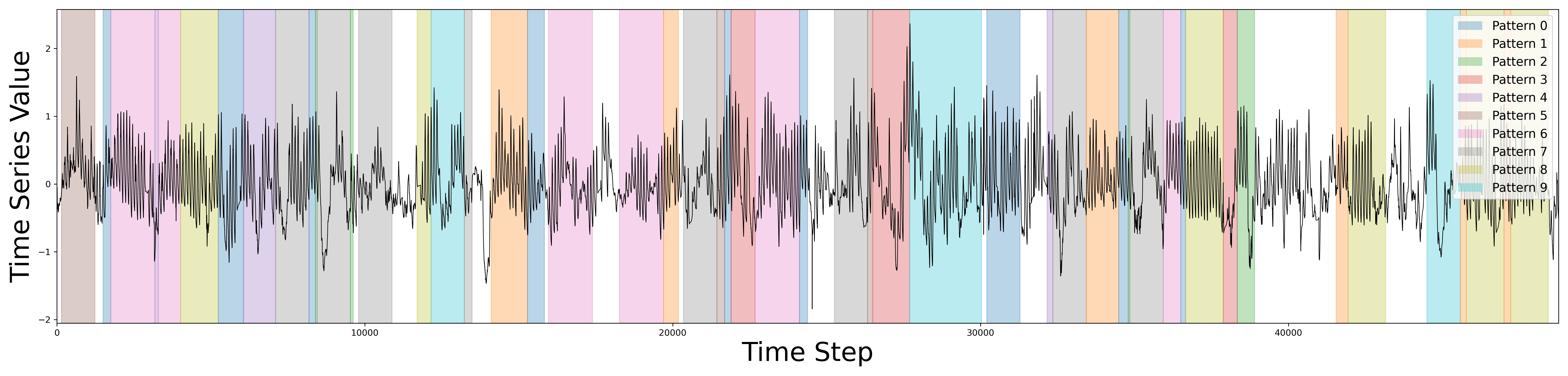}%
    }
    
    \caption{Visualization of dominant contextual motifs extracted from different ETT datasets.}
    \label{fig:motif_extraction_results}
\end{figure*}
To provide a qualitative understanding of the patterns discovered by our Contextual Motif Extraction module, Fig.~\ref{fig:motif_extraction_results} visualizes the results on representative sensors from four ETT datasets. Each panel displays the whole historical time series, with colored overlays highlighting the temporal locations of recurring dominant motifs. 

It is visually evident that our method successfully identifies a rich library of motifs with diverse characteristics. For instance, in the ETTh1 and ETTh2 datasets, we capture both short, spiky patterns and broader, smoother cyclical behaviors. Similarly, in the more fine-grained ETTm1 and ETTm2 datasets, it discerns high-frequency oscillatory patterns alongside lower-frequency trends. The colored regions demonstrate that these discovered motifs are not isolated events but recur throughout the series, validating our core hypothesis that time series are governed by such representative, long-contextual motifs. This visualization underscores the module's ability to automatically and effectively build a comprehensive motif library, which forms the foundation for our downstream forecasting task.

% ETTh1_96_96
\begin{figure*}[h!]
    \centering
    
    % --- Row 1 ---
    \subfloat[\textcolor{red}{\textbf{CometNet (Ours)}}]{%
        \includegraphics[width=0.33\textwidth]{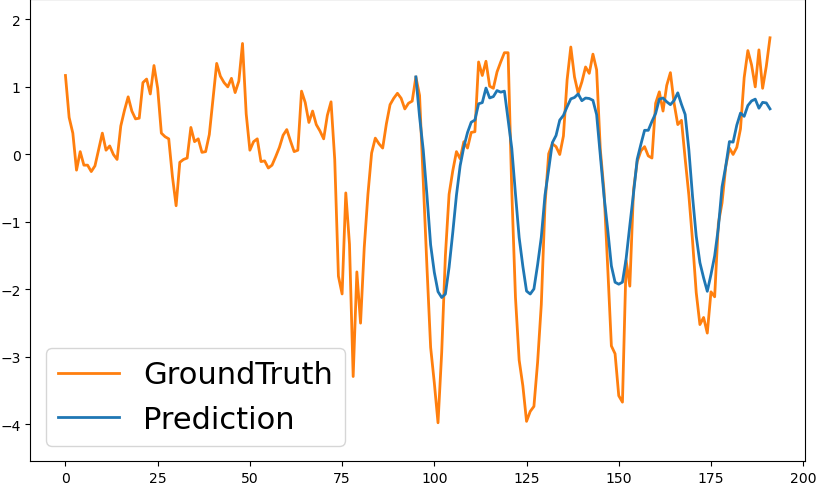}%
    }
    \hfill % Horizontal space between figures
    \subfloat[\textbf{TimeMixer++}]{%
        \includegraphics[width=0.33\textwidth]{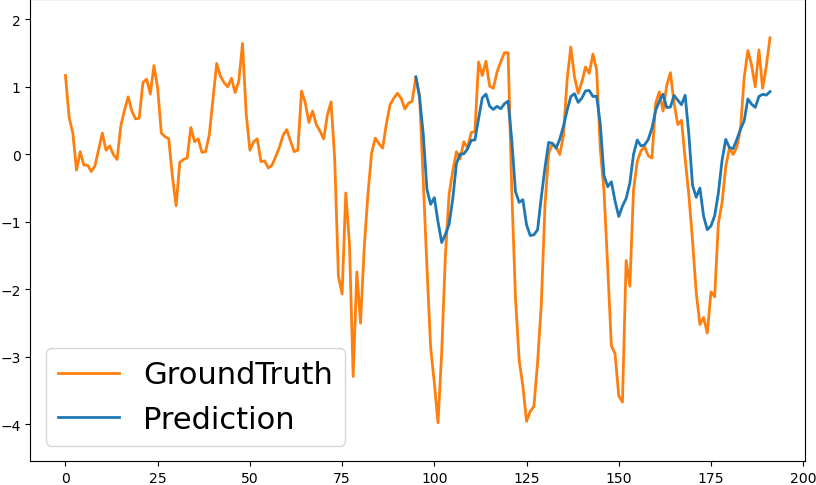}%
    }
    \hfill
    \subfloat[\textbf{iTransformer}]{%
        \includegraphics[width=0.33\textwidth]{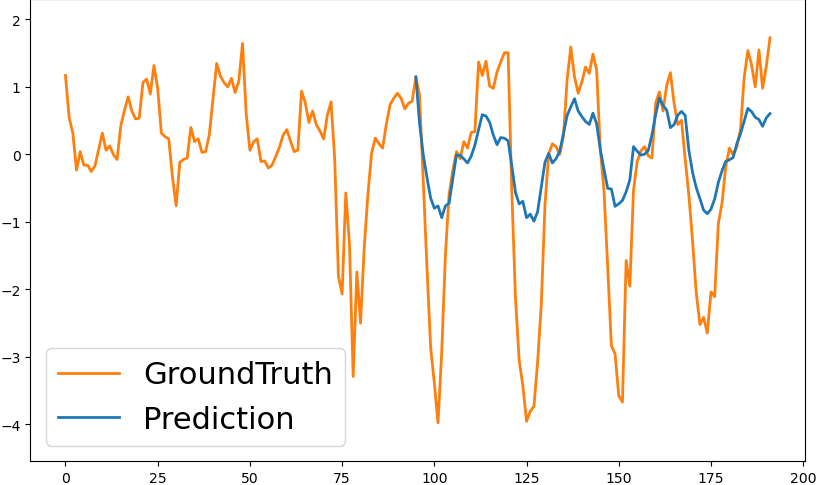}%
    }
    
    \vspace{0.5cm} % Vertical space between rows
    
    % --- Row 2 ---
    \subfloat[\textbf{PatchMLP}]{%
        \includegraphics[width=0.33\textwidth]{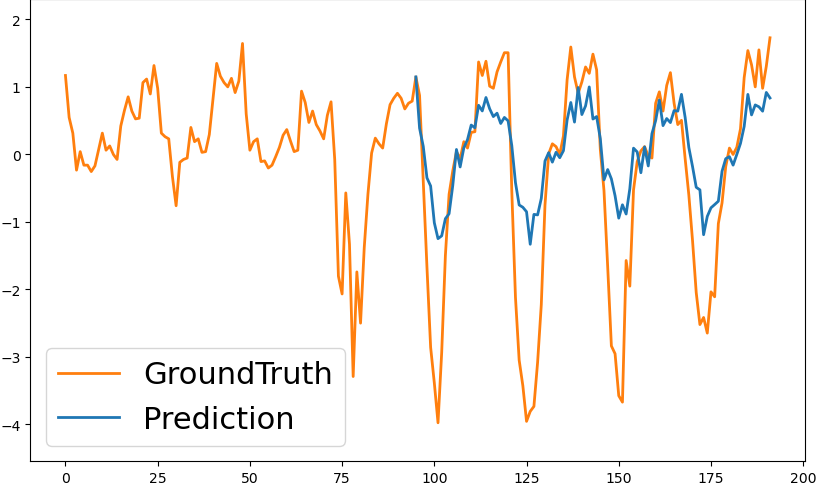}%
    }
    \hfill
    \subfloat[\textbf{PatchTST}]{%
        \includegraphics[width=0.33\textwidth]{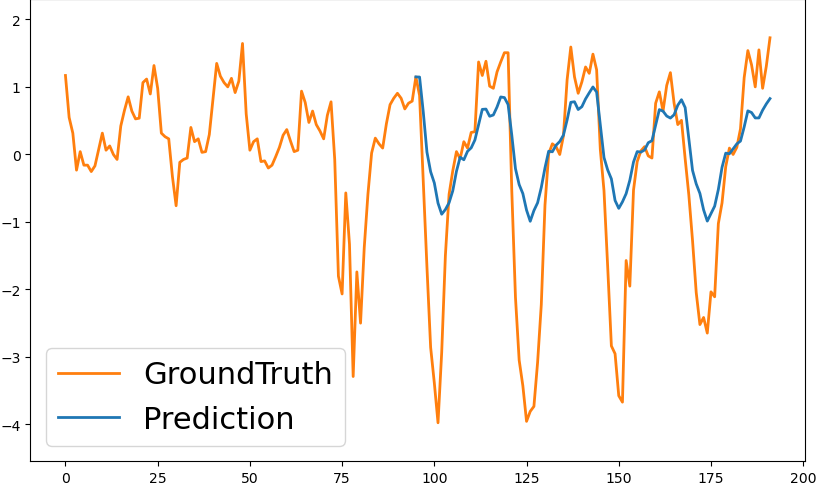}%
    }
    \hfill
    \subfloat[\textbf{DLinear}]{%
        \includegraphics[width=0.33\textwidth]{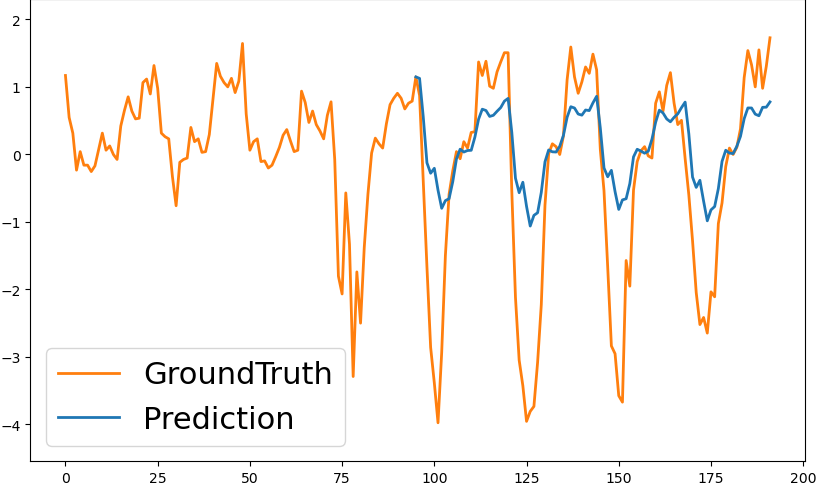}%
    }
    
    \caption{Visualization of input-96-predict-96 results on the ETTh1 dataset.}
    \label{fig:etth1_96}
\end{figure*}

% ETTh1_96_192
\begin{figure*}[h!]
    \centering
    
    % --- Row 1 ---
    \subfloat[\textcolor{red}{\textbf{CometNet (Ours)}}]{%
        \includegraphics[width=0.33\textwidth]{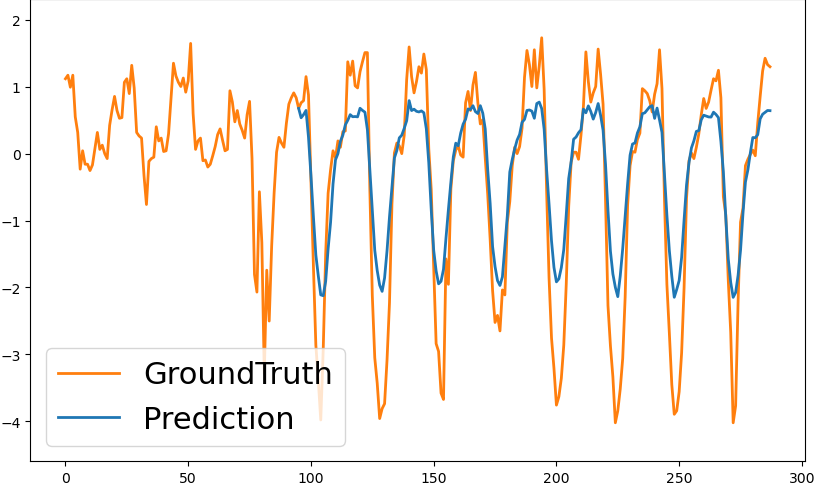}%
    }
    \hfill % Horizontal space between figures
    \subfloat[\textbf{TimeMixer++}]{%
        \includegraphics[width=0.33\textwidth]{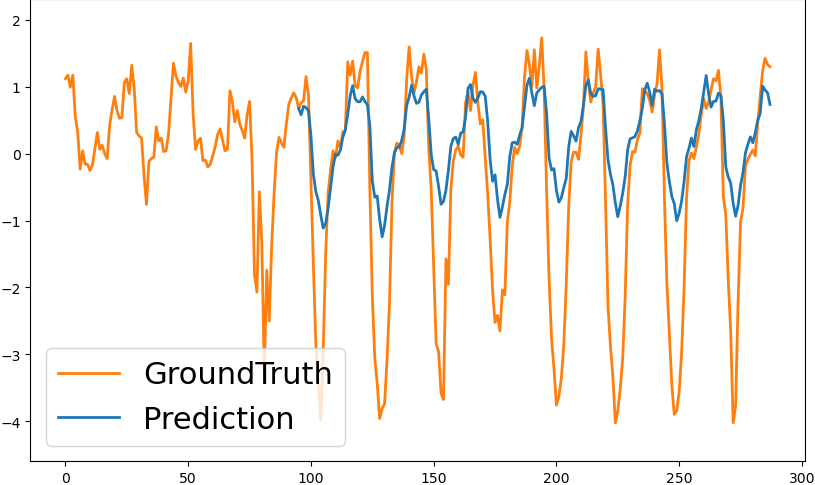}%
    }
    \hfill
    \subfloat[\textbf{iTransformer}]{%
        \includegraphics[width=0.33\textwidth]{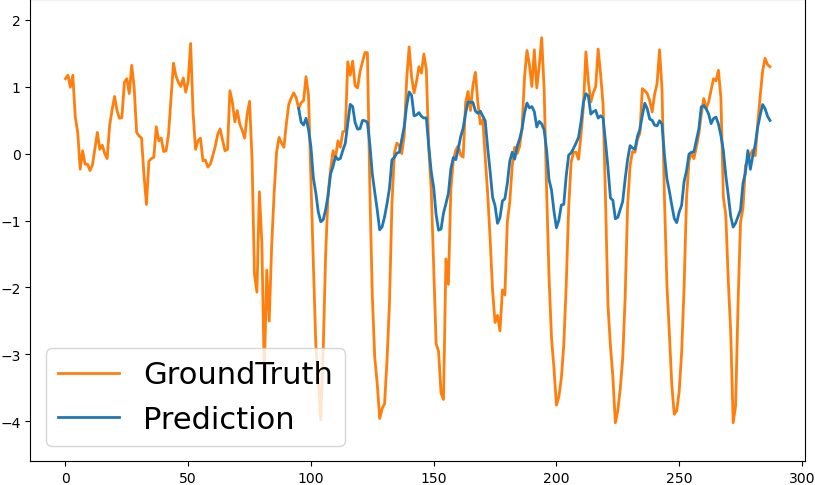}%
    }
    
    \vspace{0.5cm} % Vertical space between rows
    
    % --- Row 2 ---
    \subfloat[\textbf{PatchMLP}]{%
        \includegraphics[width=0.33\textwidth]{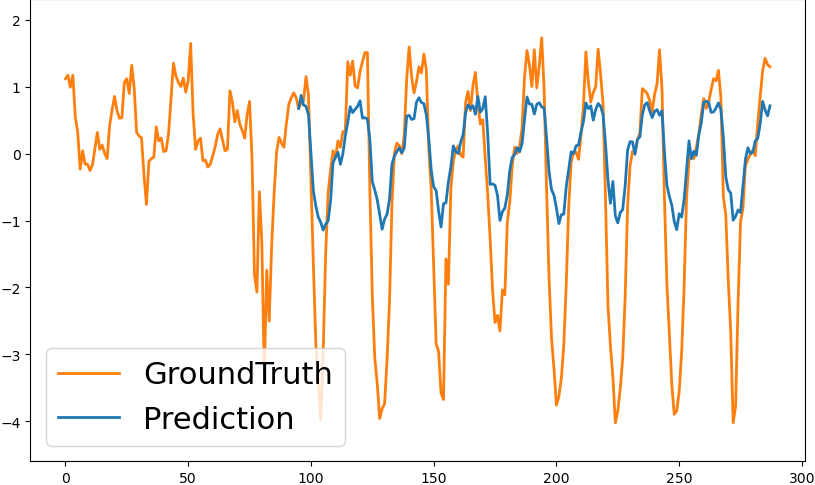}%
    }
    \hfill
    \subfloat[\textbf{PatchTST}]{%
        \includegraphics[width=0.33\textwidth]{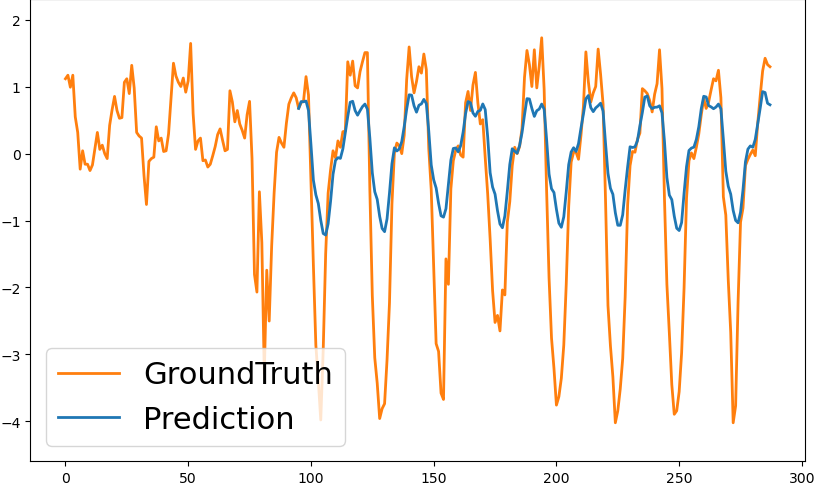}%
    }
    \hfill
    \subfloat[\textbf{DLinear}]{%
        \includegraphics[width=0.33\textwidth]{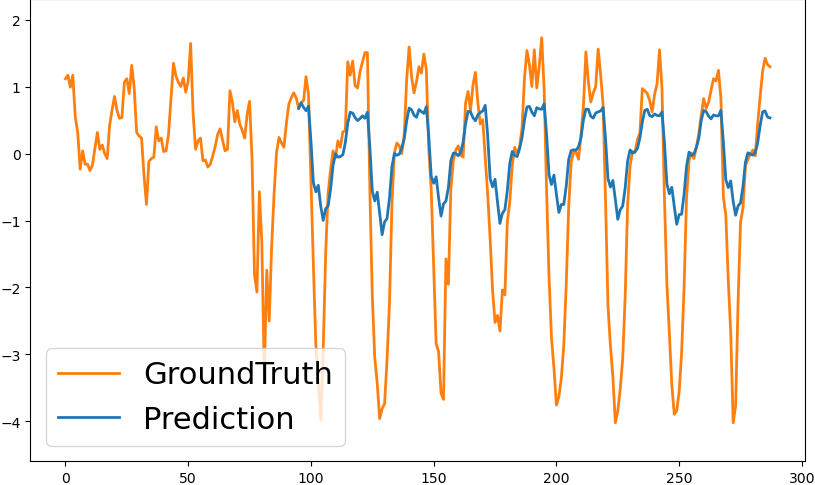}%
    }
    
    \caption{Visualization of input-96-predict-192 results on the ETTh1 dataset.}
    \label{fig:etth1_192}
\end{figure*}

% ETTh1_96_336
\begin{figure*}[h!]
    \centering
    
    % --- Row 1 ---
    \subfloat[\textcolor{red}{\textbf{CometNet (Ours)}}]{%
        \includegraphics[width=0.33\textwidth]{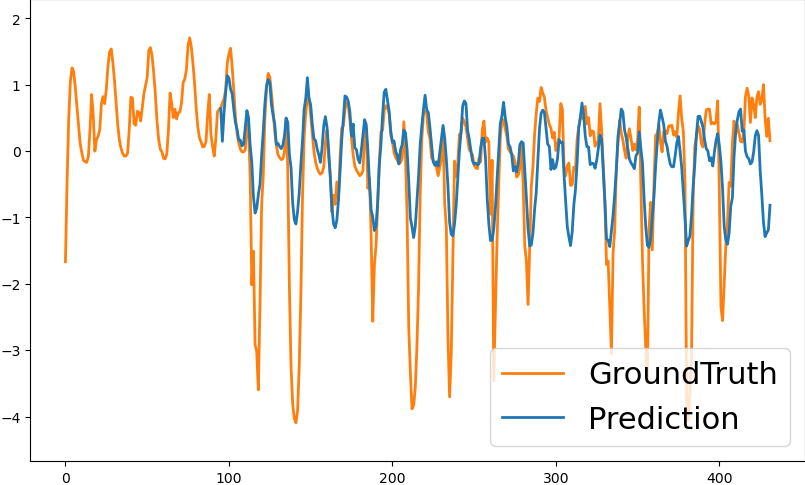}%
    }
    \hfill % Horizontal space between figures
    \subfloat[\textbf{TimeMixer++}]{%
        \includegraphics[width=0.33\textwidth]{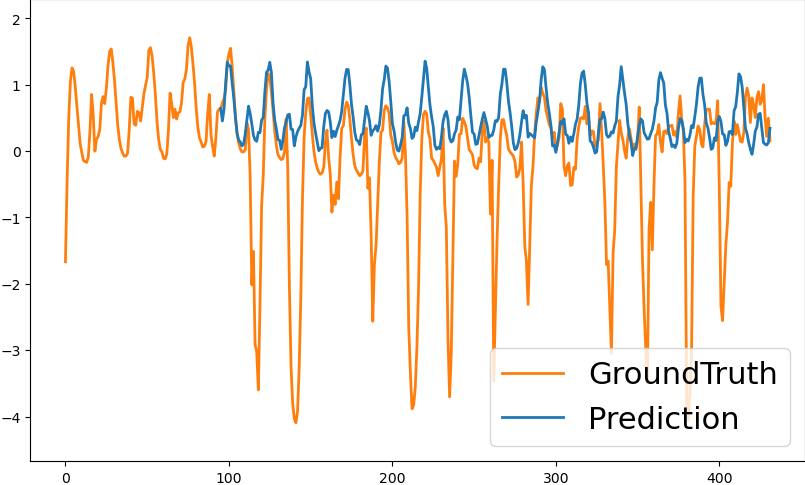}%
    }
    \hfill
    \subfloat[\textbf{iTransformer}]{%
        \includegraphics[width=0.33\textwidth]{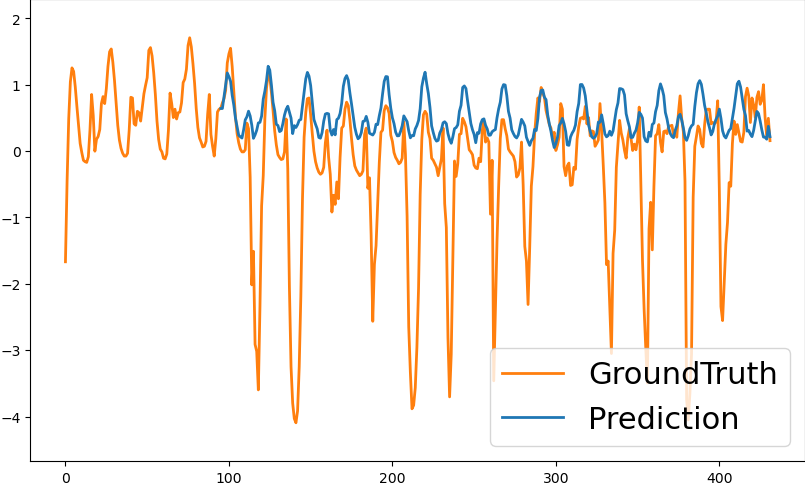}%
    }
    
    \vspace{0.5cm} % Vertical space between rows
    
    % --- Row 2 ---
    \subfloat[\textbf{PatchMLP}]{%
        \includegraphics[width=0.33\textwidth]{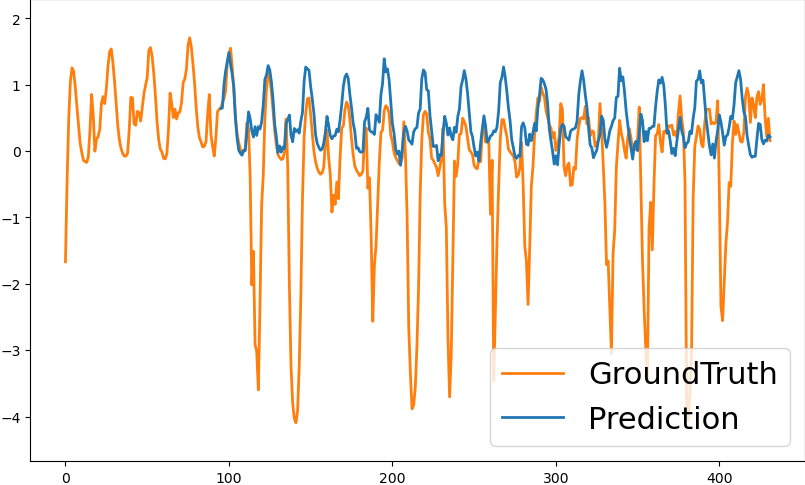}%
    }
    \hfill
    \subfloat[\textbf{PatchTST}]{%
        \includegraphics[width=0.33\textwidth]{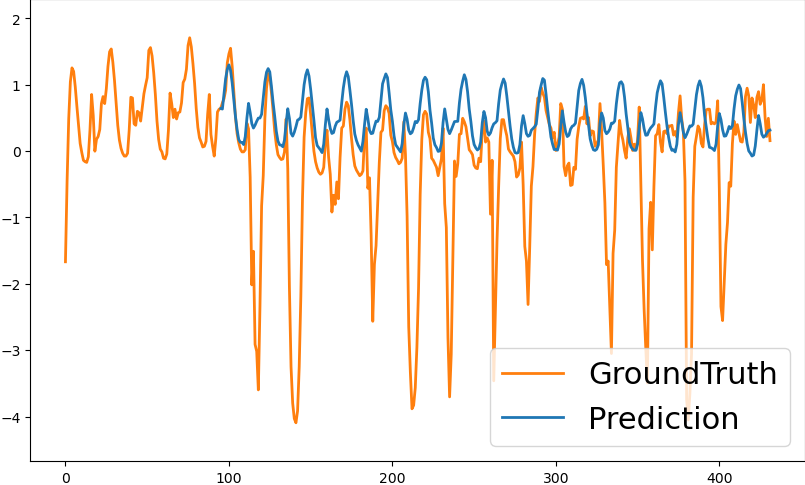}%
    }
    \hfill
    \subfloat[\textbf{DLinear}]{%
        \includegraphics[width=0.33\textwidth]{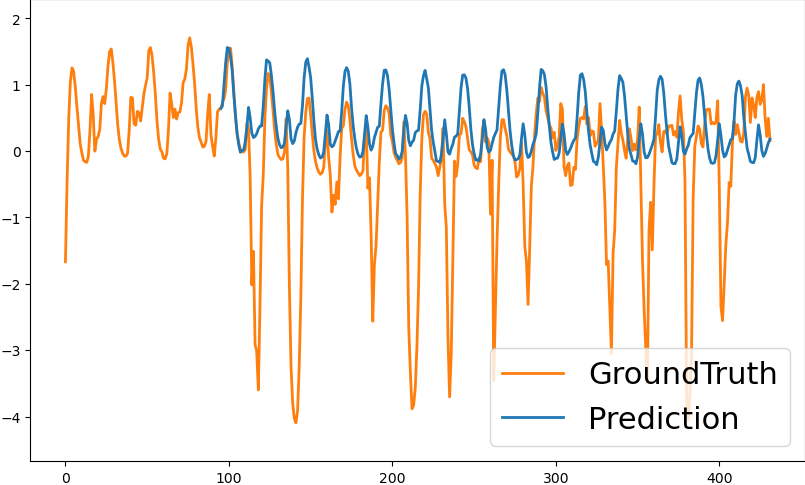}%
    }
    
    \caption{Visualization of input-96-predict-336 results on the ETTh1 dataset.}
    \label{fig:etth1_336}
\end{figure*}

% ETTh1_96_720
\begin{figure*}[h!]
    \centering
    
    % --- Row 1 ---
    \subfloat[\textcolor{red}{\textbf{CometNet (Ours)}}]{%
        \includegraphics[width=0.33\textwidth]{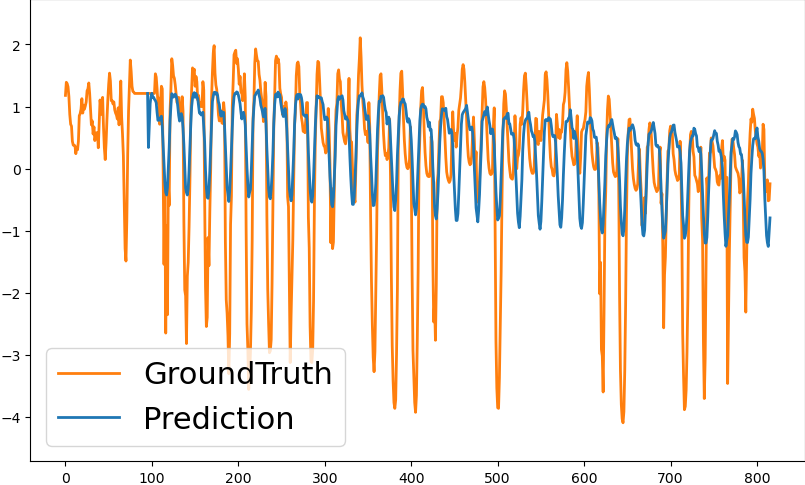}%
    }
    \hfill % Horizontal space between figures
    \subfloat[\textbf{TimeMixer++}]{%
        \includegraphics[width=0.33\textwidth]{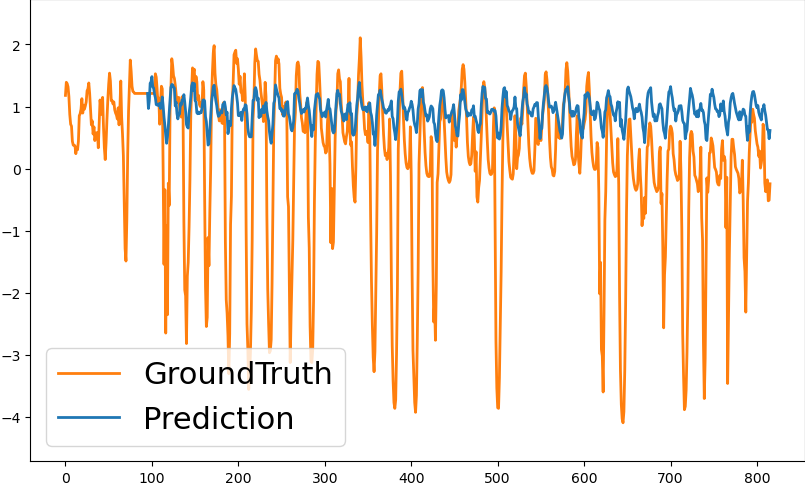}%
    }
    \hfill
    \subfloat[\textbf{iTransformer}]{%
        \includegraphics[width=0.33\textwidth]{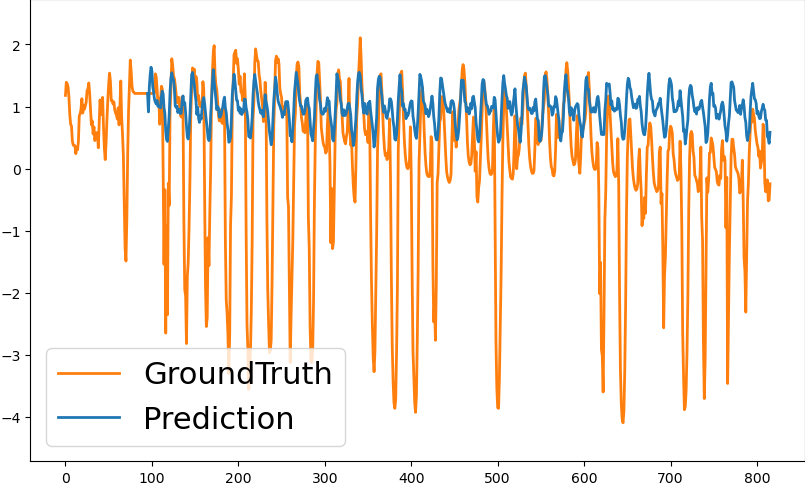}%
    }
    
    \vspace{0.5cm} % Vertical space between rows
    
    % --- Row 2 ---
    \subfloat[\textbf{PatchMLP}]{%
        \includegraphics[width=0.33\textwidth]{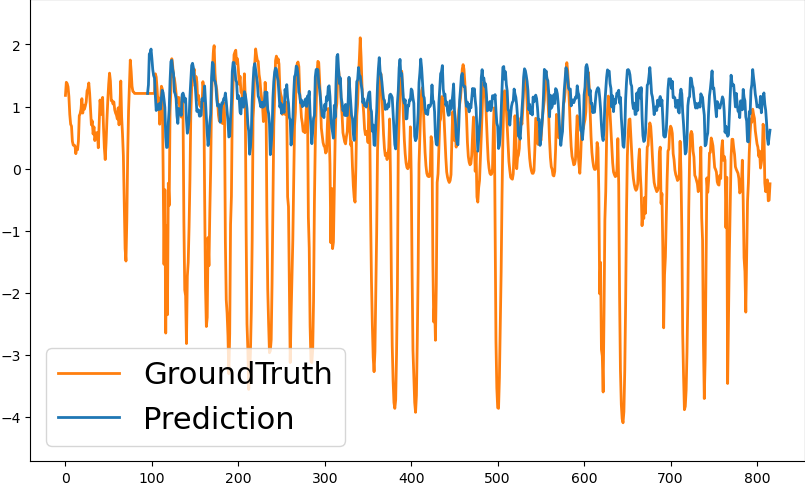}%
    }
    \hfill
    \subfloat[\textbf{PatchTST}]{%
        \includegraphics[width=0.33\textwidth]{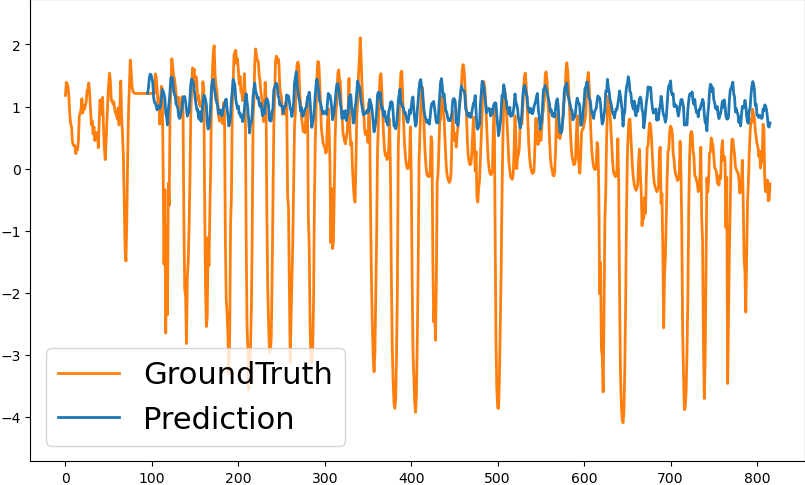}%
    }
    \hfill
    \subfloat[\textbf{DLinear}]{%
        \includegraphics[width=0.33\textwidth]{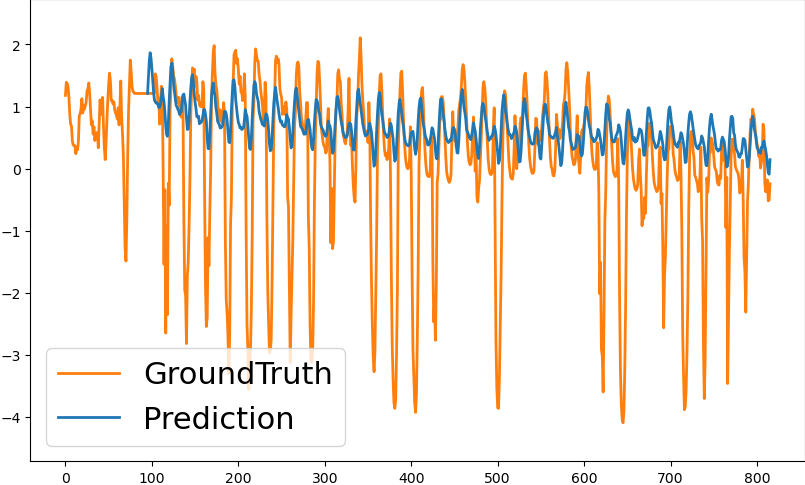}%
    }
    
    \caption{Visualization of input-96-predict-720 results on the ETTh1 dataset.}
    \label{fig:etth1_720}
\end{figure*}

% ETTh2_96_96
\begin{figure*}[h!]
    \centering
    
    % --- Row 1 ---
    \subfloat[\textcolor{red}{\textbf{CometNet (Ours)}}]{%
        \includegraphics[width=0.33\textwidth]{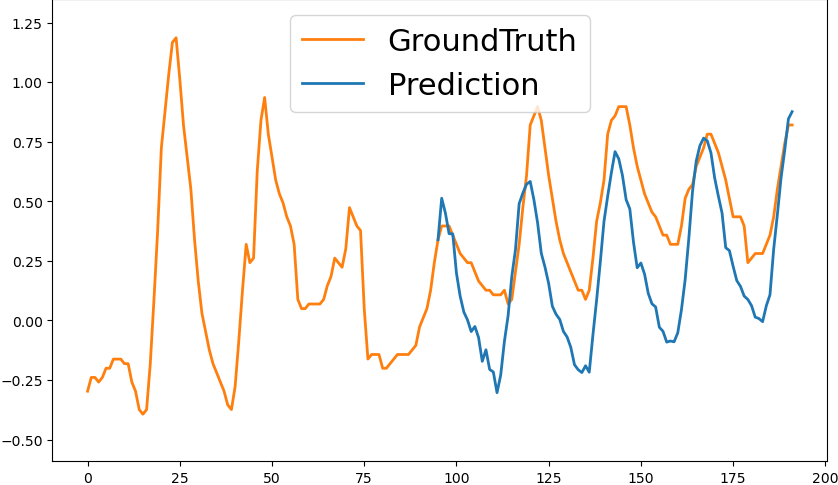}%
    }
    \hfill % Horizontal space between figures
    \subfloat[\textbf{TimeMixer++}]{%
        \includegraphics[width=0.33\textwidth]{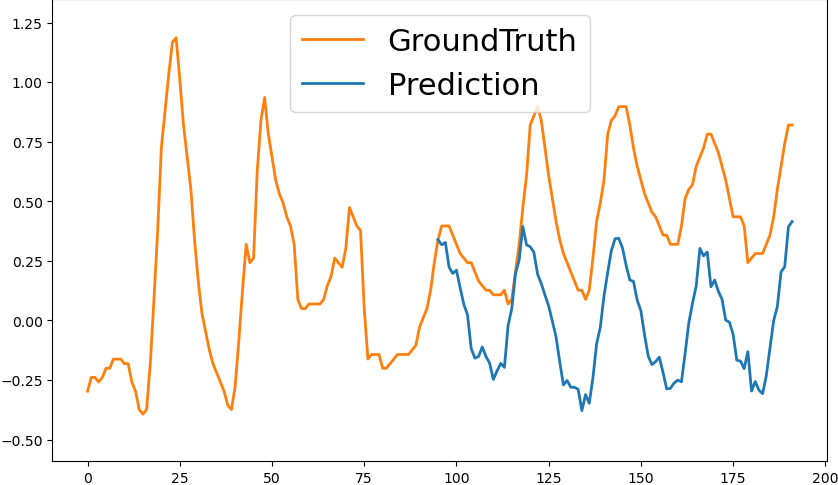}%
    }
    \hfill
    \subfloat[\textbf{iTransformer}]{%
        \includegraphics[width=0.33\textwidth]{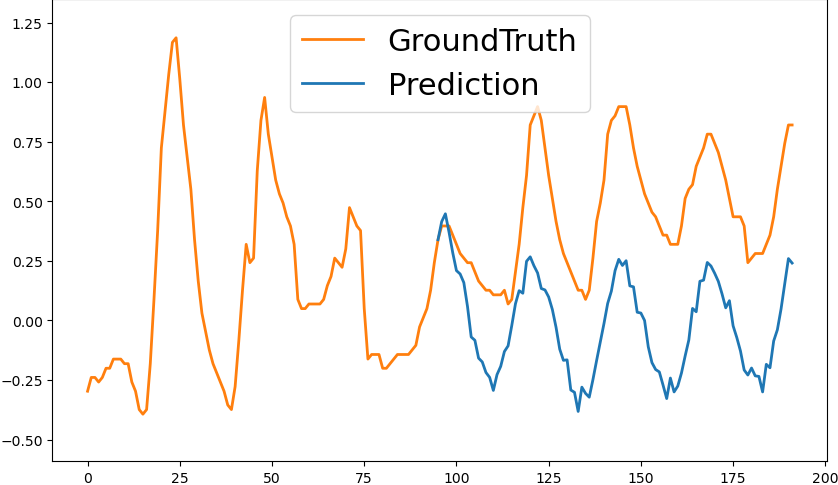}%
    }
    
    \vspace{0.5cm} % Vertical space between rows
    
    % --- Row 2 ---
    \subfloat[\textbf{PatchMLP}]{%
        \includegraphics[width=0.33\textwidth]{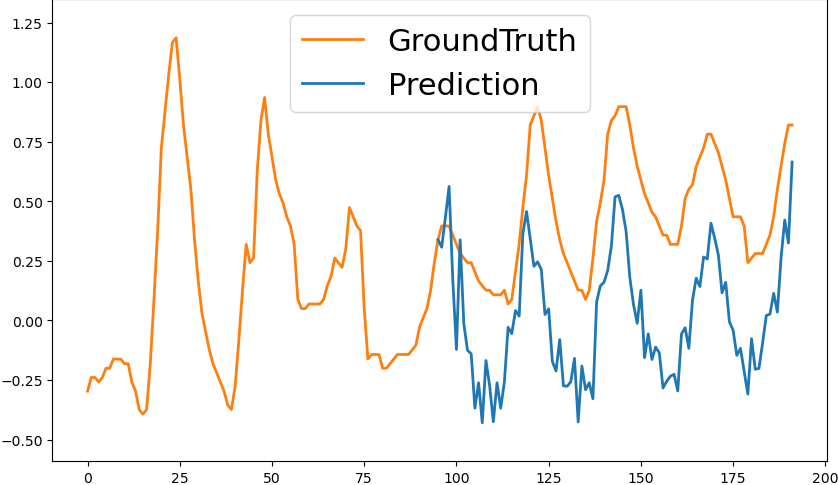}%
    }
    \hfill
    \subfloat[\textbf{PatchTST}]{%
        \includegraphics[width=0.33\textwidth]{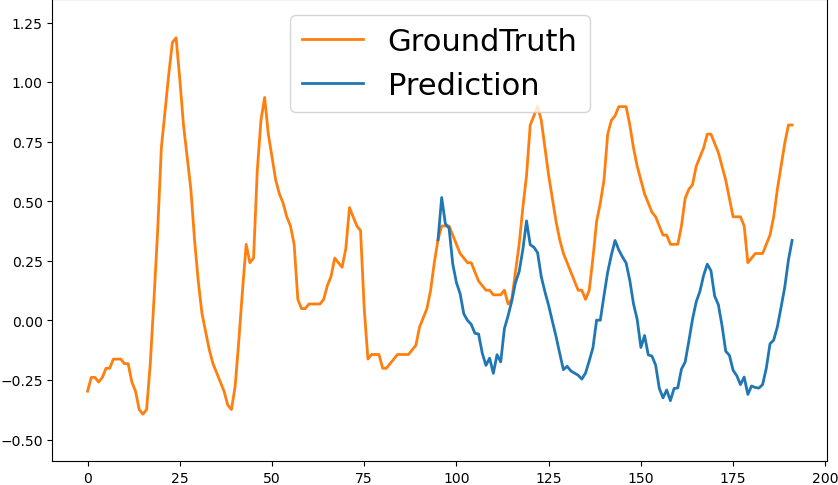}%
    }
    \hfill
    \subfloat[\textbf{DLinear}]{%
        \includegraphics[width=0.33\textwidth]{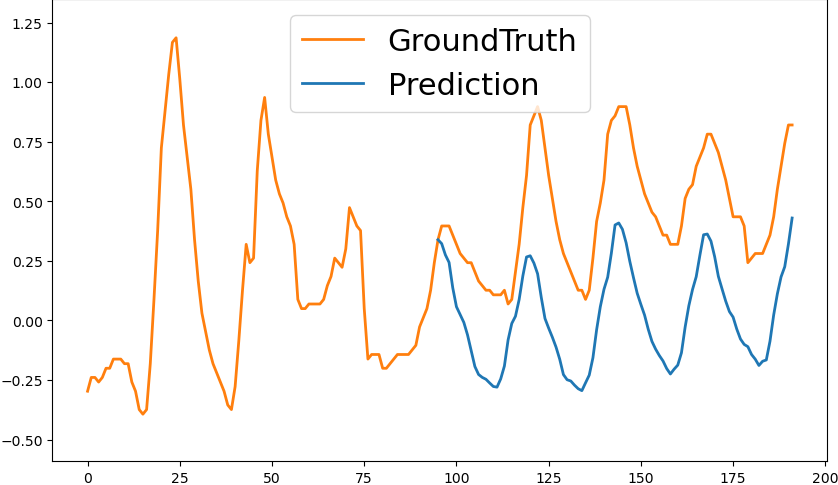}%
    }
    
    \caption{Visualization of input-96-predict-96 results on the ETTh2 dataset.}
    \label{fig:etth2_96}
\end{figure*}

% ETTh2_96_192
\begin{figure*}[h!]
    \centering
    
    % --- Row 1 ---
    \subfloat[\textcolor{red}{\textbf{CometNet (Ours)}}]{%
        \includegraphics[width=0.33\textwidth]{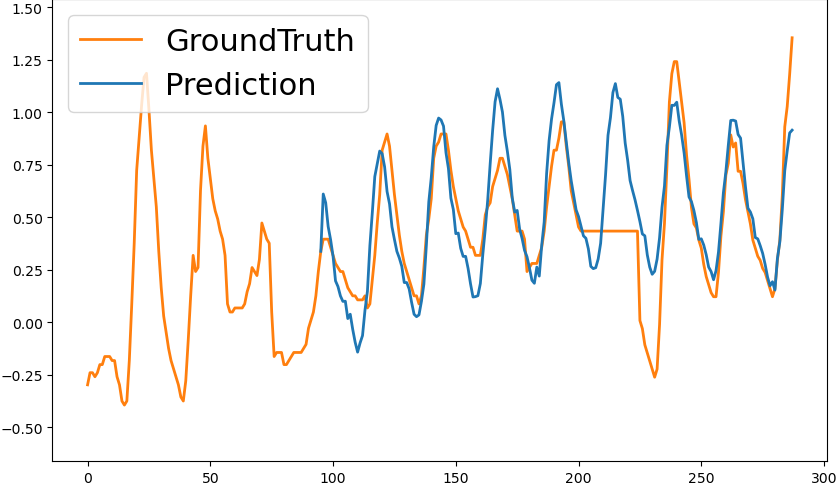}%
    }
    \hfill % Horizontal space between figures
    \subfloat[\textbf{TimeMixer++}]{%
        \includegraphics[width=0.33\textwidth]{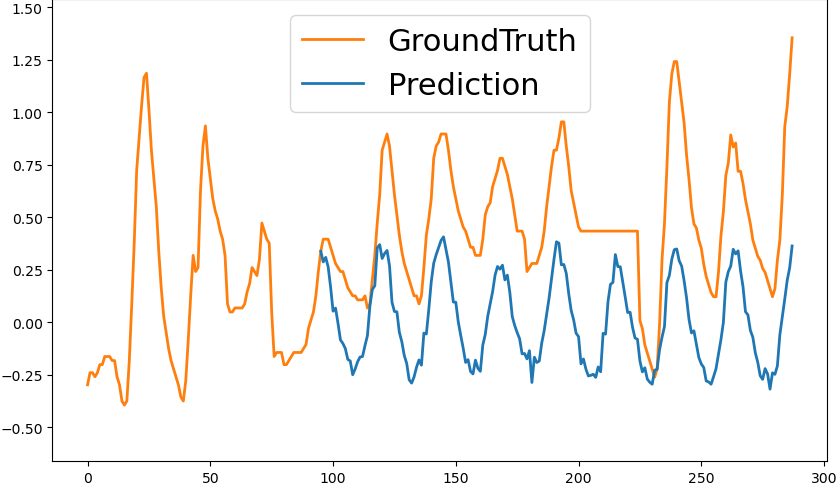}%
    }
    \hfill
    \subfloat[\textbf{iTransformer}]{%
        \includegraphics[width=0.33\textwidth]{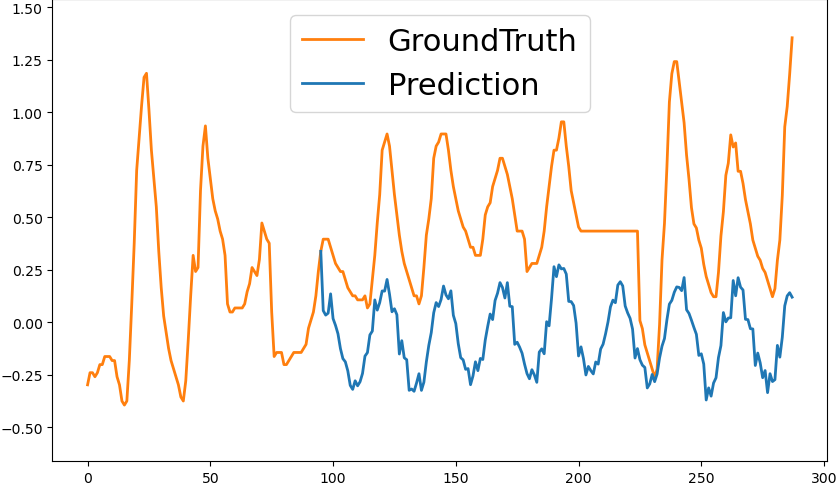}%
    }
    
    \vspace{0.5cm} % Vertical space between rows
    
    % --- Row 2 ---
    \subfloat[\textbf{PatchMLP}]{%
        \includegraphics[width=0.33\textwidth]{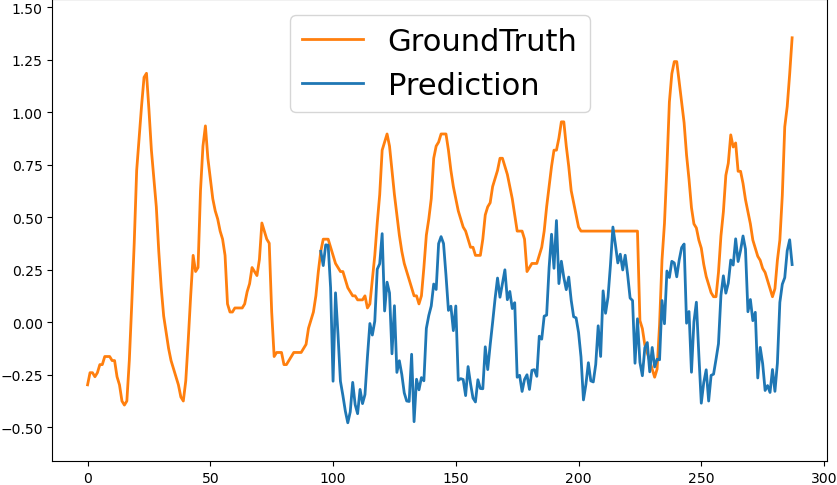}%
    }
    \hfill
    \subfloat[\textbf{PatchTST}]{%
        \includegraphics[width=0.33\textwidth]{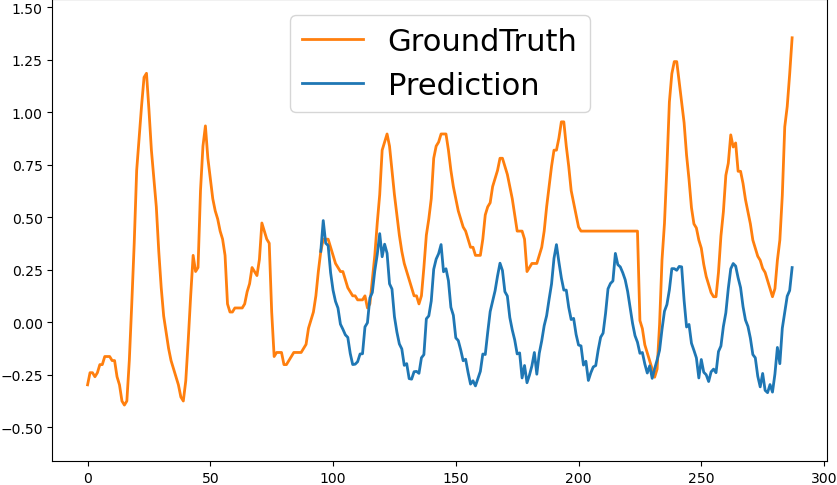}%
    }
    \hfill
    \subfloat[\textbf{DLinear}]{%
        \includegraphics[width=0.33\textwidth]{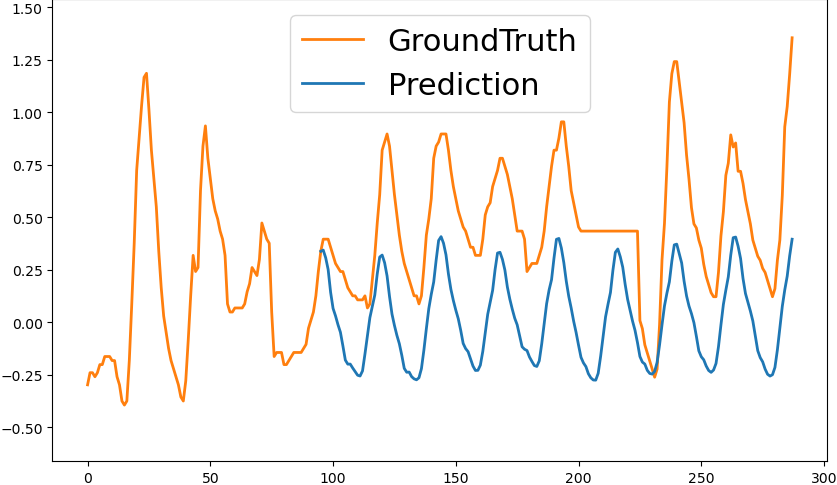}%
    }
    
    \caption{Visualization of input-96-predict-192 results on the ETTh2 dataset.}
    \label{fig:etth1_192}
\end{figure*}

% ETTh2_96_336
\begin{figure*}[h!]
    \centering
    
    % --- Row 1 ---
    \subfloat[\textcolor{red}{\textbf{CometNet (Ours)}}]{%
        \includegraphics[width=0.33\textwidth]{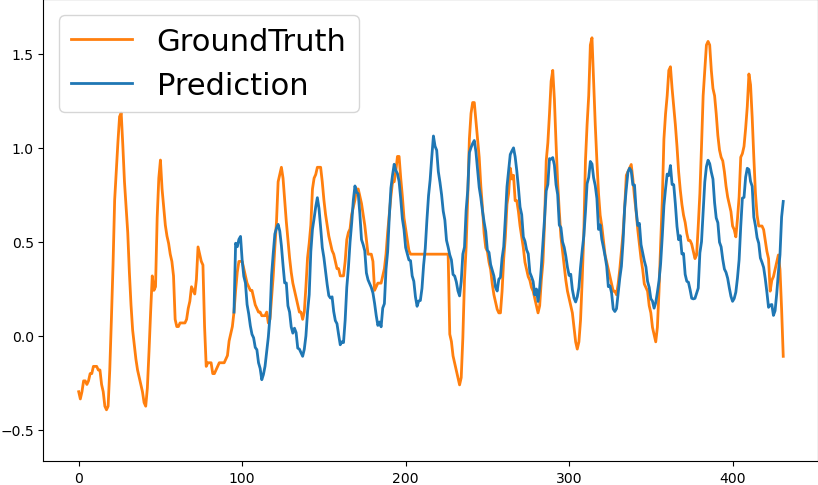}%
    }
    \hfill % Horizontal space between figures
    \subfloat[\textbf{TimeMixer++}]{%
        \includegraphics[width=0.33\textwidth]{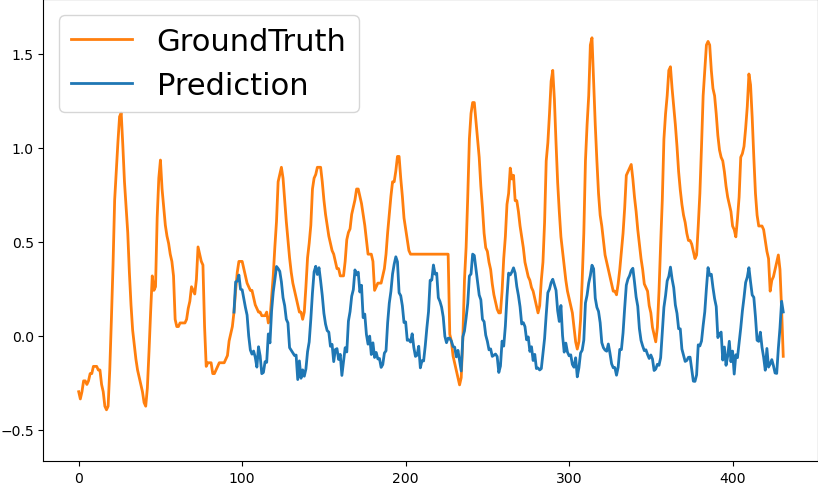}%
    }
    \hfill
    \subfloat[\textbf{iTransformer}]{%
        \includegraphics[width=0.33\textwidth]{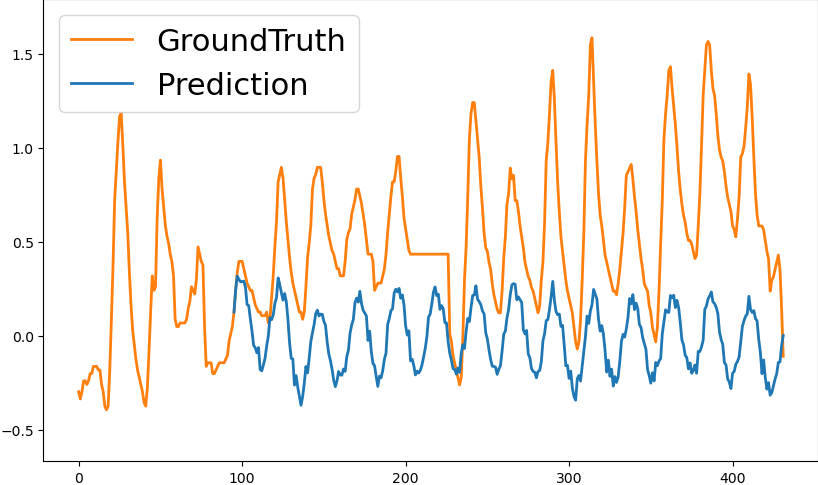}%
    }
    
    \vspace{0.5cm} % Vertical space between rows
    
    % --- Row 2 ---
    \subfloat[\textbf{PatchMLP}]{%
        \includegraphics[width=0.33\textwidth]{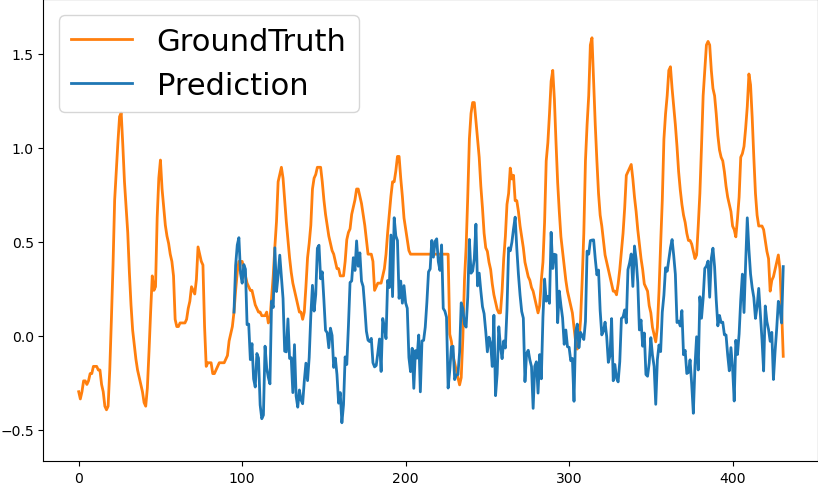}%
    }
    \hfill
    \subfloat[\textbf{PatchTST}]{%
        \includegraphics[width=0.33\textwidth]{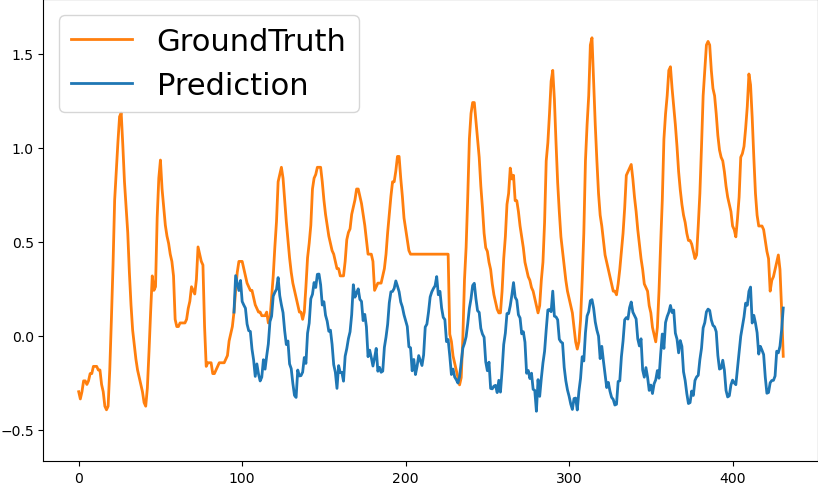}%
    }
    \hfill
    \subfloat[\textbf{DLinear}]{%
        \includegraphics[width=0.33\textwidth]{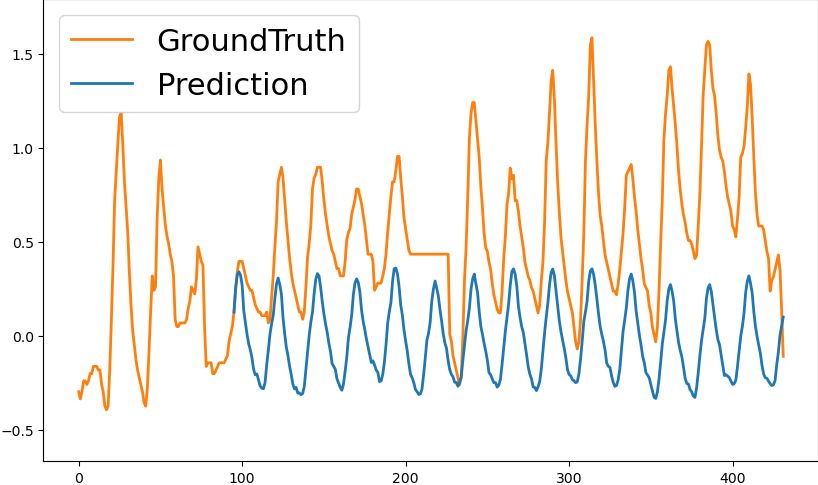}%
    }
    
    \caption{Visualization of input-96-predict-336 results on the ETTh2 dataset.}
    \label{fig:etth2_336}
\end{figure*}

% ETTh2_96_720
\begin{figure*}[h!]
    \centering
    
    % --- Row 1 ---
    \subfloat[\textcolor{red}{\textbf{CometNet (Ours)}}]{%
        \includegraphics[width=0.33\textwidth]{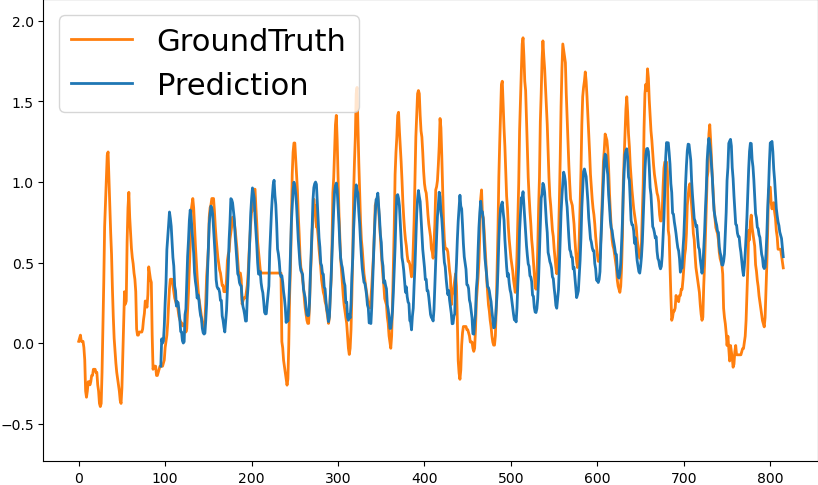}%
    }
    \hfill % Horizontal space between figures
    \subfloat[\textbf{TimeMixer++}]{%
        \includegraphics[width=0.33\textwidth]{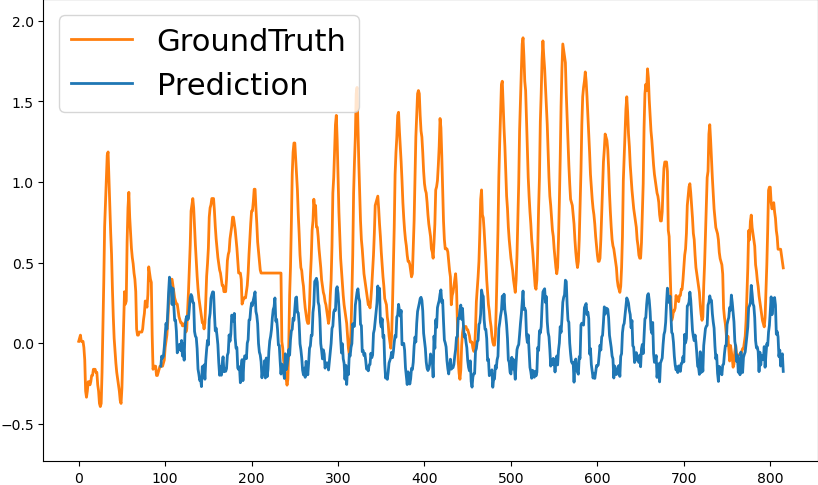}%
    }
    \hfill
    \subfloat[\textbf{iTransformer}]{%
        \includegraphics[width=0.33\textwidth]{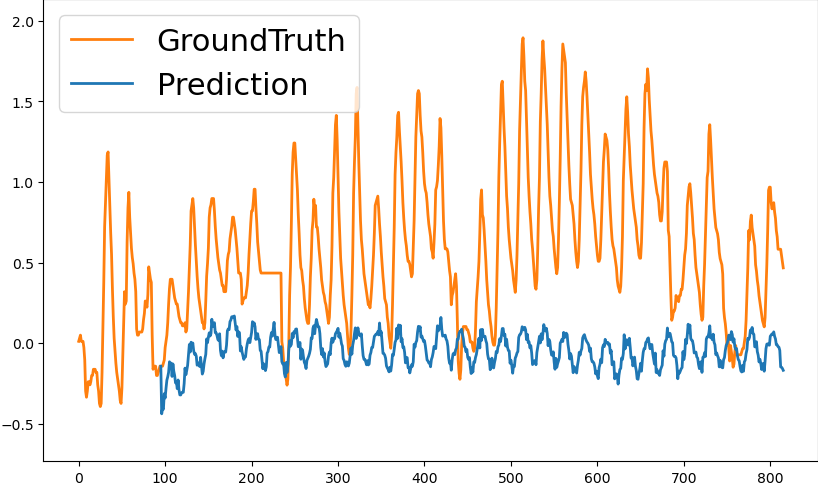}%
    }
    
    \vspace{0.5cm} % Vertical space between rows
    
    % --- Row 2 ---
    \subfloat[\textbf{PatchMLP}]{%
        \includegraphics[width=0.33\textwidth]{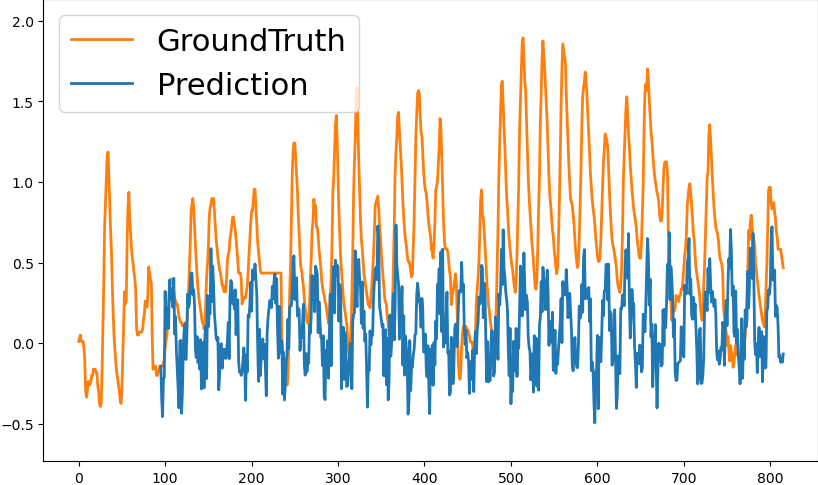}%
    }
    \hfill
    \subfloat[\textbf{PatchTST}]{%
        \includegraphics[width=0.33\textwidth]{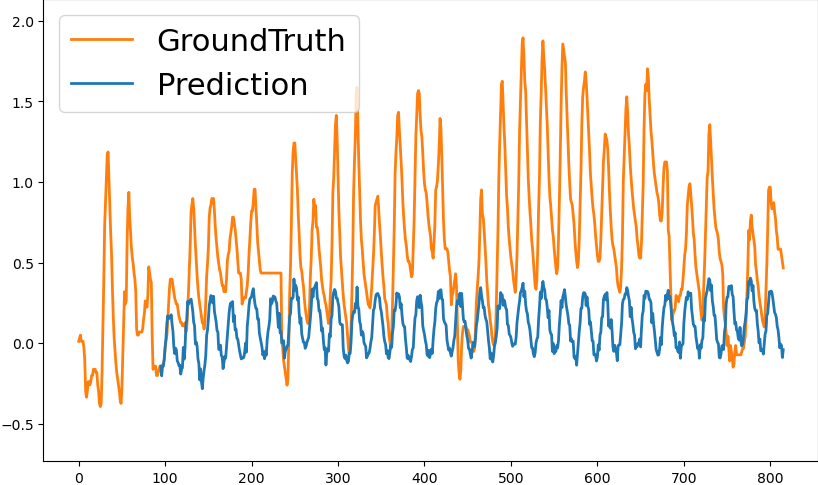}%
    }
    \hfill
    \subfloat[\textbf{DLinear}]{%
        \includegraphics[width=0.33\textwidth]{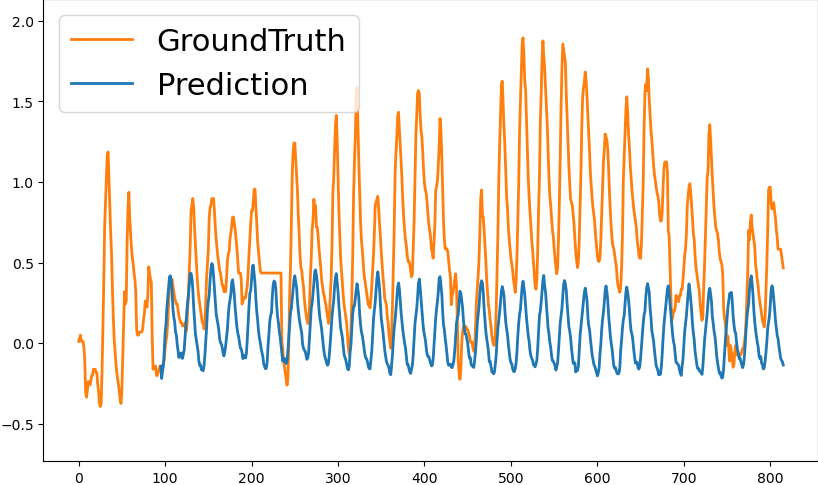}%
    }
    
    \caption{Visualization of input-96-predict-720 results on the ETTh2 dataset.}
    \label{fig:etth2_720}
\end{figure*}

% ETTm1_96_96
\begin{figure*}[h!]
    \centering
    
    % --- Row 1 ---
    \subfloat[\textcolor{red}{\textbf{CometNet (Ours)}}]{%
        \includegraphics[width=0.33\textwidth]{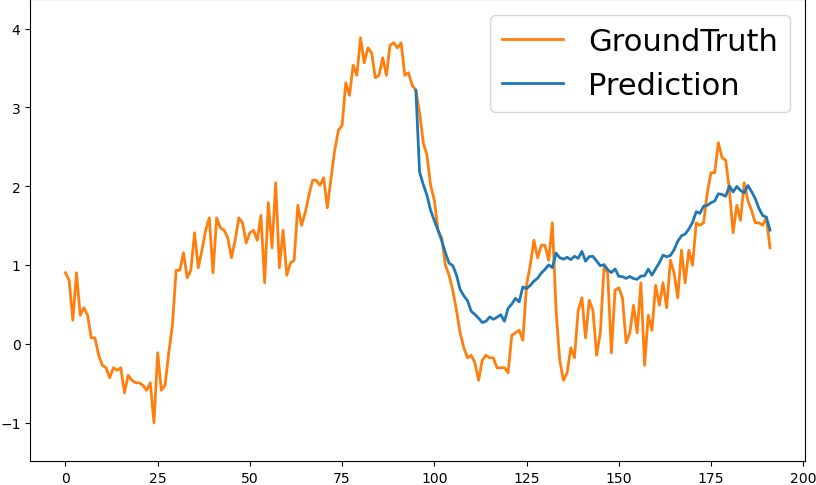}%
    }
    \hfill % Horizontal space between figures
    \subfloat[\textbf{TimeMixer++}]{%
        \includegraphics[width=0.33\textwidth]{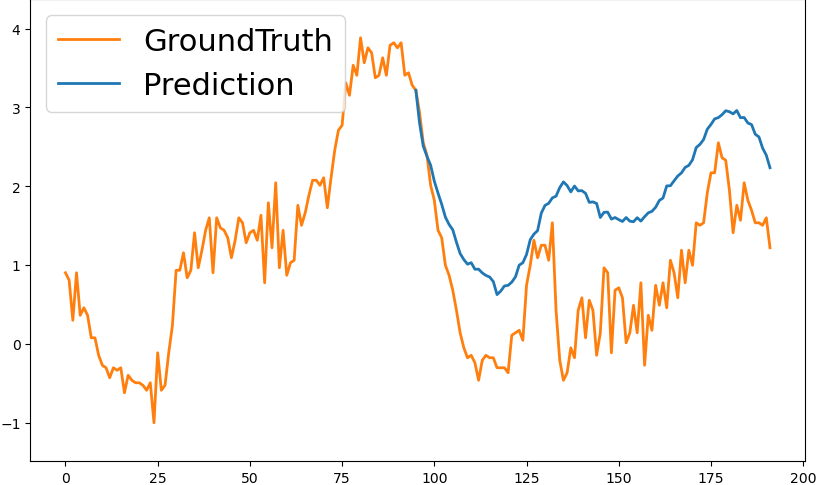}%
    }
    \hfill
    \subfloat[\textbf{iTransformer}]{%
        \includegraphics[width=0.33\textwidth]{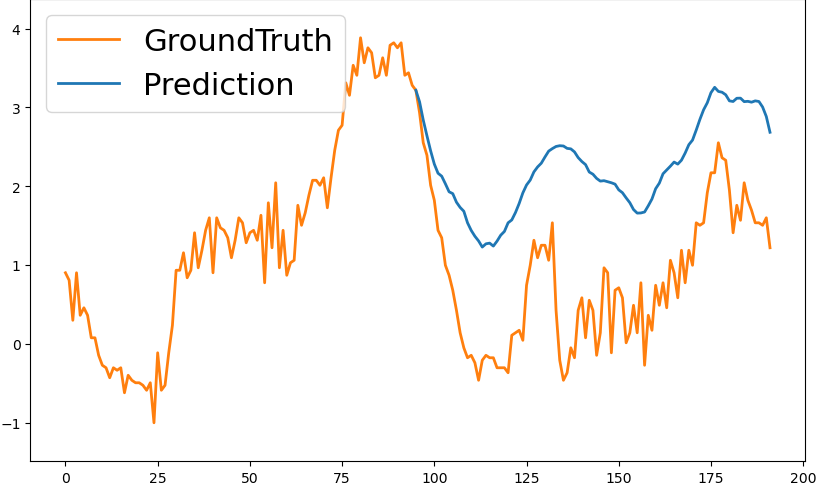}%
    }
    
    \vspace{0.5cm} % Vertical space between rows
    
    % --- Row 2 ---
    \subfloat[\textbf{PatchMLP}]{%
        \includegraphics[width=0.33\textwidth]{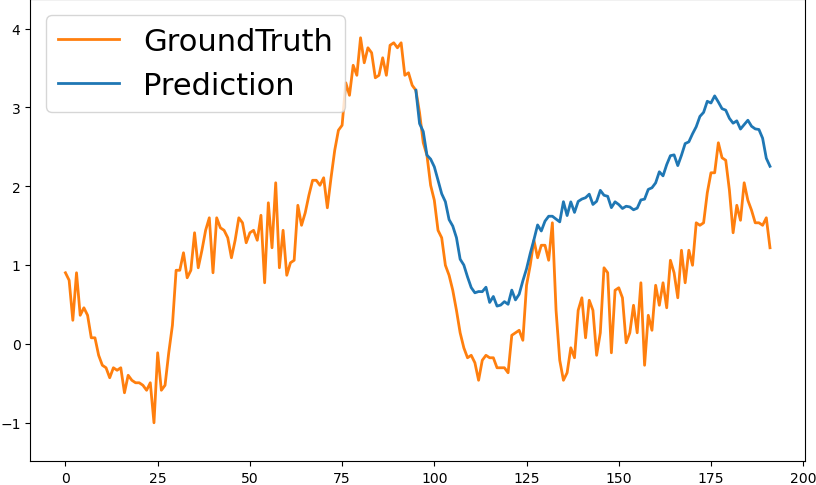}%
    }
    \hfill
    \subfloat[\textbf{PatchTST}]{%
        \includegraphics[width=0.33\textwidth]{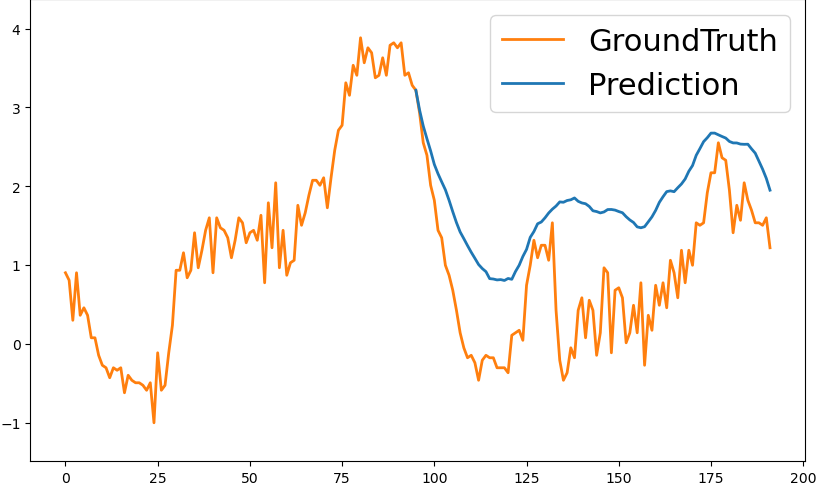}%
    }
    \hfill
    \subfloat[\textbf{DLinear}]{%
        \includegraphics[width=0.33\textwidth]{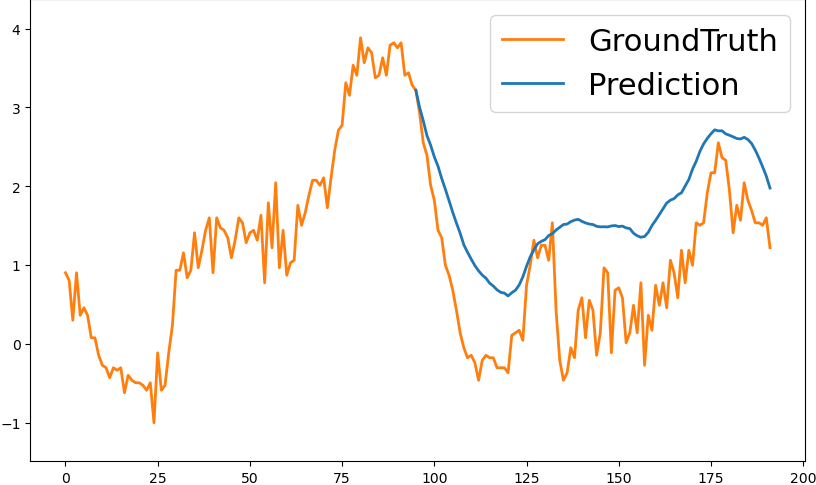}%
    }
    
    \caption{Visualization of input-96-predict-96 results on the ETTm1 dataset.}
    \label{fig:ettm1_96}
\end{figure*}

% ETTm1_96_192
\begin{figure*}[h!]
    \centering
    
    % --- Row 1 ---
    \subfloat[\textcolor{red}{\textbf{CometNet (Ours)}}]{%
        \includegraphics[width=0.33\textwidth]{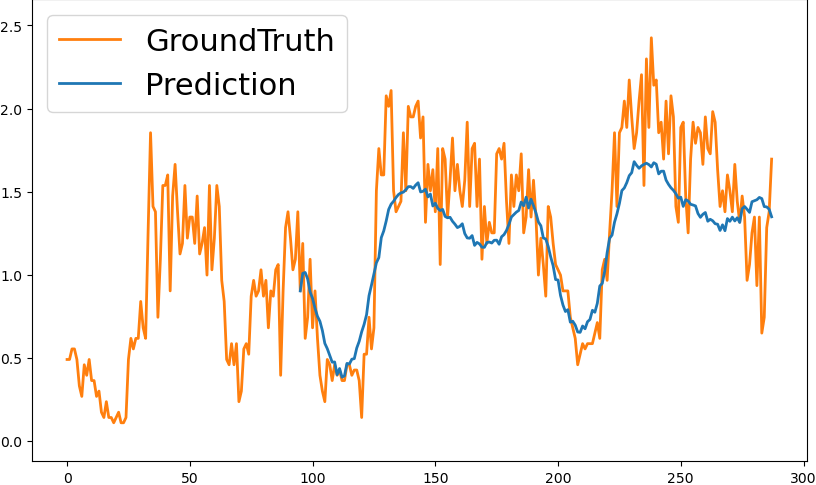}%
    }
    \hfill % Horizontal space between figures
    \subfloat[\textbf{TimeMixer++}]{%
        \includegraphics[width=0.33\textwidth]{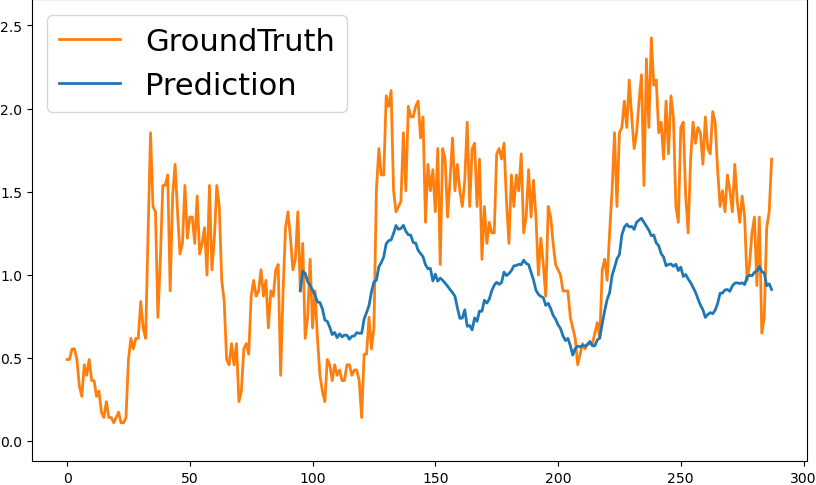}%
    }
    \hfill
    \subfloat[\textbf{iTransformer}]{%
        \includegraphics[width=0.33\textwidth]{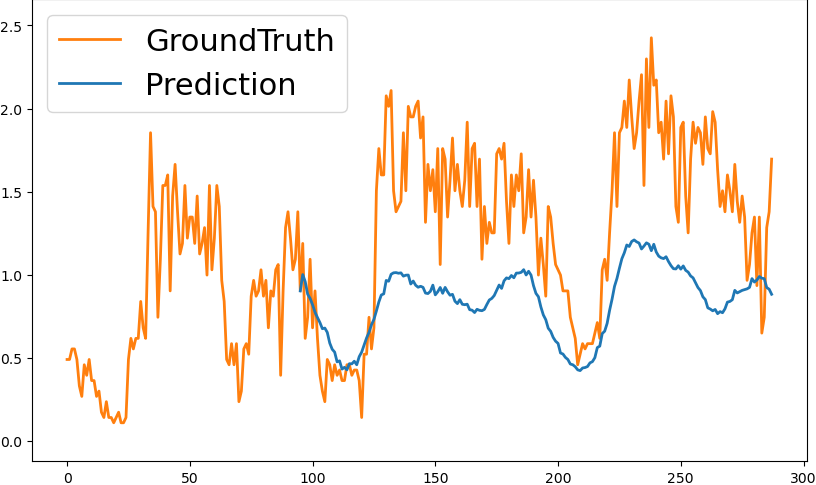}%
    }
    
    \vspace{0.5cm} % Vertical space between rows
    
    % --- Row 2 ---
    \subfloat[\textbf{PatchMLP}]{%
        \includegraphics[width=0.33\textwidth]{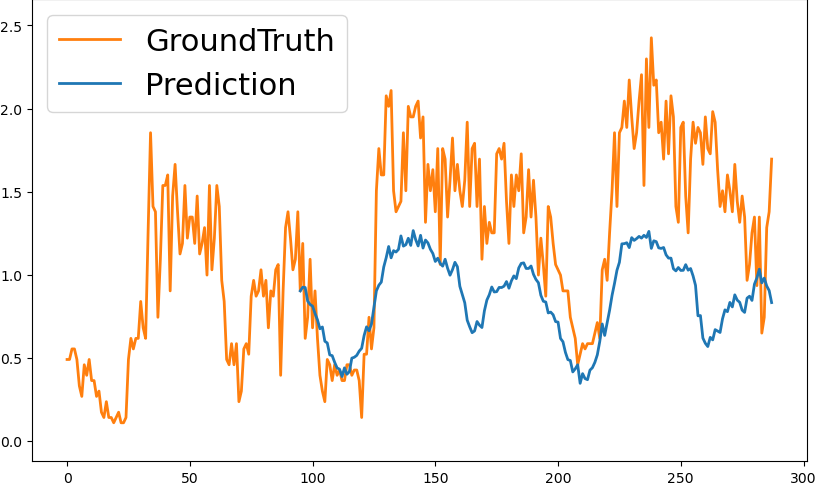}%
    }
    \hfill
    \subfloat[\textbf{PatchTST}]{%
        \includegraphics[width=0.33\textwidth]{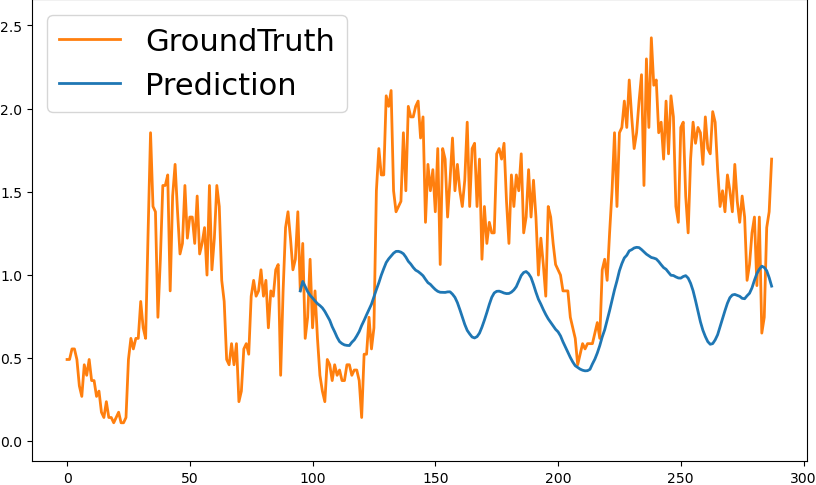}%
    }
    \hfill
    \subfloat[\textbf{DLinear}]{%
        \includegraphics[width=0.33\textwidth]{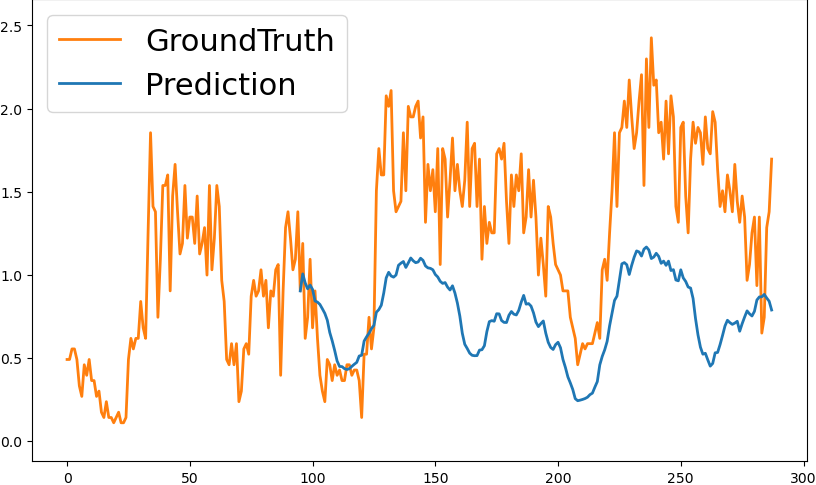}%
    }
    
    \caption{Visualization of input-96-predict-192 results on the ETTm1 dataset.}
    \label{fig:ettm1_192}
\end{figure*}

% ETTm1_96_336
\begin{figure*}[h!]
    \centering
    
    % --- Row 1 ---
    \subfloat[\textcolor{red}{\textbf{CometNet (Ours)}}]{%
        \includegraphics[width=0.33\textwidth]{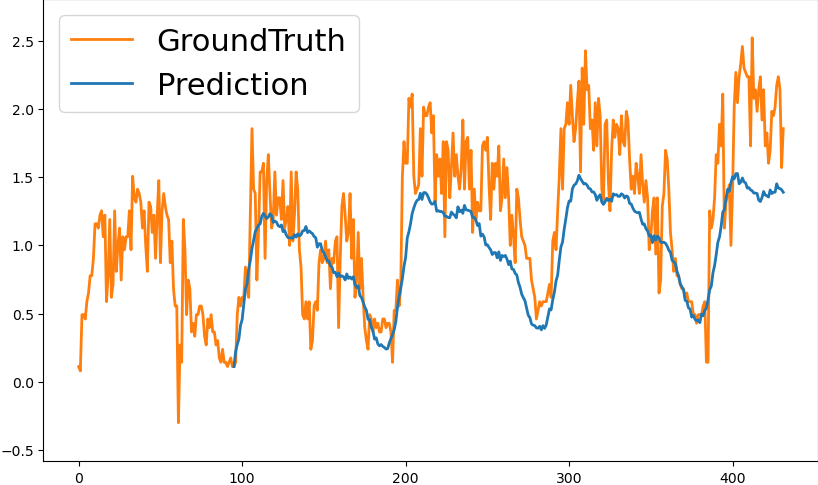}%
    }
    \hfill % Horizontal space between figures
    \subfloat[\textbf{TimeMixer++}]{%
        \includegraphics[width=0.33\textwidth]{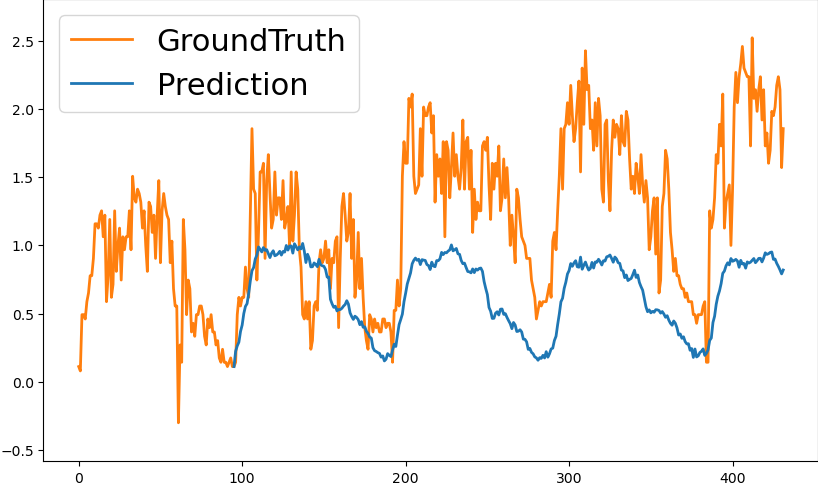}%
    }
    \hfill
    \subfloat[\textbf{iTransformer}]{%
        \includegraphics[width=0.33\textwidth]{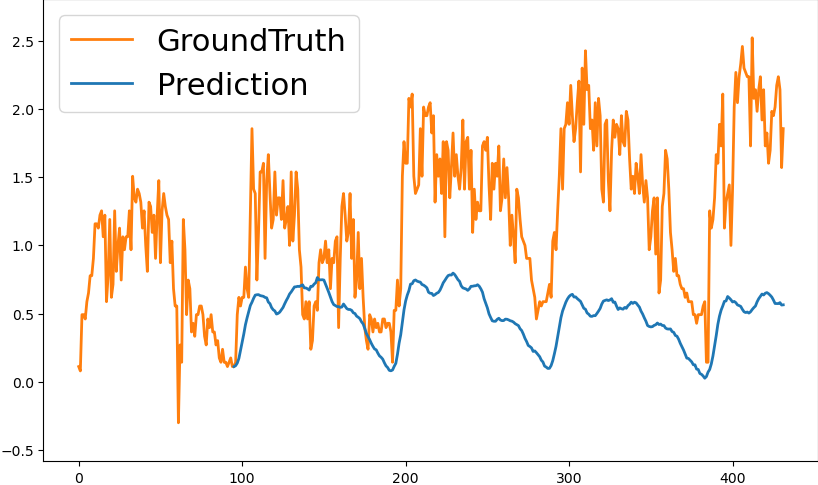}%
    }
    
    \vspace{0.5cm} % Vertical space between rows
    
    % --- Row 2 ---
    \subfloat[\textbf{PatchMLP}]{%
        \includegraphics[width=0.33\textwidth]{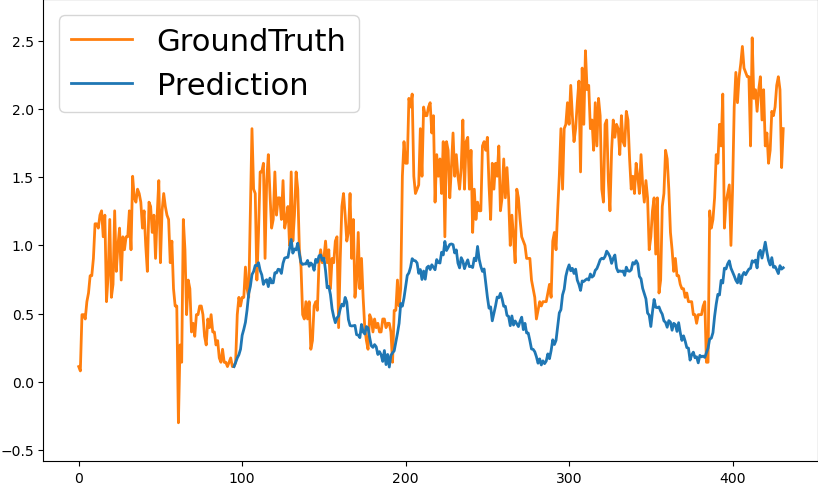}%
    }
    \hfill
    \subfloat[\textbf{PatchTST}]{%
        \includegraphics[width=0.33\textwidth]{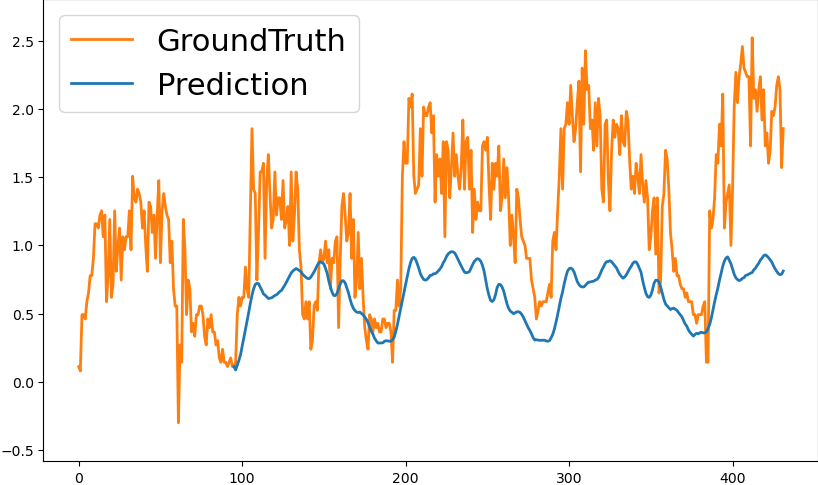}%
    }
    \hfill
    \subfloat[\textbf{DLinear}]{%
        \includegraphics[width=0.33\textwidth]{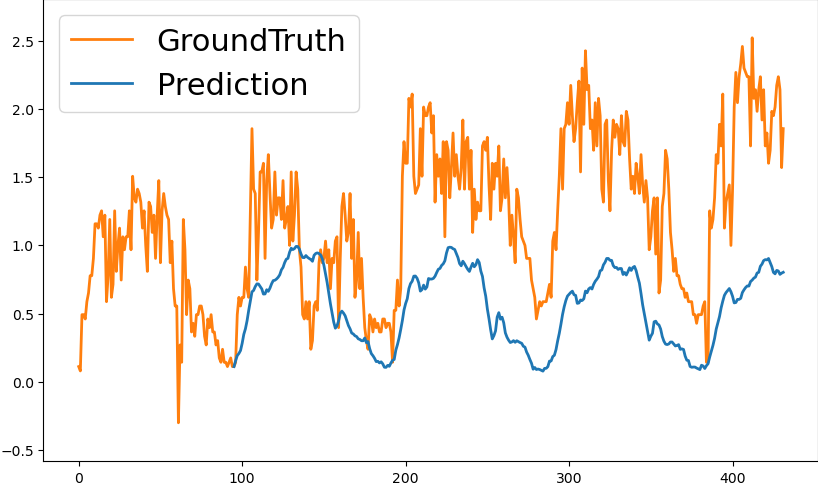}%
    }
    
    \caption{Visualization of input-96-predict-336 results on the ETTm1 dataset.}
    \label{fig:ettm1_336}
\end{figure*}

% ETTm1_96_720
\begin{figure*}[h!]
    \centering
    
    % --- Row 1 ---
    \subfloat[\textcolor{red}{\textbf{CometNet (Ours)}}]{%
        \includegraphics[width=0.33\textwidth]{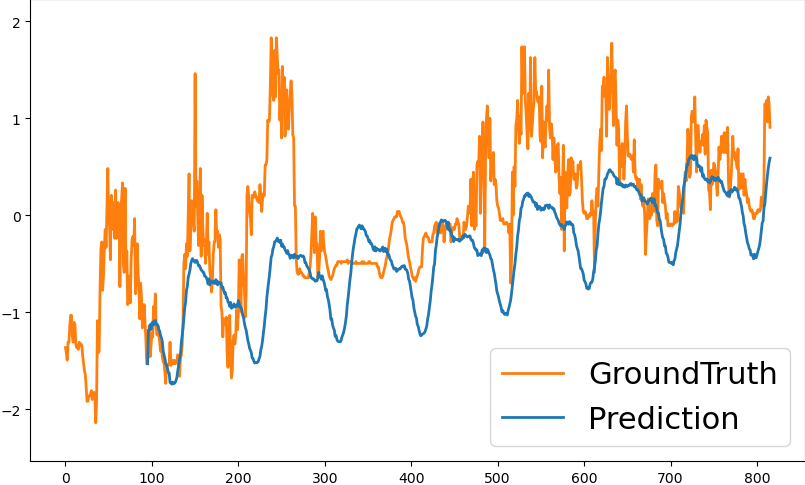}%
    }
    \hfill % Horizontal space between figures
    \subfloat[\textbf{TimeMixer++}]{%
        \includegraphics[width=0.33\textwidth]{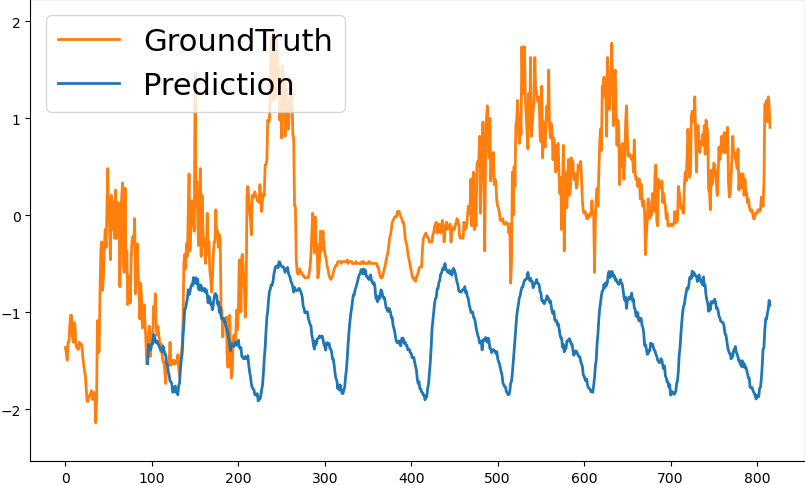}%
    }
    \hfill
    \subfloat[\textbf{iTransformer}]{%
        \includegraphics[width=0.33\textwidth]{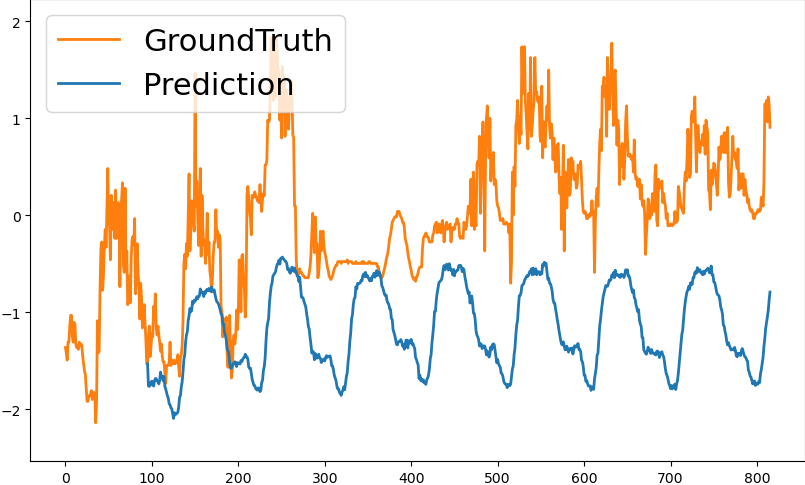}%
    }
    
    \vspace{0.5cm} % Vertical space between rows
    
    % --- Row 2 ---
    \subfloat[\textbf{PatchMLP}]{%
        \includegraphics[width=0.33\textwidth]{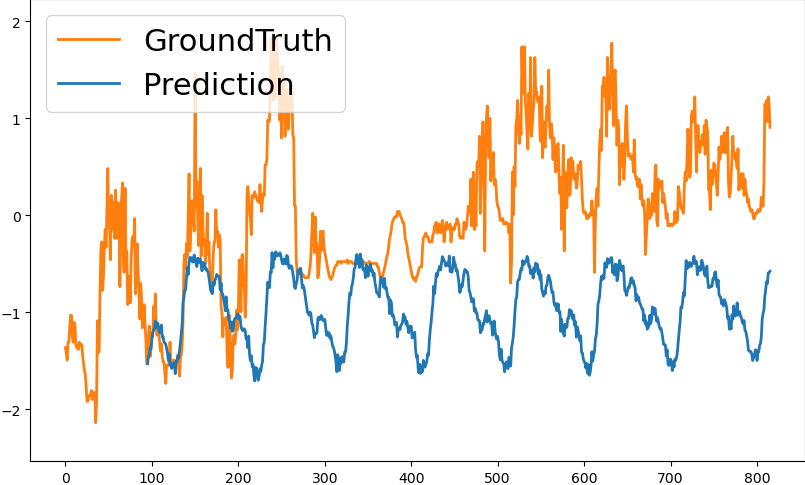}%
    }
    \hfill
    \subfloat[\textbf{PatchTST}]{%
        \includegraphics[width=0.33\textwidth]{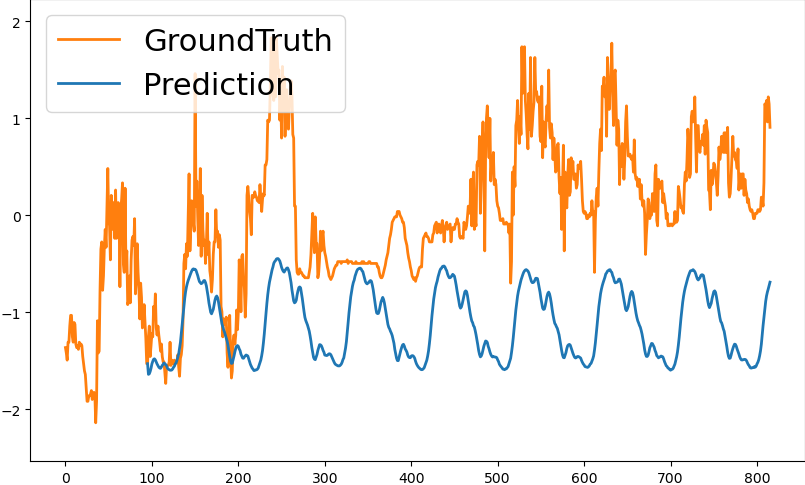}%
    }
    \hfill
    \subfloat[\textbf{DLinear}]{%
        \includegraphics[width=0.33\textwidth]{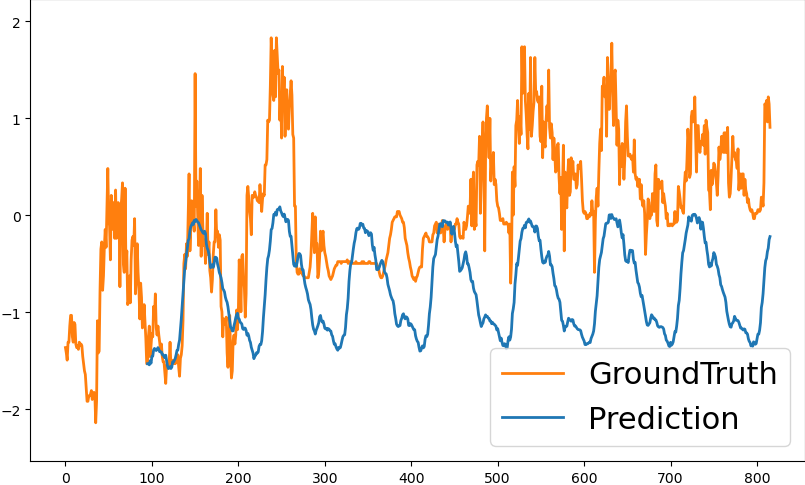}%
    }
    
    \caption{Visualization of input-96-predict-720 results on the ETTm1 dataset.}
    \label{fig:ettm1_720}
\end{figure*}

% ETTm2_96_96
\begin{figure*}[h!]
    \centering
    
    % --- Row 1 ---
    \subfloat[\textcolor{red}{\textbf{CometNet (Ours)}}]{%
        \includegraphics[width=0.33\textwidth]{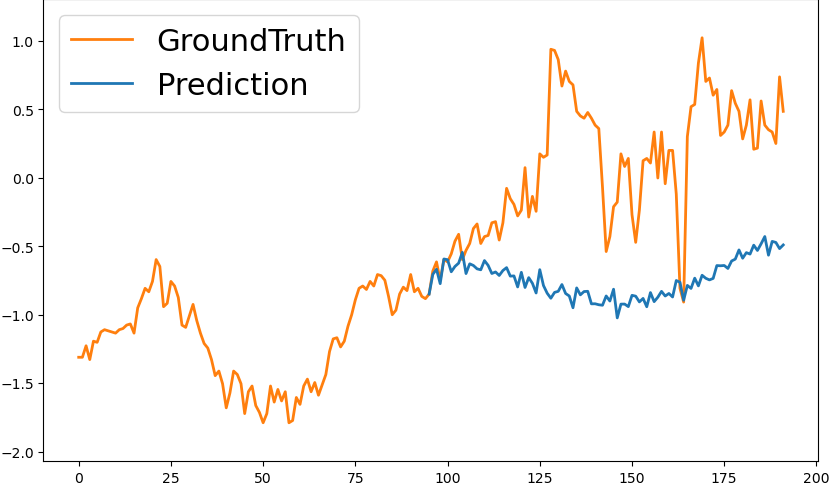}%
    }
    \hfill % Horizontal space between figures
    \subfloat[\textbf{TimeMixer++}]{%
        \includegraphics[width=0.33\textwidth]{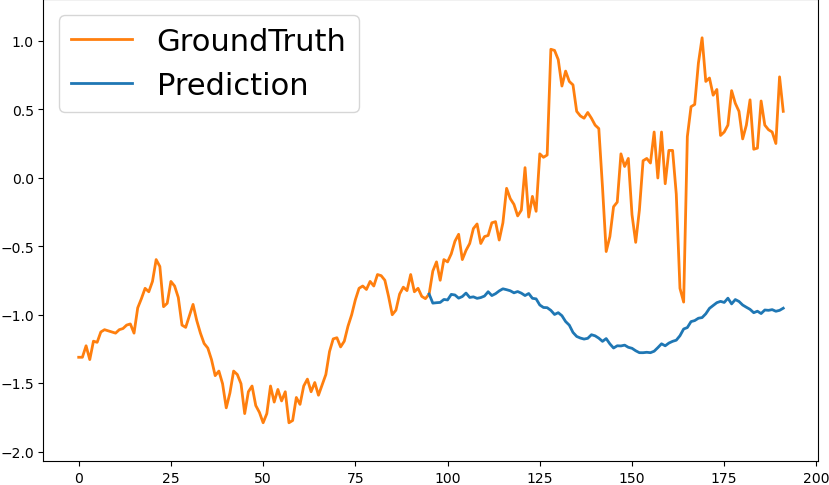}%
    }
    \hfill
    \subfloat[\textbf{iTransformer}]{%
        \includegraphics[width=0.33\textwidth]{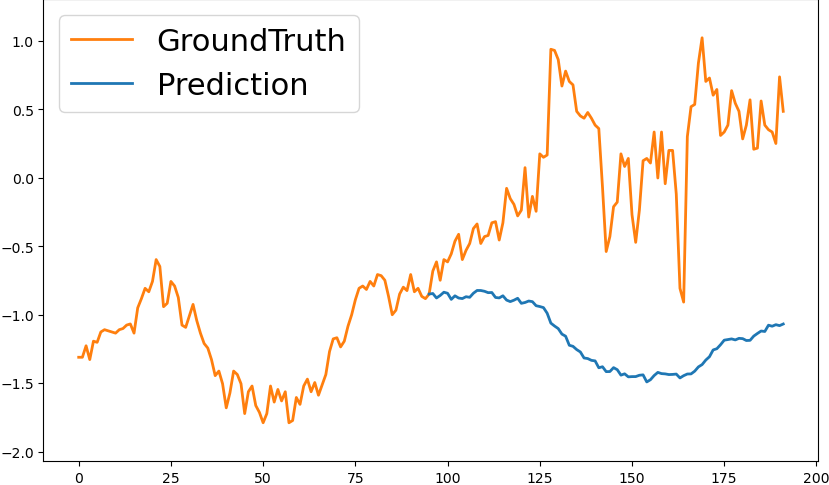}%
    }
    
    \vspace{0.5cm} % Vertical space between rows
    
    % --- Row 2 ---
    \subfloat[\textbf{PatchMLP}]{%
        \includegraphics[width=0.33\textwidth]{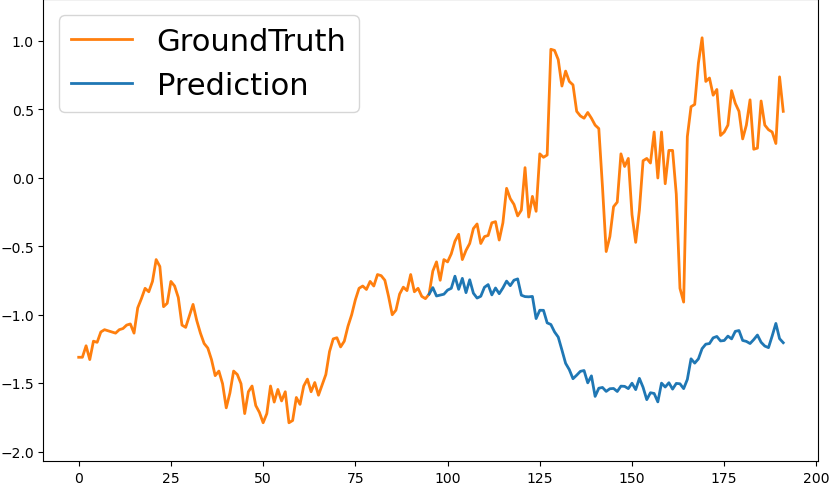}%
    }
    \hfill
    \subfloat[\textbf{PatchTST}]{%
        \includegraphics[width=0.33\textwidth]{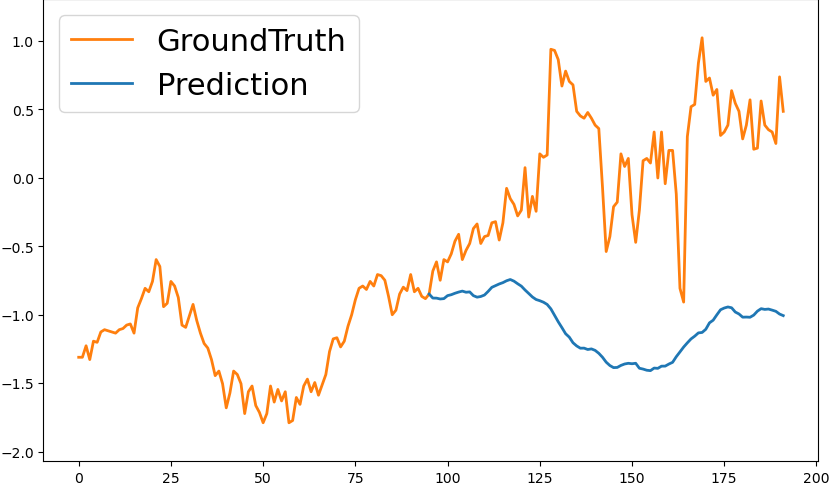}%
    }
    \hfill
    \subfloat[\textbf{DLinear}]{%
        \includegraphics[width=0.33\textwidth]{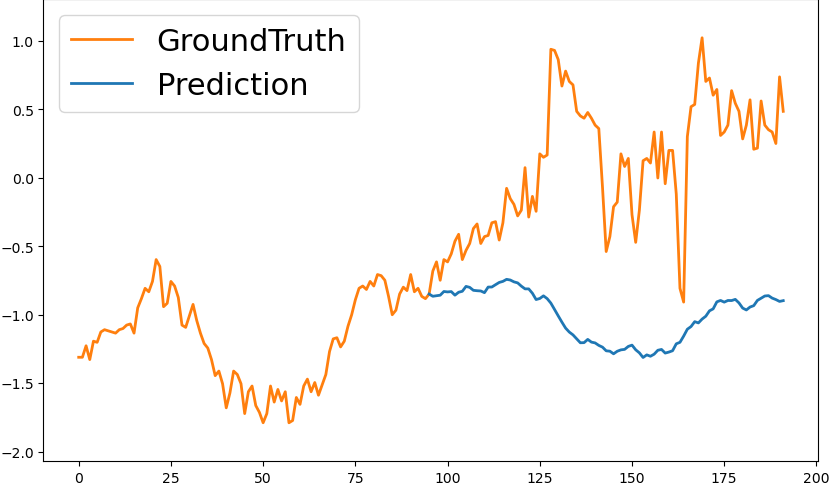}%
    }
    
    \caption{Visualization of input-96-predict-96 results on the ETTm2 dataset.}
    \label{fig:ettm2_96}
\end{figure*}

% ETTm2_96_192
\begin{figure*}[h!]
    \centering
    
    % --- Row 1 ---
    \subfloat[\textcolor{red}{\textbf{CometNet (Ours)}}]{%
        \includegraphics[width=0.33\textwidth]{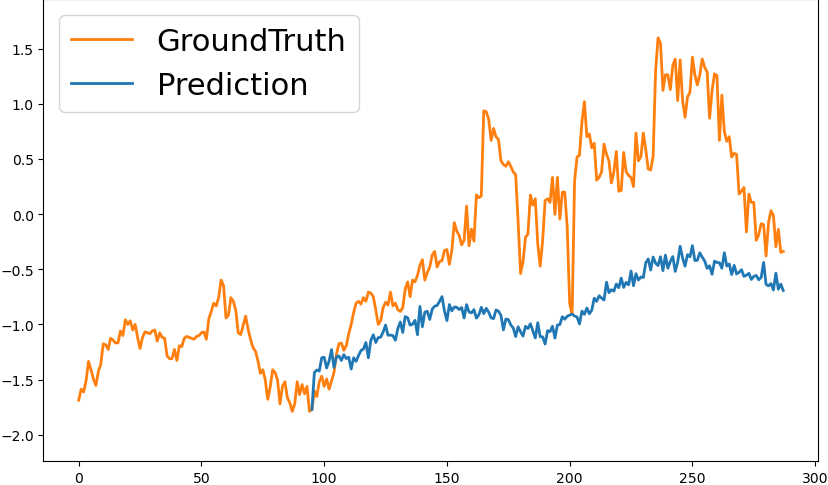}%
    }
    \hfill % Horizontal space between figures
    \subfloat[\textbf{TimeMixer++}]{%
        \includegraphics[width=0.33\textwidth]{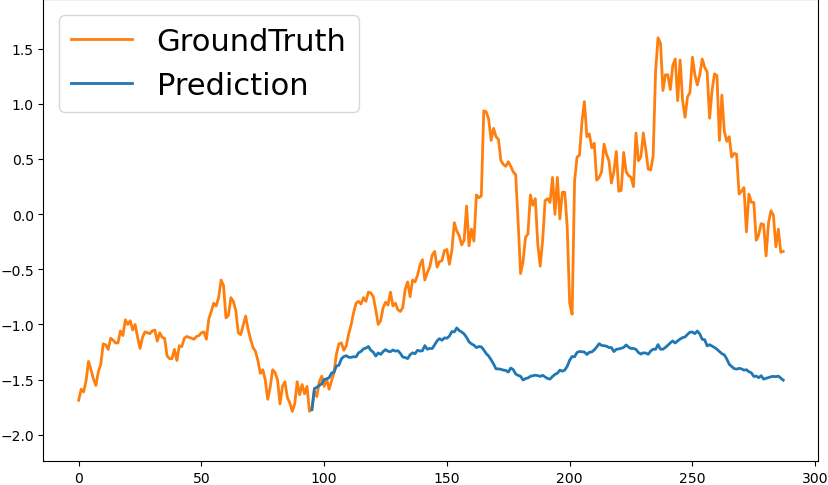}%
    }
    \hfill
    \subfloat[\textbf{iTransformer}]{%
        \includegraphics[width=0.33\textwidth]{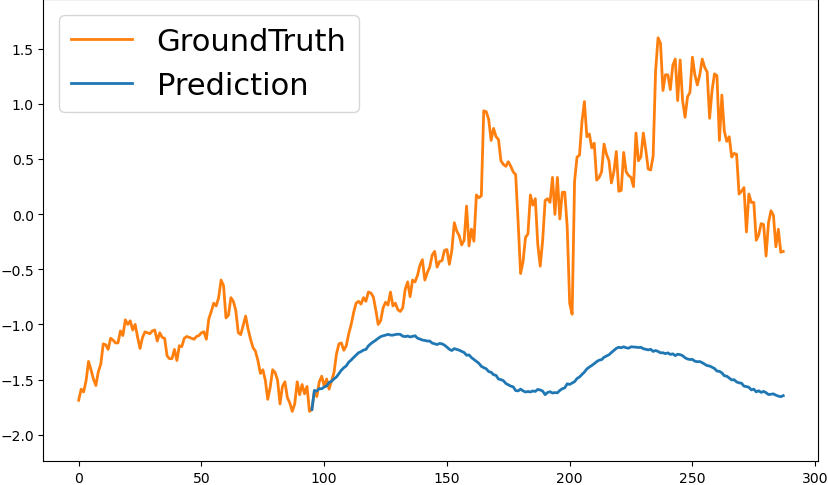}%
    }
    
    \vspace{0.5cm} % Vertical space between rows
    
    % --- Row 2 ---
    \subfloat[\textbf{PatchMLP}]{%
        \includegraphics[width=0.33\textwidth]{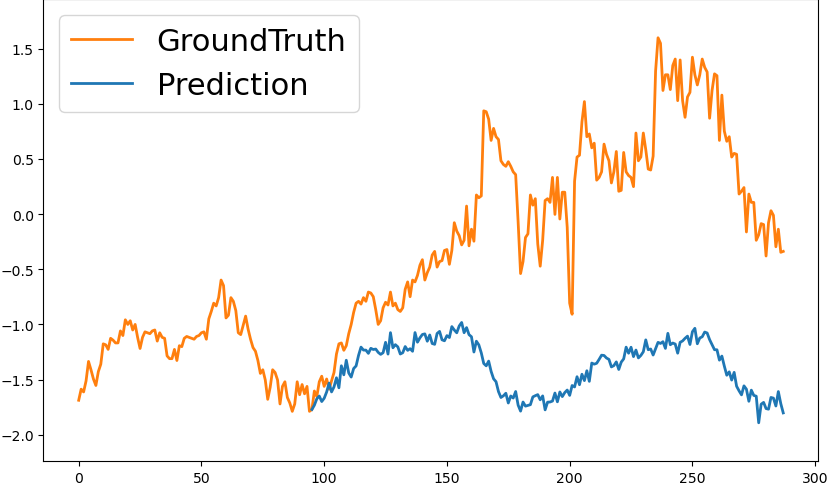}%
    }
    \hfill
    \subfloat[\textbf{PatchTST}]{%
        \includegraphics[width=0.33\textwidth]{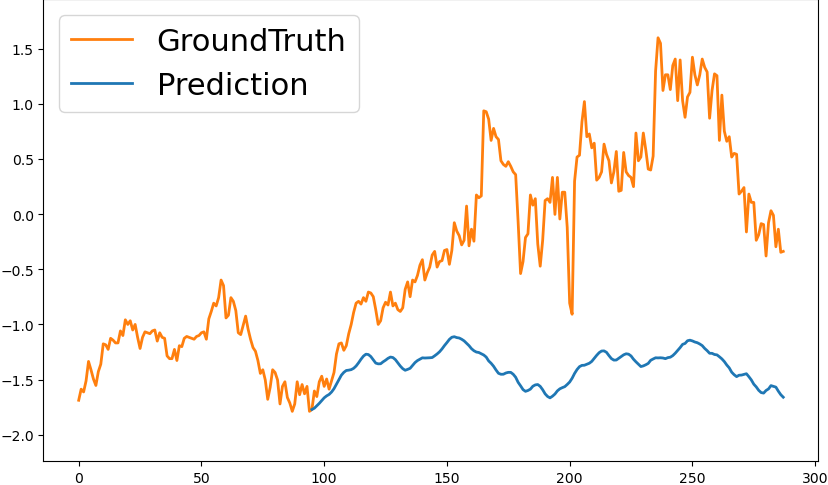}%
    }
    \hfill
    \subfloat[\textbf{DLinear}]{%
        \includegraphics[width=0.33\textwidth]{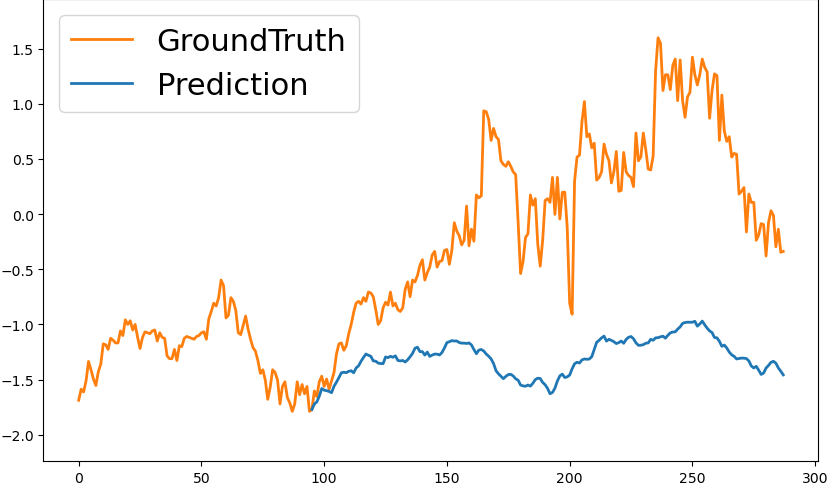}%
    }
    
    \caption{Visualization of input-96-predict-192 results on the ETTm2 dataset.}
    \label{fig:ettm2_192}
\end{figure*}

% ETTm2_96_336
\begin{figure*}[h!]
    \centering
    
    % --- Row 1 ---
    \subfloat[\textcolor{red}{\textbf{CometNet (Ours)}}]{%
        \includegraphics[width=0.33\textwidth]{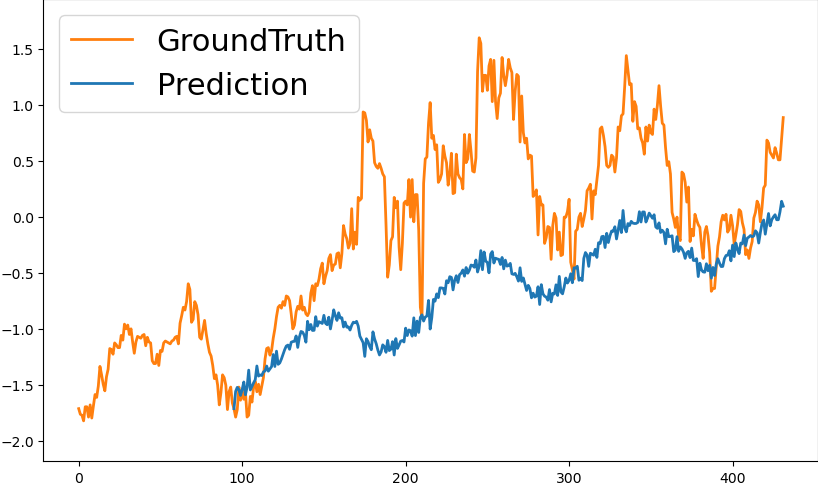}%
    }
    \hfill % Horizontal space between figures
    \subfloat[\textbf{TimeMixer++}]{%
        \includegraphics[width=0.33\textwidth]{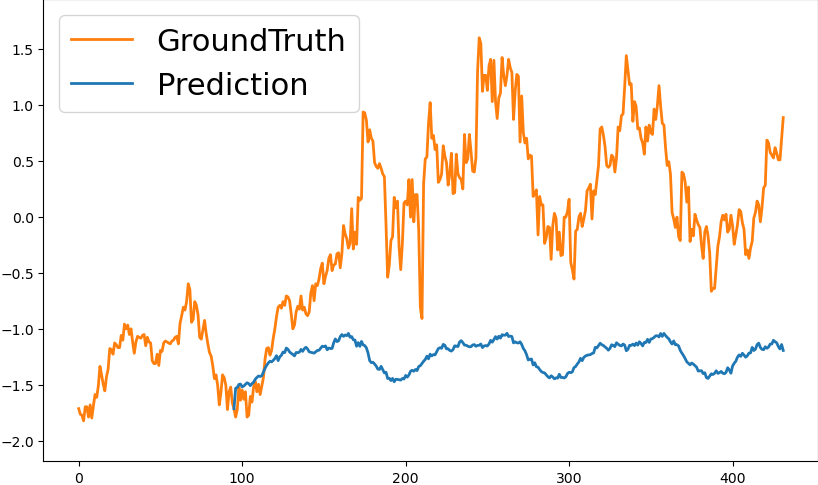}%
    }
    \hfill
    \subfloat[\textbf{iTransformer}]{%
        \includegraphics[width=0.33\textwidth]{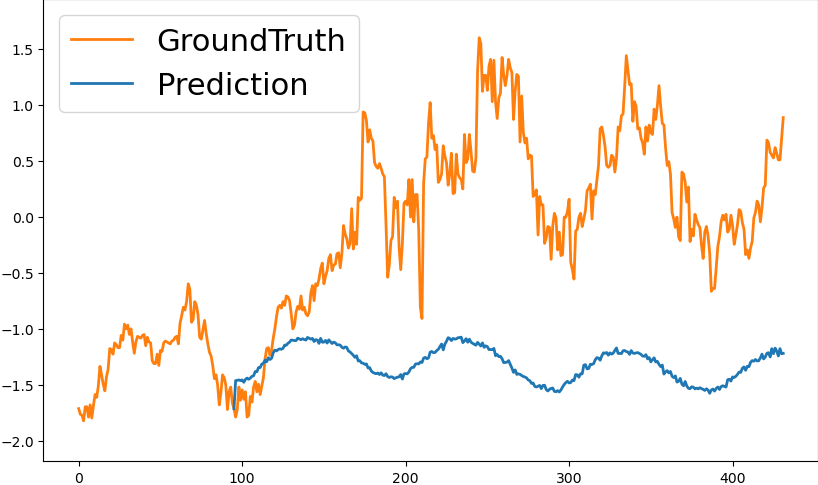}%
    }
    
    \vspace{0.5cm} % Vertical space between rows
    
    % --- Row 2 ---
    \subfloat[\textbf{PatchMLP}]{%
        \includegraphics[width=0.33\textwidth]{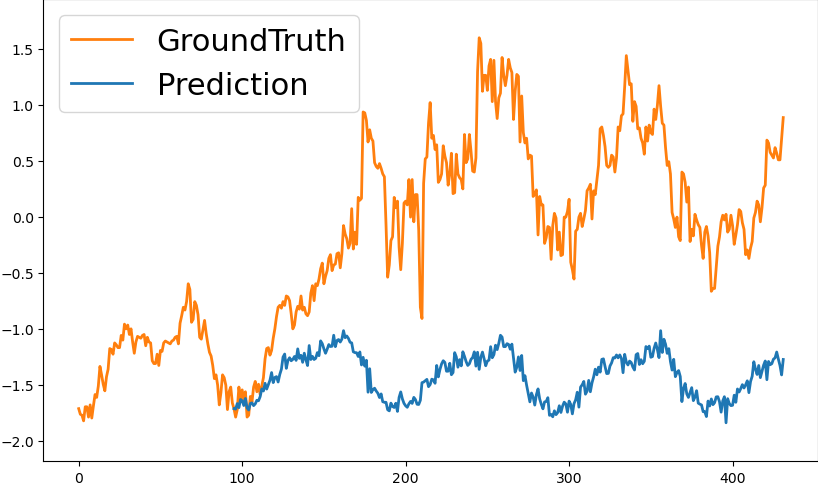}%
    }
    \hfill
    \subfloat[\textbf{PatchTST}]{%
        \includegraphics[width=0.33\textwidth]{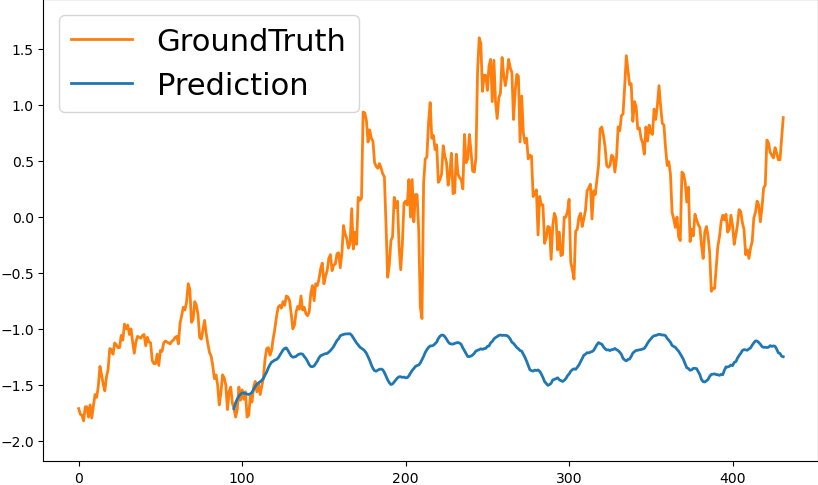}%
    }
    \hfill
    \subfloat[\textbf{DLinear}]{%
        \includegraphics[width=0.33\textwidth]{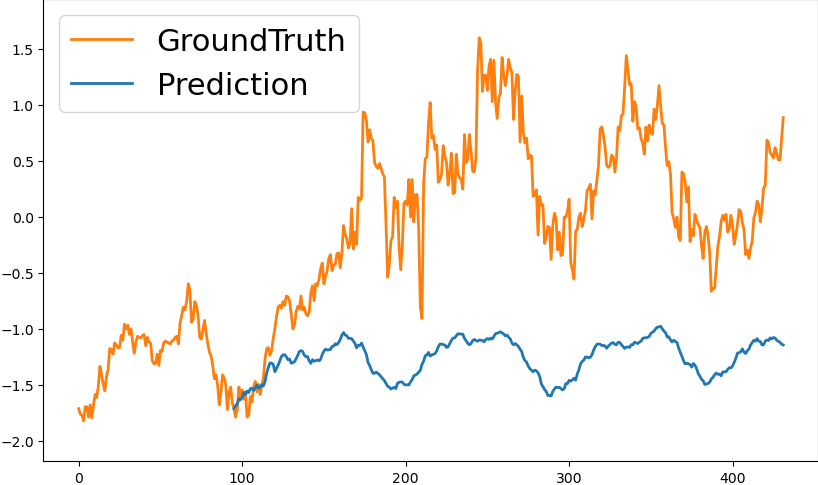}%
    }
    
    \caption{Visualization of input-96-predict-336 results on the ETTm2 dataset.}
    \label{fig:ettm2_336}
\end{figure*}

% ETTm2_96_720
\begin{figure*}[h!]
    \centering
    
    % --- Row 1 ---
    \subfloat[\textcolor{red}{\textbf{CometNet (Ours)}}]{%
        \includegraphics[width=0.33\textwidth]{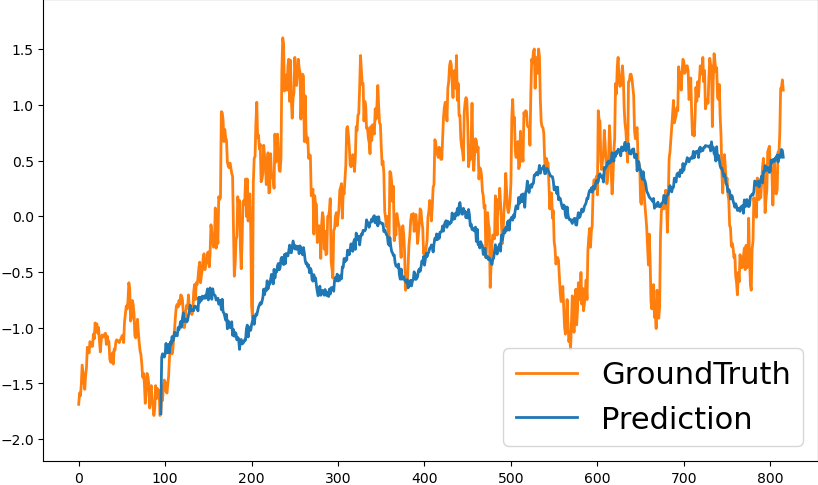}%
    }
    \hfill % Horizontal space between figures
    \subfloat[\textbf{TimeMixer++}]{%
        \includegraphics[width=0.33\textwidth]{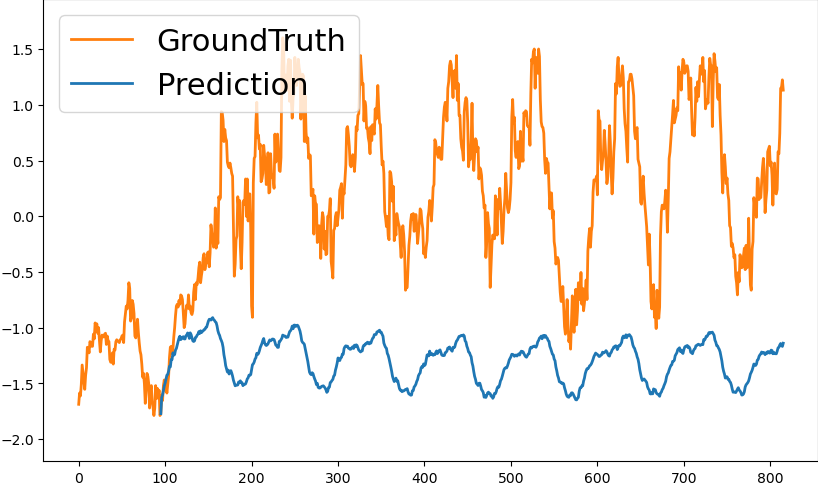}%
    }
    \hfill
    \subfloat[\textbf{iTransformer}]{%
        \includegraphics[width=0.33\textwidth]{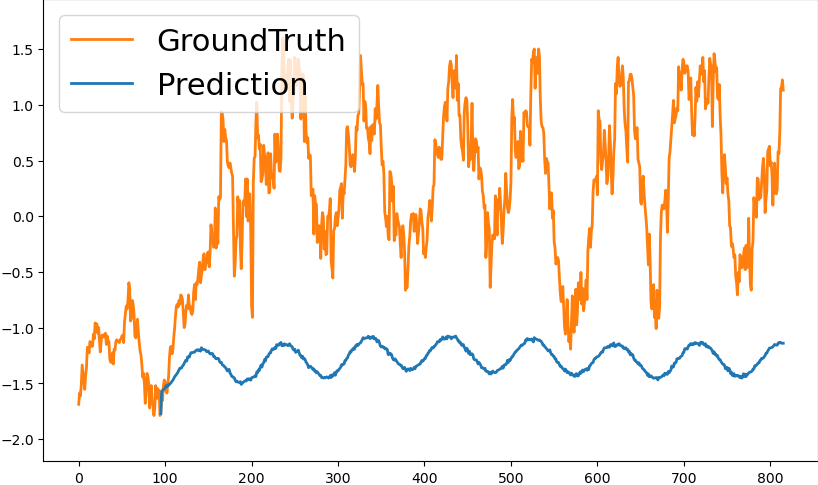}%
    }
    
    \vspace{0.5cm} % Vertical space between rows
    
    % --- Row 2 ---
    \subfloat[\textbf{PatchMLP}]{%
        \includegraphics[width=0.33\textwidth]{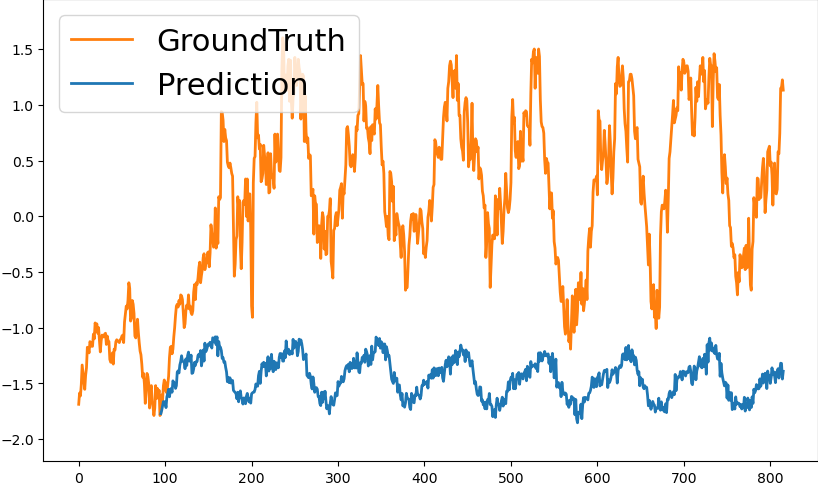}%
    }
    \hfill
    \subfloat[\textbf{PatchTST}]{%
        \includegraphics[width=0.33\textwidth]{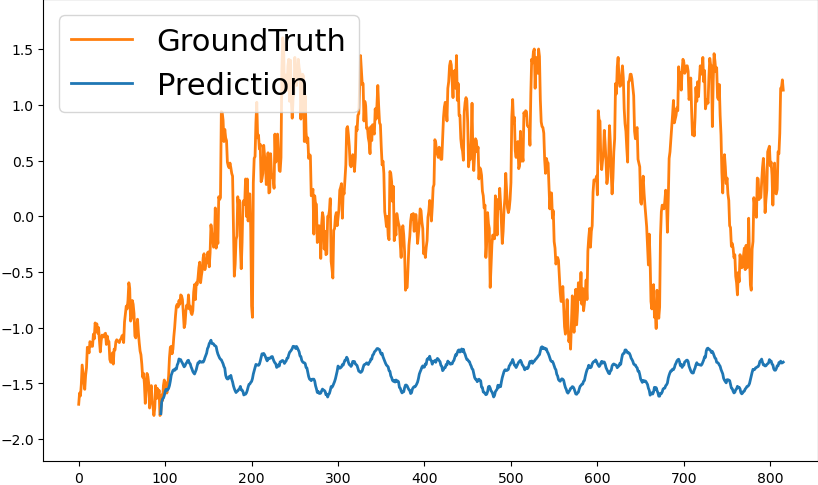}%
    }
    \hfill
    \subfloat[\textbf{DLinear}]{%
        \includegraphics[width=0.33\textwidth]{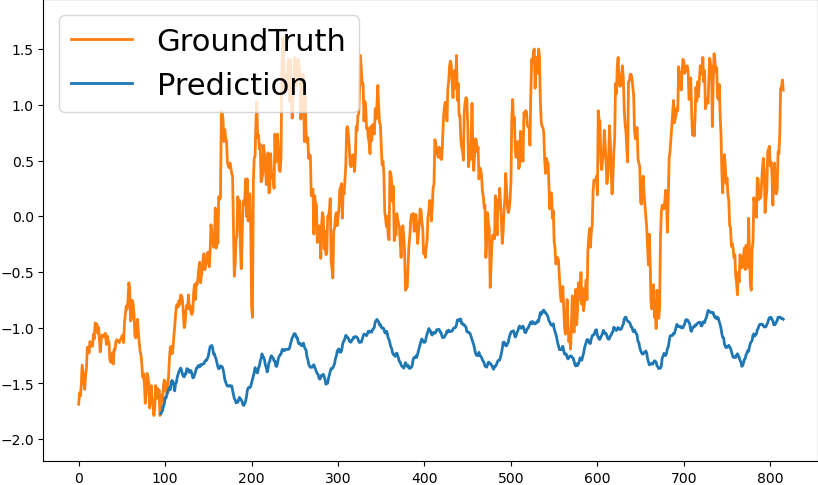}%
    }
    
    \caption{Visualization of input-96-predict-720 results on the ETTm2 dataset.}
    \label{fig:ettm2_720}
\end{figure*}

\subsection{Prediction Results Visualization}
\label{appendix:prediction_visualization}
To provide a qualitative and intuitive comparison of forecasting performance, we present a comprehensive set of prediction showcases in Fig.~\ref{fig:etth1_96} through Fig.~\ref{fig:ettm2_720}. These figures visualize the prediction results of CometNet against five strong baselines---TimeMixer++~\cite{wang2025timemixer++}, iTransformer~\cite{liu2024itransformer}, PatchMLP~\cite{tang2025patchmlp}, PatchTST~\cite{nie2023patchtst}, and DLinear~\cite{zeng2023transformers},across all four ETT datasets and for various forecast horizons, ranging from 96 to 720 steps.

A consistent pattern emerges from these visualizations. While most models perform reasonably well on shorter horizons (e.g., 96 steps), their ability to capture the true underlying dynamics of the time series degrades significantly as the prediction length increases. Baseline models often produce flattened, simplistic forecasts that miss crucial turning points and underestimate the volatility of the ground truth, a clear manifestation of the receptive field bottleneck. In contrast, CometNet consistently generates predictions that more accurately follow the long-term trend and complex variations of the actual data, even at the most challenging 720-step horizon. For instance, as seen in Fig.~\ref{fig:etth1_720} and Fig.~\ref{fig:etth2_720}, our model successfully anticipates the overall upward or downward trends, whereas other models' predictions tend to lag or diverge. This visual evidence strongly corroborates our quantitative results, highlighting CometNet's superior capability in long-term forecasting, which we attribute to its effective leveraging of contextual motifs.

\end{document}